\documentclass[apj]{emulateapj}

\newcommand{\Msun}{\mbox{\,$M_{\odot}$}}

\def\spose#1{\hbox to 0pt{#1\hss}}
\def\simlt{\mathrel{\spose{\lower 3pt\hbox{$\mathchar"218$}}
     \raise 2.0pt\hbox{$\mathchar"13C$}}}
\def\simgt{\mathrel{\spose{\lower 3pt\hbox{$\mathchar"218$}}
     \raise 2.0pt\hbox{$\mathchar"13E$}}}
\slugcomment{Accepted for publication in ApJ}
\usepackage{epsf}
\usepackage{graphicx}
\usepackage{natbib}
\usepackage{url}
\font\smcap=cmcsc10

\newcommand{\kms}{\,km~s$^{-1}$}

\newcommand{\nai}{Na\,{\smcap i}}
\newcommand{\caii}{Ca\,{\smcap ii}}

\newcommand{\vio}{$(V-I)_0$}
\newcommand{\ivi}{($I,\,V-I$)}

\newcommand{\feh}{$\rm[Fe/H]$}
\newcommand{\afe}{$\rm[\alpha/Fe]$}
\newcommand{\fehp}{$\rm[Fe/H]_{phot}$}
\newcommand{\fehs}{$\rm[Fe/H]_{CaT}$}

\newcommand{\olkhd}{$\langle L_i\rangle$}

\newcommand{\rproj}{$R_{\rm proj}$}

\newcommand{\sigvkcc}{$\sigma_{v,\rm kcc}$}

\shorttitle{Metallicity Profile of M31's Stellar Halo}
\shortauthors{Gilbert et~al.}

\begin{document}
\bibliographystyle{apj}

\title{
Global Properties of M31's Stellar Halo from the SPLASH Survey: II. Metallicity Profile\footnotemark[*] }

\footnotetext[*]{The data presented herein were obtained at the W.M. Keck Observatory,
which is operated as a scientific partnership among the California
Institute of Technology, the University of California and the National
Aeronautics and Space Administration. The Observatory was made
possible by the generous financial support of the W.M. Keck
Foundation.}

\author{
Karoline~M.~Gilbert\altaffilmark{1,2},
Jason~S.~Kalirai\altaffilmark{1,2},
Puragra~Guhathakurta\altaffilmark{3},
Rachael~L.~Beaton\altaffilmark{4},
Marla~C.~Geha\altaffilmark{5},
Evan~N.~Kirby\altaffilmark{6},
Steven~R.~Majewski\altaffilmark{4},
Richard~J.~Patterson\altaffilmark{4},
Erik~J.~Tollerud\altaffilmark{5,7},
James~S.~Bullock\altaffilmark{8},
Mikito~Tanaka\altaffilmark{9},
Masashi~Chiba\altaffilmark{9}
}

\email{
kgilbert@stsci.edu}

\altaffiltext{1}{Space Telescope Science Institute, Baltimore, MD 21218, USA}
\altaffiltext{2}{Center for Astrophysical Sciences, Johns Hopkins University, Baltimore, MD, 21218}
\altaffiltext{3}{UCO/Lick Observatory, Department of Astronomy \&
Astrophysics, University of California Santa Cruz, 1156 High Street, Santa
Cruz, California 95064.}
\altaffiltext{4}{Department of Astronomy, University of Virginia, PO Box~400325, Charlottesville, VA 22904-4325.}
\altaffiltext{5}{Astronomy Department, Yale University, New Haven, CT 06520, USA}
\altaffiltext{6}{California Institute of Technology, 1200 E.\ California Blvd., MC 249-17, Pasadena, CA 91125, USA}
\altaffiltext{7}{Hubble Fellow.}
\altaffiltext{8}{Center for Cosmology, Department of Physics and Astronomy, University of California at Irvine, Irvine, CA, 92697, USA.}
\altaffiltext{9}{Astronomical Institute, Tohoku University, Aoba-ku, Sendai 980-8578, Japan.}

\begin{abstract}
We present the metallicity distribution of red giant branch (RGB) stars in M31's stellar halo, derived from photometric metallicity estimates for over 1500 spectroscopically confirmed RGB halo stars.  The stellar sample comes from 38 halo fields observed with the Keck/DEIMOS spectrograph, ranging from 9 to 175~kpc in projected distance from M31's center, and includes 52 confirmed M31 halo stars beyond 100~kpc.   
While a wide range of metallicities is seen throughout the halo, the metal-rich peak of the metallicity distribution function becomes significantly less prominent with increasing radius.  The metallicity profile of M31's stellar halo shows a continuous gradient from 9 to $\sim 100$~kpc, with a magnitude of $\sim -0.01$~dex kpc$^{-1}$.  The stellar velocity distributions in each field are used to identify stars that are likely associated with tidal debris features.   The removal of tidal debris features does not significantly alter the metallicity gradient in M31's halo: a gradient is maintained in fields spanning 10 to 90 kpc.   We analyze the halo metallicity profile, as well as the relative metallicities of stars associated with tidal debris features and the underlying halo population, in the context of current simulations of stellar halo formation.  We argue that the large scale gradient in M31's halo implies M31 accreted at least one relatively massive progenitor in the past, while the field to field variation seen in the metallicity profile indicates that multiple smaller progenitors are likely to have contributed substantially to M31's outer halo.  

\end{abstract}

\keywords{galaxies: halo --- galaxies: individual (M31) --- stars: kinematics --- techniques:
spectroscopic}

\section{Introduction}\label{sec:intro}
 
\setcounter{footnote}{9}

The sparse stellar environments and long mixing times of galactic halos provide an opportunity to infer a galaxy's past merger history from observations of its current state.  Stellar halos are at least partially comprised of stars stripped from satellite galaxies during their infall and disruption.  Simulations of the build-up of stellar halos of spiral galaxies such as the Milky Way and Andromeda (M31) predict a significant spread in properties should be observed across stellar halos, since the structure of a stellar halo is determined by the galaxy's individual merger history \citep{bullock2005,font2006,font2008,johnston2008,zolotov2009,zolotov2010,cooper2010,font2011}.

Observations of the stellar halos of the Milky Way and M31 illustrate the influence of individual merger histories on the properties of stellar halos.  The halos of both galaxies show ample evidence of the tidal disruption of smaller systems in the form of abundant substructures, such as streams and shells \citep[e.g.,][]{ibata1994,ibata2001,ferguson2002,majewski2003,yanny2003,ibata2005}.
However, despite the similar overall size and morphology of the Milky Way and M31 there are several notable differences in the properties of their stellar halos.  

Early work on the inner region of M31's stellar halo found that M31's spheroid had a stellar density $\sim 10\times$ higher \citep{reitzel1998} and was more metal-rich than a comparable location in the Milky Way's halo \citep[$10\times$ more metal-rich at $R\sim 7$~kpc;][]{mould1986,durrell1994,rich1996}.  Later studies confirmed this result, and showed no evidence of a gradient in metallicity in M31's stellar halo out to $R=30$~kpc \citep{durrell2001,durrell2004,bellazzini2003}.  Moreover, in early studies the surface brightness profile of the metal-rich inner spheroid was found to be well fit as an extension of M31's bulge, with an $R^{1/4}$ law or Sersic profile \citep{pritchet1994,durrell2004,irwin2005}, in contrast to the power-law density profile found in the Milky Way's halo \citep[e.g.,][]{yanny2000,juric2008,carollo2010,ivezic2012,feltzing2013}. 

Large surveys of M31's stellar halo have since characterized its properties over drastically larger distances.  The SPLASH (Spectroscopic and Photometric Landscape of Andromeda's Stellar Halo) survey has obtained photometric and spectroscopic observations of lines of sight in all quadrants of M31.  The range of projected distance from M31's center of the fields is \rproj\,$\sim 4$ to 225~kpc, and they target M31's inner regions, smooth halo fields, tidal debris features, and dwarf galaxy population.  The PAndAS survey \citep[Pan-Andromeda Archaeological Survey;][]{mcconnachie2009} has significantly extended a previous imaging survey \citep{ibata2001a,ibata2007} to obtain contiguous photometric observations of M31's stellar halo to 
projected distances of $\sim 150$~kpc in all directions, including an extension to M33, as well as spectroscopy in isolated lines of sight in M31's disk, halo, halo substructures, and dwarf galaxies.  Early results from these teams found the first evidence of a spatially extended, low-metallicity stellar halo in M31 \citep{guhathakurta2005,irwin2005,kalirai2006halo,chapman2006}, similar to that seen in the MW, and also found evidence of an extended, metal-rich disk structure \citep{ibata2005}.

Analysis of the extended surface brightness profile of M31's halo has found that the outer regions are well described by a power-law with an index in the range $-1.75$ to $-2.5$ \citep{guhathakurta2005,irwin2005,ibata2007,tanaka2010,courteau2011,gilbert2012,ibata2014}, with an extent out to at least $175$~kpc \citep{gilbert2012}.  \citet{dorman2013} performed a structural decomposition of M31's surface brightness distribution over the largest radial range to date, $4 <$\rproj$< 225$~kpc, using a compilation of datasets including stellar kinematics from SPLASH, HST imaging of M31's bulge and disk from the PHAT survey \citep{dalcanton2012}, and $I$-band imaging.  They found that M31's stellar halo can be described as a cored power-law over its full extent.  \citet{dorman2013} found no structural distinction between the `inner' and `outer' halo, as had been inferred from most previous analyses.  

The literature has been divided over whether a metallicity gradient exists in M31's stellar halo.  Most existing studies either cover limited radial ranges or are based on small numbers of fields.   In addition to limited spatial sampling, measurements of the metallicity profile have been complicated by the prominence of the Giant Southern Stream, an extensive tidal debris structure that covers a large portion of the south quadrant of M31's stellar halo and much of the inner regions \citep{ibata2001,fardal2007,gilbert2007,fardal2012}.  

The studies of \citet{kalirai2006halo}, \citet{chapman2006}, and \citet{koch2008} each analyzed samples of spectroscopically confirmed M31 stars obtained with multi-object spectroscopy using the DEIMOS instrument on the Keck~II telescope.  \citet{kalirai2006halo} and \citet{koch2008} both found evidence of a gradient in M31's stellar halo.  The \citet{kalirai2006halo} work analyzed 27 multi-object slitmasks ranging from 12 to 165 kpc in projection along the southeast minor axis and in the southern quadrant of M31's stellar halo, and the \citet{koch2008} work analyzed 23 of the \citet{kalirai2006halo} multi-object slitmasks plus an additional thirteen multi-object slitmasks in the same region.  \citet{chapman2006} measured the mean metallicity from stacked spectra of halo stars in 23 multi-object slitmasks, primarily within \rproj\,$\le 40$~kpc of M31's center and near M31's major axis, but also including two masks near the minor axis at \rproj\,$\sim 60$~kpc.  They used the velocity of the stars to remove M31 disk stars from the sample, using the results of \citet{ibata2005}.  They observed a significant population of metal-poor halo stars in their fields, but did not see compelling evidence of a gradient in metallicity with radius.   

The studies of \citet{richardson2009}, \citet{tanaka2010}, and \citet{ibata2014} analyzed photometric datasets in M31's stellar halo.  \citet{richardson2009} analyzed {\it Hubble Space Telescope}/Advanced Camera for Surveys images of nine lines of sight from $20\le$\,\rproj\,$\le 60$~kpc, spread over all four quadrants of M31's stellar halo; they also did not find evidence of a gradient.   Meanwhile, the Subaru/Suprime-Cam survey presented by \citet{tanaka2010}, which imaged mostly contiguous fields from 20 to 90 kpc on the southeast minor axis and from 15 to 100 kpc on the northwest minor axis, found evidence of a gradient on the northwest minor axis but not on the southeast minor axis, which is crossed by multiple tidal debris features (Figure~\ref{fig:roadmap}).    
Most recently, \citet{ibata2014} utilized the full PAndAS imaging survey to investigate the structural properties of M31's stellar halo.  They found that the full PAndAS dataset supported a metallicity gradient in M31's stellar halo, declining from \feh\,$=-0.7$ at \rproj\,$=30$~kpc to \feh\,$=-1.5$ at \rproj\,$=150$~kpc.  However, due to severe contamination from foreground MW stars, they had to impose color cuts on the photometric sample that excluded most of the metal-rich (\feh\,$>-0.5$) RGB stars, and also excluded a fraction of stars with $-1.0<$\,\feh\,$<-0.5$.  Thus the metal-rich end of the metallicity distribution function had to be estimated from isochrone modeling.  Each of the above analyses used different methods to identify M31 halo stars and to measure metallicities, precluding a simple compilation of the previously published data.  

In this paper, we investigate the metallicity profile of M31's stellar halo using our extensive spectroscopic survey.  The first paper of this series \citep{gilbert2012} analyzed the surface brightness profile of M31's stellar halo using counts of spectroscopically confirmed M31 RGB stars in 38 fields spaced throughout M31's stellar halo.  Spectra in each field were obtained with the DEIMOS spectrograph on the Keck~II telescope as part of SPLASH.  The fields cover a range of projected distances of 9\,--\,175~kpc from M31's center, target all four quadrants of M31's halo, and include lines of sight targeting tidal debris features and dwarf satellite galaxies as well as `smooth' halo fields.  \citet{gilbert2012} found that M31's halo followed a power-law profile with power-law index $-2.2\pm 0.2$ to at least 175~kpc \citep[$\sim 2/3$ of M31's estimated virial radius;][]{seigar2008}.   We identified tidal debris features in half of the fields within 90~kpc of M31's center, and used the kinematical distribution of M31 RGB stars to statistically subtract stars associated with tidal debris features in individual fields.  This allowed us to analyze the aggregate effect of tidal debris on the surface brightness profile.    

In this contribution, we analyze the metallicity distribution of the stellar halo RGB stars presented by \citet{gilbert2012} and confirm a metallicity gradient in M31's stellar halo, as found by \citet{kalirai2006halo} and \citet{koch2008}. 
The \citet{kalirai2006halo} work verified the previous metal-rich results in the inner regions of the galaxy (e.g., \feh $= -0.5$ at $R < 20$\,--\,30~kpc) and showed that the population is almost $10\times$ more metal-poor in the outskirts of the halo (\feh $= -1.5 \pm 0.1$ at $R\sim 150$~kpc).  Those results were based on 261 spectroscopically confirmed M31 stars in twelve fields, including 47 stars beyond 60~kpc and just 4 stars beyond 100~kpc.  \citet{koch2008} also found a significant metallicity gradient in M31's halo, based on an extension of the \citet{kalirai2006halo} dataset as discussed above.   
In this paper, we characterize the change in the metallicity distribution of M31 halo stars as a function of distance, based on a much larger data set with over 1500 spectroscopically confirmed M31 halo stars in 38 fields (106 multi-object slit masks).\footnote{The fields presented here include 34 of the 36 masks in \citet{koch2008} (and by extension, the 27 masks in \citet{kalirai2006halo}).} 
 Notably, we now have a sample of 206 spectroscopically confirmed M31 halo stars beyond 60~kpc, 52 of which are beyond 100 kpc.  

The paper is organized as follows.  Section \ref{sec:data} summarizes the data reduction and sample selection.  Section \ref{sec:met_est}  describes the photometric and spectroscopic \feh\ estimates.  Section \ref{sec:met_halo} presents the metallicity distribution functions (MDFs) and metallicity profile of stars in M31's halo.  Section~\ref{sec:litcompare} discusses our results in the context of previous estimates of the metallicity of M31's halo.  Section~\ref{sec:sims} interprets the results within the context of our current understanding of stellar halo formation.  Section \ref{sec:subst_v_smooth} analyzes the metallicities of stars associated with tidal debris features in relation to stars associated with M31's spatially diffuse, dynamically hot spheroid.  Finally, Section \ref{sec:conclusion} summarizes our conclusions.   

A distance modulus of 24.47 is assumed for all conversions of
angular to physical units 
\citep[corresponding to a distance to M31 of 783~kpc;][]{stanek1998}.  Unless stated otherwise, all distances from M31's center refer to projected distance.
 
\section{Data}\label{sec:data}

\begin{figure}[tb!]
\plotone{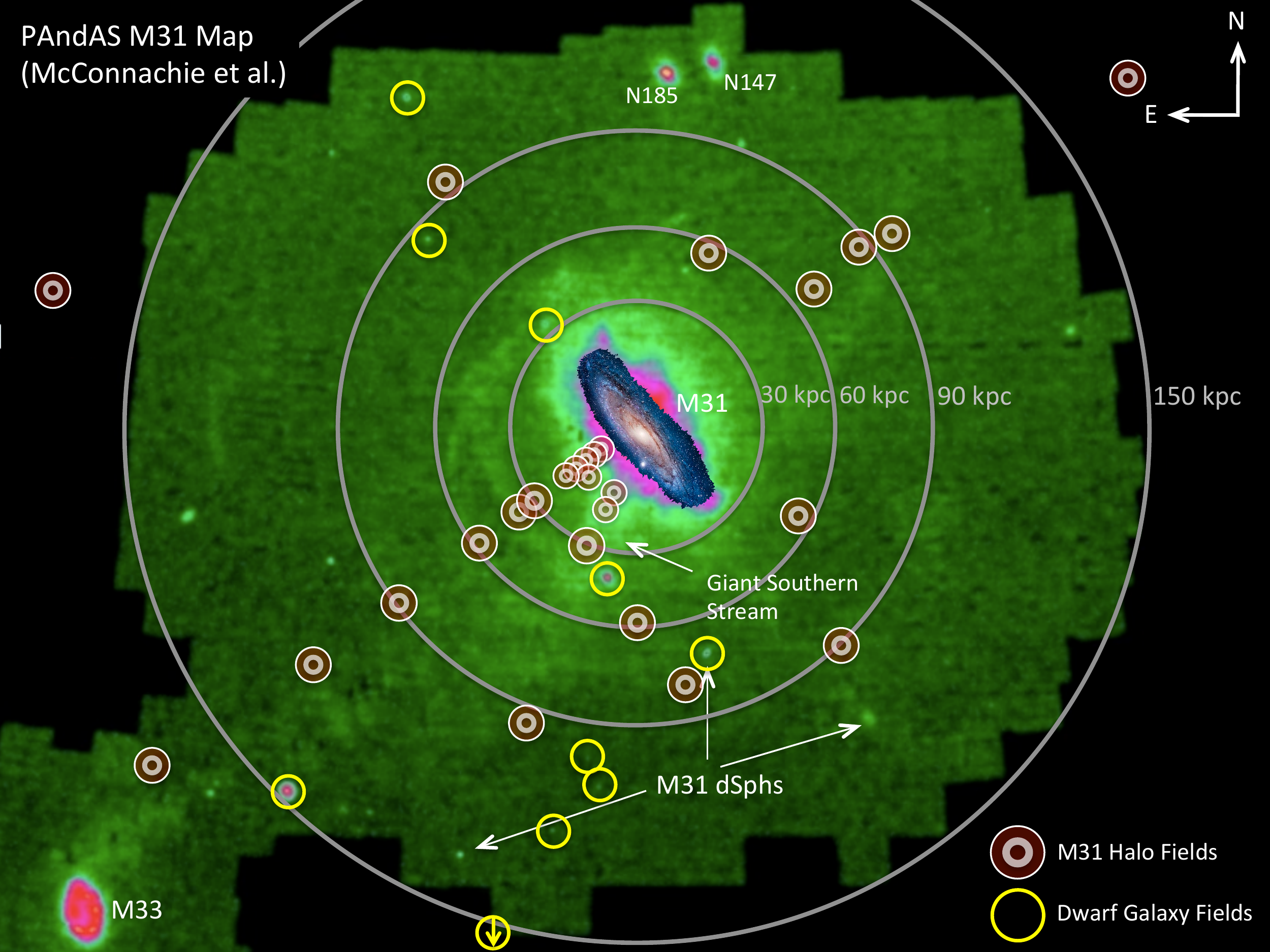}
\caption{
Locations of the spectroscopic fields overlaid on the PAndAS starcount map \citep{mcconnachie2009}.  Spectroscopic fields designed primarily to target M31's dwarf galaxies are denoted by open yellow circles; the remaining fields target M31's halo.  Exact locations and orientations of the spectroscopic slitmasks can be found in Figure~1 of \citet{gilbert2012}. 
}
\label{fig:roadmap}
\end{figure}

The data analyzed in this paper come from observations of 
106 multi-object spectroscopic slitmasks spread over 
38 fields [PIs J.~S.~Bullock, P. Guhathakurta, and R.~M.~Rich (21 masks)], 
which span a large range in azimuth and 
projected distance from the center of M31 
(Figure~\ref{fig:roadmap}).
Below we briefly summarize the photometric (Section~\ref{sec:phot}) and 
spectroscopic (Section~\ref{sec:spec}) observations and data reduction and 
the identification of the M31 RGB star sample 
(Section~\ref{sec:cleansample}).   
A more detailed description of the data reduction and sample selection techniques were 
presented by \citet{gilbert2012} and references therein.

\subsection{Photometric Observations}\label{sec:phot}

The photometric data came from several sources, and were obtained in multiple
filter sets \citep[Table 1;][]{gilbert2012}. 
The majority of the fields  (\rproj\,$>30$~kpc) utilized observations taken 
with the Mosaic Camera on the Kitt Peak National Observatory (KPNO) 
4-m Mayall telescope.\footnote{Kitt Peak National
Observatory of the National Optical Astronomy Observatory is operated by the
Association of Universities for Research in Astronomy, Inc., under cooperative
agreement with the National Science Foundation.}
Images were obtained in the Washington 
system $M$ and $T_2$ filters and the intermediate-width DDO51 filter 
\citep{ostheimer2003,beaton2014aas}.  The photometric calibration for each field was solved iteratively, and is based on  observations of Washington system standard fields SA98, SA110, and SA114 \citep{geisler1990}.  The root-mean-square errors from the photometric calibration are included in the final photometric error for each star, but are a very small fraction of the total photometric error ($\sim  2$\%). 
The photometrically calibrated $M$ and $T_2$ magnitudes were transformed
to Johnson-Cousins  $V$ and $I$ magnitudes using the transformation equations of 
\citet{majewski2000}.  The errors in the color transformation are expected to be on the order of $\sim 0.014$ magnitudes \citep{majewski2000}, which is also small fraction ($\sim 10$\%) of the typical photometric errors.    

The innermost fields (\rproj\,$<30$~kpc) utilized
observations we obtained with the MegaCam instrument on 
the 3.6-m Canada-France-Hawaii Telescope (CFHT)\footnote{MegaPrime/MegaCam 
is a joint project of CFHT and CEA/DAPNIA, 
at the Canada-France-Hawaii Telescope
which is operated by the National Research Council of Canada, the Institut
National des Science de l'Univers of the Centre National de la Recherche
Scientifique of France, and the University of Hawaii.}  \citep{kalirai2006gss}.
Images were obtained with the $g'$ and $i'$ filters.  Observed magnitudes 
were photometrically calibrated and transformed to Johnson-Cousins $V$ and $I$ magnitudes
using observations of Landolt photometric standard stars.

Finally, the photometry in field `d10' was derived from 
$V$ and $I$ images obtained with the William Herschel Telescope \citep{zucker2007}.  
Photometry for fields `streamE' and `streamF' was derived from  $V$ and $I$
images obtained with the SuprimeCam instrument on the Subaru Telescope \citep{tanaka2010}.

A comparison of the photometric measurements from the different telescopes is a useful exercise.  There are overlapping fields between the Mayall/Mosaic and Subaru/SuprimeCam imaging surveys.  \citet{tanaka2010} compared the KPNO and Subaru photometry in the overlapping fields, and found good agreement between the photometric measurements from the two surveys (see their Section 2.4.2 and Figure 7).

There is no overlap between our CFHT/MegaCam data and the Mayall/Mosaic data.  In the absence of overlapping photometric observations, ideally one would compare the CMD distributions and resulting metallicity distributions of two fields that are expected to have very similar intrinsic properties.  However, given tidal debris features, the observed metallicity gradient, and the spread in properties that is observed at any given radius in M31's halo, doing this with any two halo fields is problematic.  However, we have multiple spectroscopic masks targeting the bright eastern edge of the Giant Southern Stream: f207 and H13s (based on CFHT/MegaCam data), and a3 (based on Mayall/Mosaic data).  While the metallicity of the stream has been shown to vary across the width of the stream \citep{ibata2007,gilbert2009gss}, no significant gradient is seen along the length of the stream \citep{mcconnachie2003,gilbert2009gss}.  This comparison has already been done by \citet[][Figure 15 and Section 4.1.2]{gilbert2009gss}.   The derived metallicity distributions, based on photometric measurements (Section~\ref{sec:met_est_phot}) of stars within 2$\sigma_{\rm v}$ of the mean velocity of the Giant Southern Stream, are very similar between the three fields.  After correcting for the variation in distance along the length of the stream, the median metallicities of the Giant Southern Stream in the three fields agree to within ~0.05 dex, indicating that any systematic offsets between the photometric calibrations of the MegaCam and Mosaic photometry are not significantly affecting our results.

\subsection{Spectroscopic Observations}\label{sec:spec}
We obtained spectra 
with the DEIMOS spectrograph on the Keck~II 10-m telescope.  
The multi-object spectroscopic slitmasks were designed using the above photometric catalogs.  Stars 
with colors and magnitudes consistent with RGB stars at the distance of M31
were assigned the highest priority for inclusion on the spectroscopic masks.   When 
available, DDO51 photometry was used to prioritize stars with the highest probability
of being RGB stars: red giants occupy a distinct region in the ($M-T_2$, $M-$DDO51) color-color diagram \citep{majewski2000}.

The 1200~line~mm$^{-1}$ grating, which has a dispersion of $\rm0.33~\AA$~pix$^{-1}$, 
was used for all observations, yielding a spectral wavelength range of $\lambda\lambda\sim$~6450\,--\,9150\AA.   
The nominal exposure time per mask was one hour; slight modifications to
the exposure time were made for particularly good or poor observing conditions.  

Spectra were reduced using modified 
versions of the {\tt spec2d} and {\tt spec1d} software developed 
at the University of California, Berkeley \citep{newman2013}. 
These routines perform the standard reduction steps of
flat-fielding, night-sky emission line removal, extraction of the two-dimensional 
spectra ({\tt spec2d}) and cross-correlation of the resulting 
one-dimensional spectra with template spectra to determine redshift ({\tt spec1d}).  
We have supplemented the standard {\tt spec1d} 
template library with stellar spectra of cool stars obtained with DEIMOS \citep{simon2007}.

\subsection{Identification of M31 RGB halo stars}\label{sec:cleansample}

\begin{figure}
\plotone{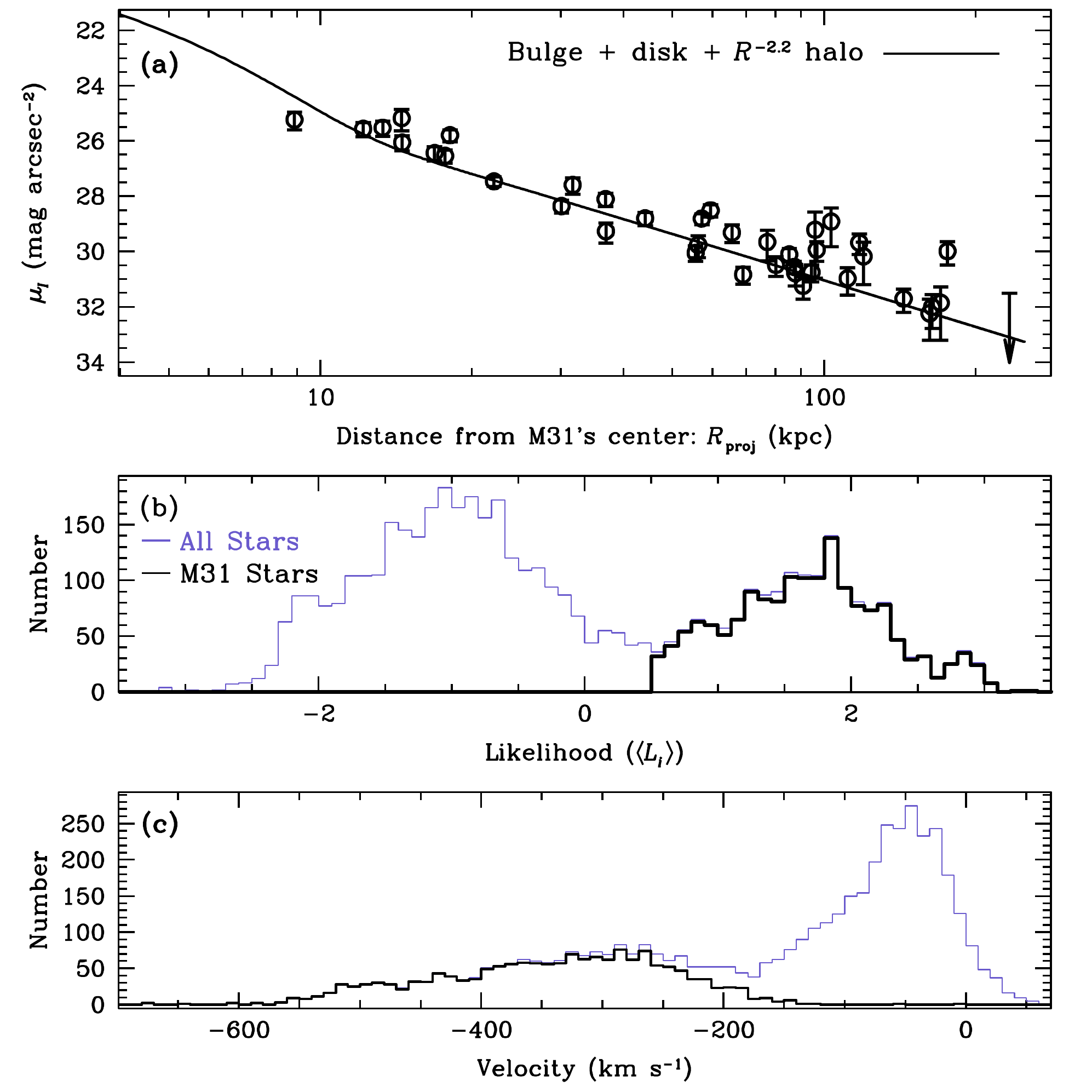}
\caption{
Properties of the M31 halo spectroscopic sample.  (a) Surface brightness profile of M31's stellar halo based on counts of spectroscopically confirmed M31 red giants \citep{gilbert2012}. 
(b) Likelihood distribution of the 4721 stars in the spectroscopic sample (blue histogram).  Stars with \olkhd $ > 0$ are more likely to be M31 red giants (dashed histogram), while stars with \olkhd $< 0$ are more likely to be MW dwarf stars; the two populations present a clear bimodal distribution in likelihood space.  The analysis presented here is restricted to stars with \olkhd $ >0.5$ (black histogram), which are 3 times more probable to be M31 RGB stars than MW dwarf stars.  (c) Velocity distribution of stellar sources in the spectroscopic sample.  Line types are the same as in panel (b).  M31's systemic velocity is $v_{\rm M31}=-300$~\kms.
}
\label{fig:sb}
\end{figure}

The stellar spectra obtained from the DEIMOS observations 
consist of both red giants at the distance of M31 and MW foreground contaminants.
Although the spectroscopic mask design was driven by stars
most likely to be M31 red giants, lower-priority targets were still placed on the masks as 
fillers.  In the remote outer regions of M31, MW dwarf stars 
along the line of sight dominate the observed stellar spectra even with
photometric preselection of probable RGB stars.  

We use the empirical diagnostics described by \citet{gilbert2006} to
separate M31 RGB stars from foreground MW dwarf star contaminants.  
The line-of-sight velocity distributions of the two populations overlap,
thus multiple photometric and spectroscopic properties are used to 
probabilistically separate the two populations.   We determine the probability an individual 
star is an M31 red giant or MW dwarf star in each of the following
empirical diagnostics: (1) line-of-sight velocity ($v_{\rm los}$), (2) photometric probability
of being a red giant based on location in the ($M-T_2$, $M-$DDO51) color-color 
diagram (when available; Table~1 of \citet{gilbert2012}), (3) the equivalent width 
of the \nai\ absorption line (surface-gravity and temperature sensitive) 
versus \vio\ color, (4) position in the \ivi\ color-magnitude diagram (CMD) and (5) spectroscopic 
(based on the EW of the \caii\ absorption line) versus photometric 
(comparison to theoretical RGB isochrones) \feh\ estimates.  We consider stars that are
3 times more probable to be M31 red giants than MW dwarfs to
be securely identified M31 RGB stars; stars that are more probable to be M31 red giants 
but with probability ratios less than 3 are considered marginally identified as M31 RGB 
stars.  The ratios of securely indentified and marginally identified M31 RGB stars and MW dwarf
stars is highly dependent on the projected distance of the field from M31's center
 \citep[see Figure~3 of][]{gilbert2012}.  For the full sample presented here, securely 
identified M31 halo stars outnumber marginally identified M31 halo stars by 
a factor of 8, and MW dwarf stars outnumber M31 halo stars by a factor of 1.6. 

\citet{gilbert2012} examined the location of our current sample of MW dwarf and M31 
RGB stars in these diagnostics and demonstrated that there are stars with high 
probabilities of being M31 RGB stars in fields as distant as 
\rproj\,$\sim 180$~kpc from M31's center.   Moreover, we argued that the level of 
MW foreground contamination in our sample of securely identified M31 red giants is small,
even in our outermost fields.  This is based on the following lines of evidence.
\begin{itemize}
\item{The distributions of stars identified as MW dwarfs and M31 red giants are centered in distinct regions of the \feh\,--\,$v_{\rm los}$ plane at all radii.}
\item{There is no trend in M31 surface brightness with Galactic latitude in our fields.  If MW disk stars significantly contaminate our M31 sample, fields closer to the Galactic disk would have systematically brighter surface brightness estimates.}
\item{The surface brightness estimates of our M31 fields continue to decrease with increasing projected distance from M31.  This indicates that our M31 sample does not reach the regime of being foreground-limited by misidentified MW stars.}
\item{Distant turn-off stars in the MW halo have relatively blue colors (\vio$<1$) in our spectroscopic sample and span a large range of velocities.  Stars identified as M31 red giants and blue MW stars have distinct kinematical distributions at all radii: the blue MW stars remain consistent with being centered at the Local Standard of Rest, while the M31 sample remains centered at M31's systemic velocity.  For individual fields, the surface brightness computed using M31 stars with $v\le -300$~\kms\ (M31's systemic velocity) is consistent with the surface brightness computed using M31 stars with $v>-300$~\kms.}
\end{itemize}  
Figure~\ref{fig:sb} displays the surface brightness profile of M31's halo based on counts of securely identified M31 RGB in our M31 fields, as well as the likelihood and line-of-sight velocity distributions of all the stars in our spectroscopic dataset.

\subsubsection{Separation of M31 Halo Stars From Dwarf Galaxy Members}\label{sec:andsatstars}
A significant portion of the outer halo spectroscopic masks targeted
dSph satellites of M31 \citep{kalirai2009,kalirai2010,tollerud2012}.  
We follow the method used by \citet{gilbert2009gss}
to identify stars in the stellar halo of M31 from stars bound to the
dwarf satellite galaxies.  Our goal is a clean (not necessarily  
complete) sample of M31 halo stars from each mask.  While completeness is a concern for studies of the stellar density, for analyzing 
metallicity distributions we are more concerned with 
getting a representative sample of M31 stars that is not biased in metallicity from contamination by dSph stars.  

We take advantage of the fact that the dwarf galaxies are compact, and therefore
typically cover only a portion of the spectroscopic slitmask, have
small velocity dispersions, and have stars that span a limited range of
[Fe/H].  Stars classified as RGB stars using the method described in
Section~\ref{sec:cleansample} that are well outside the King 
limiting radius of the dSph or well removed
from the locus occupied by dSph stars in velocity and \feh\ are 
classified as M31 halo stars.   This method
would classify any extratidal dwarf galaxy stars as M31 halo stars, however 
such stars are assumed to be rare in our sample. Indeed, the number of
stars outside the King limiting radius of the dSphs but near the locus of 
dSphs in velocity and \feh\ is small, comprising only $\sim 5$\% of the M31 halo stars
in the dSph fields. 
Moreover, it could be argued 
that extratidal dwarf galaxy stars should be counted as part of the M31 halo 
population.  
We may also remove some bonafide M31 halo stars
from our sample that happen to fall simultaneously within the limiting radius of the dSph, and 
within the velocity, and metallicity spread of the dSph stars.  However, we estimate that even in 
dSph fields with the highest halo densities,
our method of removing dSph members results in the removal of at most two or three M31 halo stars 
from the M31 sample.

\section{Chemical Abundance Measurements}\label{sec:met_est}
The final M31 sample consists of stars that are more than three times as likely to be M31 RGB stars than foreground MW dwarf stars, based on a set of five photometric and spectroscopic diagnostics (Section~\ref{sec:cleansample}).  Furthermore, in fields that 
target dwarf galaxies in M31's halo, we remove stars that are likely bound to the dwarf galaxy (Section~\ref{sec:andsatstars}).
This results in a sample of more than 1600 
M31 halo stars.  
We measure the iron abundance (\feh) for each confirmed M31 halo star with independent photometric and spectroscopic methods.   Photometric abundance measurements are made by comparison of the colors 
and magnitudes of the M31 stars within a grid of theoretical stellar isochrones of different metallicities, 
assuming a single age and $\alpha$-enhancement.  
Spectroscopic abundance measurements are made 
by transforming
the equivalent widths of the 
\caii\ triplet lines 
to an iron abundance using an empirical relation.  Below, we discuss 
each method and compare the results from the independent abundance measurements.

\subsection{Photometric Metallicities}\label{sec:met_est_phot}
\begin{figure}[tb!]
\plotone{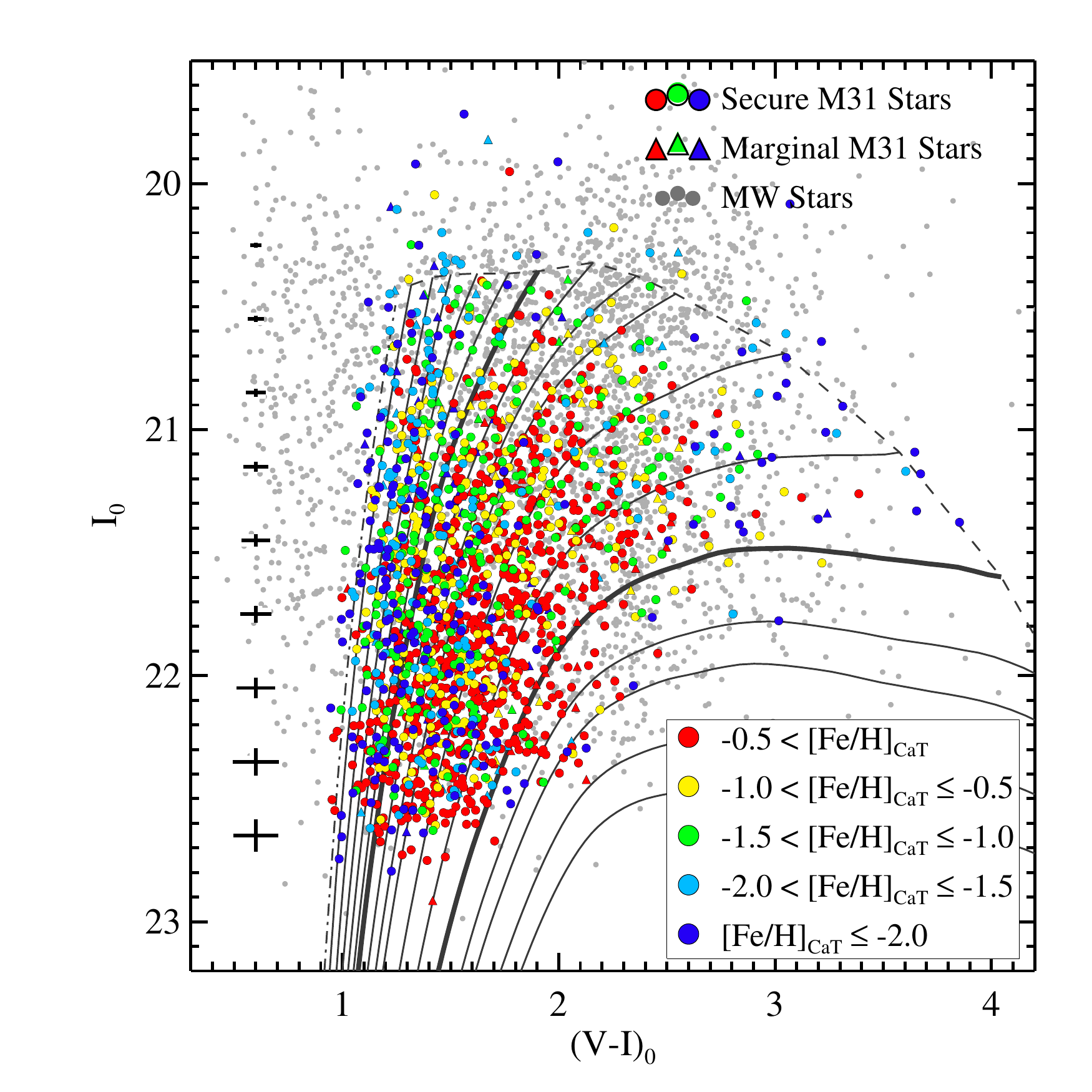}
\caption{  
Color-magnitude diagrams of spectroscopic targets with radial velocity measurements, color-coded according to spectroscopic metallicity measurements based on the equivalent width of the \caii\ triplet.  Typical photometric errors as a function of $I$ magnitude are shown on the left.   Grey curves represent isochrones for $t=10$~Gyr and \afe$=0.0$, with the most metal-poor model having \feh$=-2.3$ (left) and the most metal-rich model having \feh$=0.49$ (right).  The bold isochrones have metallicities of one-tenth Solar (\feh$=-1.0$) and Solar (\feh$=0.0$).  
Membership determination is performed using the \citet{gilbert2006} diagnostic method, summarized in Section~\ref{sec:cleansample}.  
}
\label{fig:cmd}
\end{figure}

We measure \fehp\ by comparing the CMD position of each M31 red giant with a fine grid of stellar isochrones (Figure~\ref{fig:cmd}).  We adopt the \citet{vandenberg2006} stellar evolution models as our standard isochrone set.   They contain 20 separate isochrones between the bounds of $-2.3\le$\feh$\le0.0$ and four super-solar metallicity isochrones that extend the grid up to \feh\,$=0.49$.  Our nominal \fehp\ estimates are computed assuming an age of 10~Gyr and \afe\,$=0.0$, although we explore the effect of assuming younger ages or alpha enrichment below.    

When measuring photometric metallicities, we exclude stars that are beyond the bounds of the isochrone grid (brighter than the tip of the RGB or bluer than the most metal-poor isochrone).   This reduces the size of the sample used to measure photometric metallicities by 65 stars.  Stars brighter than the tip of the RGB but classified by the diagnostics as M31 stars could be M31 halo stars closer than M31's nominal distance, or asymptotic giant branch stars in M31's halo. 
Stars bluer than the isochrone grid that are classified by the diagnostics as M31 stars could be very metal-poor halo red giants (\feh\,$<-2.3$) or red giants whose photometry is scattered blueward of the isochrones due to photometric errors (Figure~\ref{fig:cmd}).  Alternatively, they could fall blueward of the isochrone grid due to modeling errors affecting the color and magnitude of the isochrones.  Some small fraction of these stars could also be asymptotic giant branch stars.   In Section~\ref{sec:baises_bluestars}, we will assess the effect of removing the bluest stars in the sample on the measured metallicity gradient.  

\begin{figure}[tb!]
\includegraphics[width=0.5\textwidth]{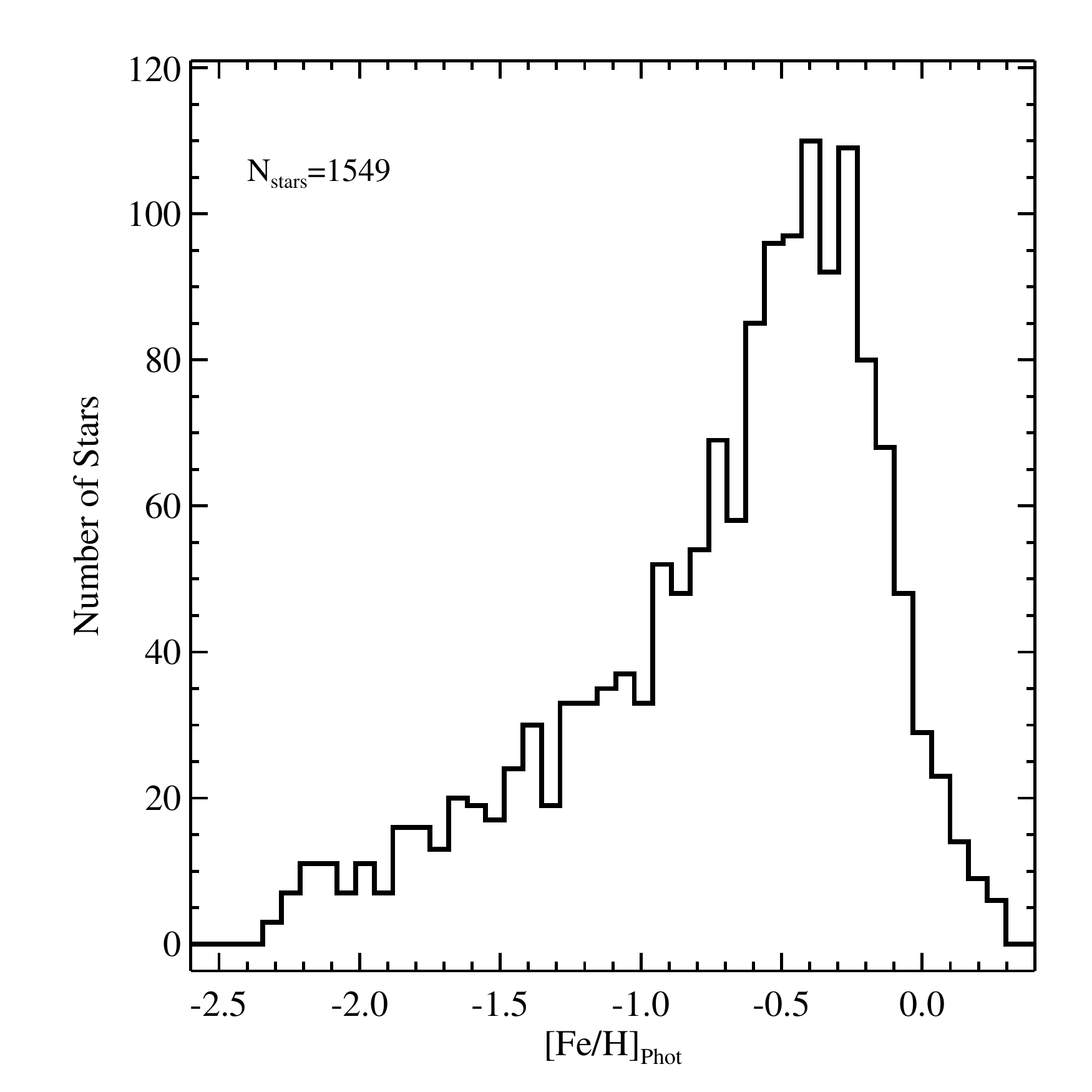}
\caption{  
Metallicity distribution function 
of the securely identified M31 red giants in our sample, based on interpolating the locations of the stars in the CMD within a fine grid of stellar isochrones for $t=10$~Gyr and \afe$=0.0$.  The MDF has a strong peak at \feh$=-0.4$ with a significant, extended tail to metal-poor values.  
}
\label{fig:mdf_all}
\end{figure}

The resulting MDF 
is shown in Figure~\ref{fig:mdf_all}, and indicates a strong peak centered at \fehp\,$\sim -0.4$, 
with a significant tail to metal-poor values.  
The mean (median) metallicity of the entire sample is 
$\langle$\fehp$\rangle=-0.69\pm 0.01$ ($-0.56\pm 0.02$). 

\subsubsection{Line of Sight Distance Variations}\label{sec:dmod_variations}
We adopt a single distance modulus \citep[24.47;][]{stanek1998} for the M31 stellar population.   However, M31's halo extends to large projected radii, meaning that along any given line of sight individual stars could be significantly closer or more distant (up to $\sim 150$~kpc) than M31's nominal distance.   A change in distance of $\pm 100$~kpc results in a shift in the median metallicity of the sample of $\pm 0.17$~dex.  The magnitude of the effect is correlated with the metallicity of the star: the derived [Fe/H] changes by $\pm 0.26$~dex for metal-poor stars ([Fe/H] $< -1.5$]) , and $\pm 0.13$~dex for metal-rich stars ([Fe/H] $\ge -0.5$].  

However, the magnitude of this effect is strongly mitigated by the fact that M31's halo is highly centrally concentrated, so that along any given line of sight a large majority of the stars can be assumed to be at a distance modulus close to that of M31.   Therefore, for all but a small percentage of stars, assuming M31's distance modulus will introduce an error in \fehp\ that is smaller than either the error introduced by the error in the star's color or the systematic error in adopting a single age and $\alpha$-enrichment (Section~\ref{sec:age_afe_variations}). 

\subsubsection{Age and \afe Variations}\label{sec:age_afe_variations}
The evolution of a star on the RGB is dependent on the combination of its metallicity, age, and enrichment in alpha elements.  For bright RGB stars, such as our M31 sample, a star's color is much more sensitive to metallicity than age (for ages $\gtrsim 5$~Gyr).  Moreover, \citeauthor{brown2008} (2007, 2008) used HST ACS imaging data that reached the main sequence turnoff to recover the full star formation histories of four locations in M31's halo.   Three of the HST fields are located along M31's minor axis, at projected distances of 11, 21 and 35 kpc from M31's center.  The fourth field, at a projected distance of 21~kpc, is located on the Giant Southern Stream.  The ACS fields are coincident with four of our spectroscopic fields (H11, f130, mask4, and H13s).  In the M31 halo fields, the star formation histories are indicative of a predominantly old population ($t=10$\,--\,12~Gyrs), but with a minority component of stars ($\lesssim 30$\%) that are intermediate aged ($t=6$\,--10~Gyr).  In the Giant Southern Stream field, 70\% of the stars are younger than 10~Gyr.  Thus, as an extreme limit we also compute photometric metallicities assuming an age of 5~Gyr for all stars; this results in a nearly uniform shift of $\Delta$\feh\,$=+0.17$~dex.   

Presently, there are no direct constraints on the $\alpha$-element abundance of stars in M31's halo, as such measurements would require higher S/N spectra than our study.  The \afe\ of M31 GCs is higher than solar \citep{colucci2009,colucci2012}, and both field stars and globular clusters in the Milky Way's halo have higher than solar \afe\ \citep[e.g.,][]{venn2004,wheeler1989,carney1996}, suggesting these populations formed early, before the ISM was enriched by the ejecta of Type Ia SN.  However, the majority of stars in M31's halo are significantly more metal-rich than the MW halo stars with enhanced [$\alpha$/Fe].  Moreover, previous studies of the metallicity of M31's stellar halo have assumed [$\alpha$/Fe]$=0$ \citep{kalirai2006halo,richardson2009,ibata2014}.   Regardless, our assumption of \afe$=0.0$ does not have a large effect on the computed metallicity scale.  Formally, the mean abundance of our entire sample of red giants shifts by $\Delta$\feh\,$=-0.17$~dex if we use \afe\,$=+0.3$, $t=10$~Gyr models.

Given that stars in M31's halo will have a range of ages and $\alpha$-element abundances, the true distribution of metallicities is likely broader than shown in Figure~\ref{fig:mdf_all}, which is derived assuming a single age and [$\alpha$/Fe].  The maximum effect per star ($\lesssim \pm 0.2$) is similar to the uncertainty in \feh\ produced by our typical photometric errors.
More problematic is a potential gradient in the age or $\alpha$-element abundance of M31's stellar halo, which will result in a systematic error in the mean metallicity with radius, and thus will affect the measured metallicity gradient.  We will explore the effect of a radial gradient in [$\alpha$/Fe] or age on the measured metallicity gradient in Section~\ref{sec:biases_measurement}.

Finally, we note that our photometric metallicity results are not strongly dependent on the stellar evolution models we have chosen.  We have also computed photometric metallicities using models from the Padova group \citep{girardi2002} and the Yale-Yonsei group \citep[Y$^2$;][]{demarque2004}.  The use of alternate models results in differences at the $\sim0.15$~dex level, on par with the systematic uncertainties resulting from choosing a single age and \afe.

\subsection{Spectroscopic Metallicities}\label{sec:met_est_spec}

\begin{figure}[tb!]
\includegraphics[width=0.45\textwidth]{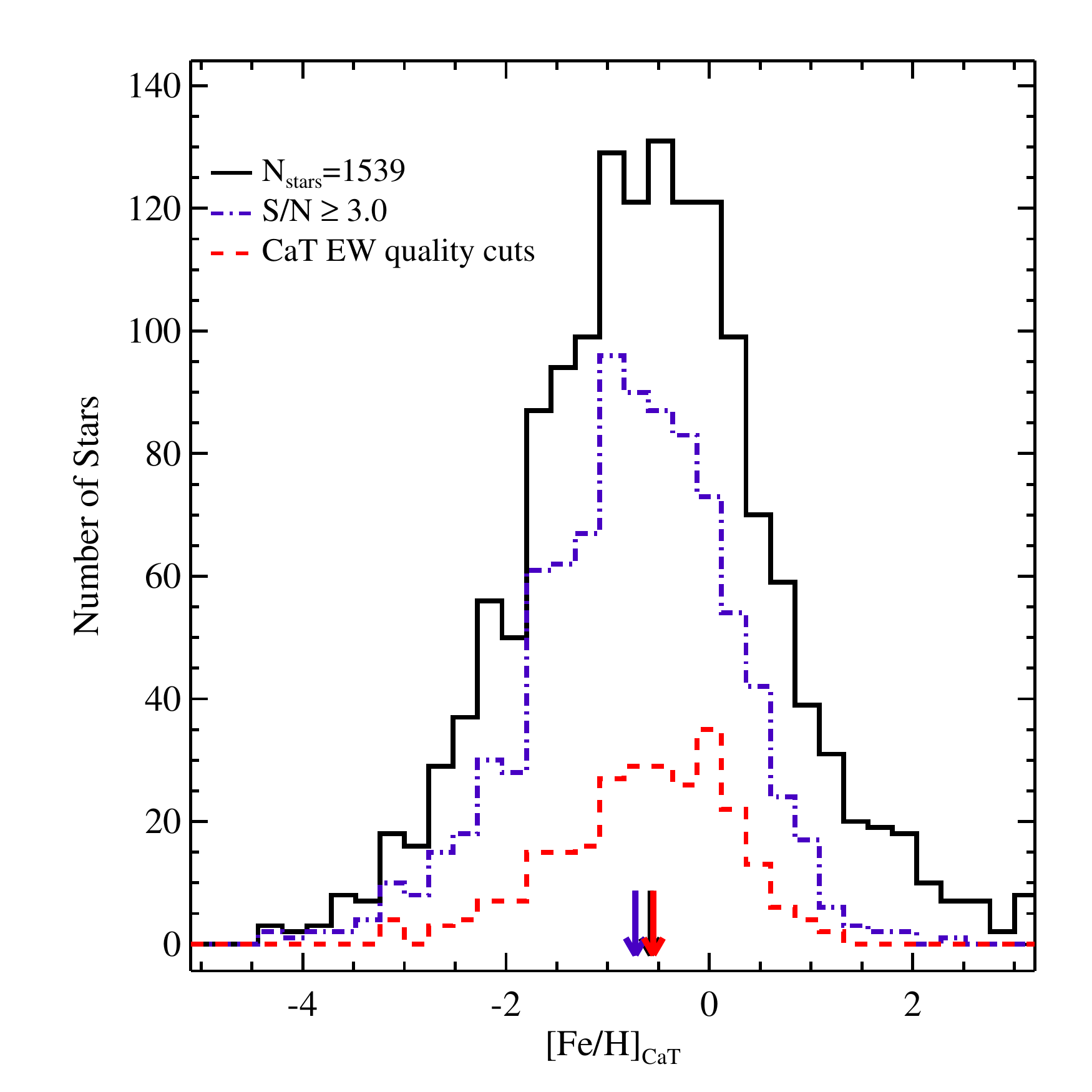}
\includegraphics[width=0.45\textwidth]{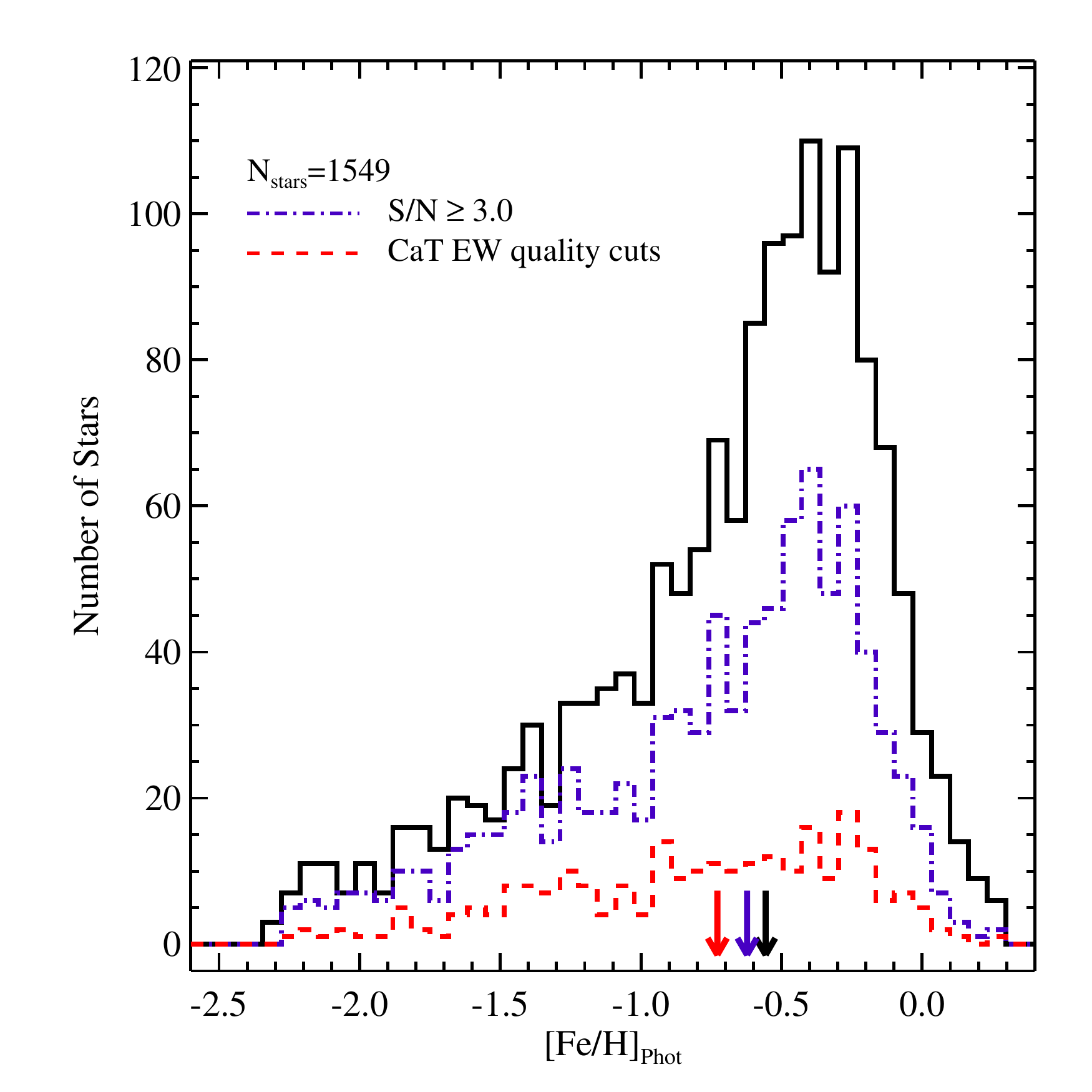}
\caption{  
{\it Top panel:} Metallicity distribution function 
of the securely identified M31 red giants in our sample, based on the equivalent width of the \caii\ triplet absorption lines (Section~\ref{sec:met_est_spec}).  Due to the low S/N of most of our spectra, the errors on any individual \fehs\ measurement are large. The spectroscopic MDF 
has a strong, relatively metal-rich peak (at \feh$\sim-0.5$) and a significant tail to metal-poor values.  The blue dot-dashed and red dashed histograms show the MDF computed when the sample is reduced to stars with S/N\,$ > 3$~pixel$^{-1}$, and when only stars that pass the quality cuts discussed in Section~\ref{sec:met_est_spec} are included.  {\it Bottom panel:}  The CMD-based metallicity distribution function for the full sample of M31 stars and the subsamples defined in the top panel.  The addition of quality cuts on the spectral measurements removes the most uncertain spectroscopic metallicity estimates, and reduces the range of the \fehs\ measurements to more physically reasonable values.   However, it also affects the CMD distribution of the stars.  The quality cuts preferentially remove the reddest RGB stars from the sample, resulting in a lower median photometric metallicity (due to a bluer median color) of the sample.   In both panels, arrows mark the median values of each MDF. 
}
\label{fig:mdf_all_spec}
\end{figure}

The chemical abundance of a star can also be derived using empirical calibrations based on the measurement of individual abundance lines in medium resolution spectra.  One advantage of spectroscopic metallicities is that they are much less dependent than photometric metallicities on the assumed age of the stellar population.  The three lines of the \caii\ triplet at 8498, 8542, and 8662 \AA\ can be mapped to an \feh\ measurement provided they are accurately characterized.  Multiple calibration relations exist in the literature \citep[e.g.,][]{armandroff1988,armandroff1991,olszewski1991,rutledge1997a,tolstoy2001,cole2004,battaglia2008,starkenburg2010}.  
We adopt the calibration presented in \citet{ho2014}, which is based on the \citet{carrera2013} calibration but modified to use the strongest two  \caii\ triplet lines at 8542 and 8662 \AA.  The \citet{carrera2013} relation is empirically calibrated using spectra of globular and open cluster stars and field stars, ranging in metallicity from $-4.0 \le $\,\feh\,$\le +0.5$.  \citet{ho2014} analyzed spectra of stars in M31's dwarf satellites; some of the data is from the same spectroscopic masks analyzed here.  Thus, their modification of the \citet{carrera2013} calibration is optimized for our data.  

The \caii\ absorption lines are the strongest features in our spectra \citep[e.g.,][]{kalirai2006halo}.  However, the spectral signal-to-noise ranges only from $\sim 2$\,--\,17\,\AA$^{-1}$, with a median S/N of 5.7\,\AA$^{-1}$, and the \caii\ absorption lines overlap with several night sky features.  As we have discussed in previous papers \citep[e.g.,][]{kalirai2006halo,kalirai2009}, this results in large \fehs\ uncertainties for any individual star: the median uncertainty in [Fe/H] is 0.68~dex for the securely identified M31 stars.  Nevertheless, the mean (median) spectroscopic metallicity of the entire sample is $\langle$\,\fehs\,$\rangle =-0.57\pm 0.04$ ($-0.58\pm 0.05$),  similar to the mean photometric metallicity (Figure~\ref{fig:mdf_all_spec}).  

\citet{ho2014} found that below a S/N of 3~pixel$^{-1}$ the \caii\ triplet equivalent-width measurement was unreliable. If we restrict the sample to the spectra with S/N per pixel $\ge 3$ (5.2 \,\AA$^{-1}$), the median error in the individual \fehs\ measurements falls to 0.5~dex.  The MDF of stars with S/N$\ge 3$~pixel$^{-1}$ is shown in Figure~\ref{fig:mdf_all_spec} (dot-dashed blue curves); this cut removes many of the unreasonably high \fehs\ estimates and reduces the median spectroscopic metallicity of the sample by 0.15~dex.  

We can further restrict the sample to stars with the highest quality \caii\ measurements.  One metric is to ensure that the measured equivalent widths of the individual \caii\ lines are consistent with the observed line ratios for the \caii\ triplet \citep[$\sim 1:0.75$ for the 8542 and 8662 \AA\ lines;][]{starkenburg2010}.   Figure~\ref{fig:mdf_all_spec} shows the MDF of stars that have 8542 and 8662 \AA\ equivalent width measurements that are within $2\sigma$ of the expected line ratios, and an \fehs\ error that is less than the median error in \fehs\ for the full sample.  Implementing these two restrictions on the full sample further reduces the number of stars with unreasonably high \fehs\ measurements but results in an almost identical median \fehs\ to the full sample, and results in a steep reduction in the number of stars in the sample. 

Any quality cuts implemented on the sample have the potential to introduce a systematic bias in the metallicity distribution.  Since all stars are observed for approximately the same amount of time (Section~\ref{sec:spec}), the largest factor in the spectral S/N is the magnitude of the star.  Thus, introducing spectral quality cuts preferentially biases the sample against the reddest RGB stars, which are the most likely to be metal-rich (bottom panel of Figure~\ref{fig:mdf_all_spec}). 

Due to the large uncertainties in our individual, \caii\ triplet-based \fehs\ measurements, the analysis in this paper focuses on the \fehp\ estimates.  However, the \fehs\ measurements provide a useful independent estimate of the mean chemical abundance of M31's stellar halo.

\subsection{Comparison of Photometric and Spectroscopic \feh\ Measurements}\label{sec:met_est_comp}
\begin{figure}[tb!]
\plotone{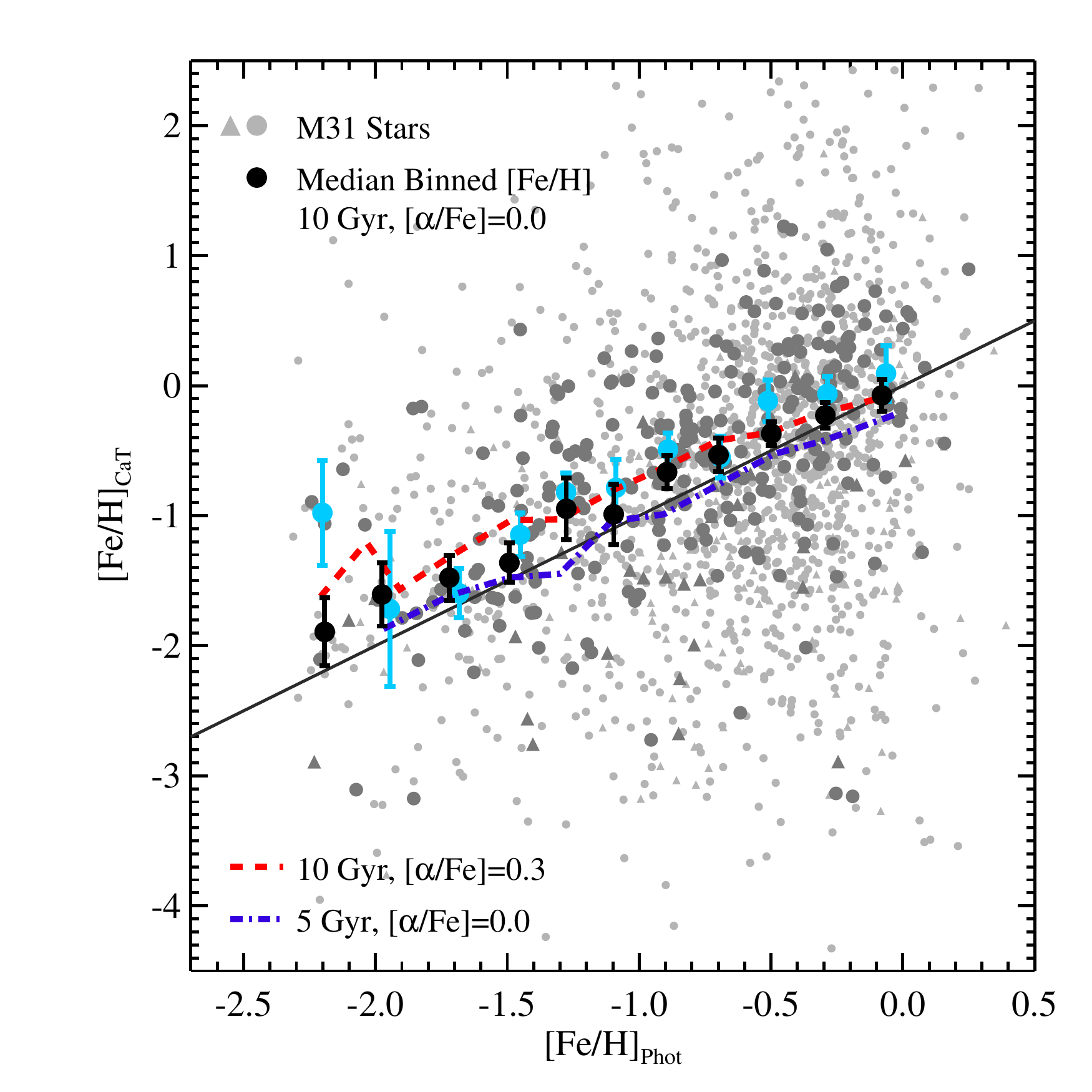}
\caption{  
Comparison of the photometric and spectroscopic metallicity estimates.  Stars securely identified as M31 red giants are depicted as filled grey circles, while stars with a marginal identification as M31 red giants are depicted by grey triangles.  The larger, darker grey points denote stars whose \caii\ triplet measurements pass the spectral quality cuts discussed in Section~\ref{sec:met_est_spec}.  The solid line denotes the one-to-one relation.  The large black points show the median \fehp\ and \fehs\ values for all securely identified M31 stars in a series of bins in \fehp; error bars denote the error in the median value.  The red dashed and blue dot-dashed lines show the running median values when alternate isochrone sets are used to compute \fehp.  The large, light blue points are the median  \fehp\ and \fehs\ values for stars that pass the quality cuts.    
Because the S/N per pixel of a typical stellar spectrum is low, individual estimates of \fehs\ have large error bars, leading to a large spread in \fehs\ values at a given \fehp.  However, in aggregate the \caii\ triplet based [Fe/H] estimates agree well with [Fe/H] estimates based on the star's position in the color-magnitude diagram.  
}
\label{fig:photvspec}
\end{figure}

The two independently determined metallicity measurements of our member M31 red giants are compared in Figure~\ref{fig:cmd}.  Points are color-coded by their \caii\ triplet-based abundance measurements, and stellar isochrones for $t=10$~Gyr, \afe$=0.0$ are overplotted.   The spectroscopic metallicities generally track the overall color distribution of points on the CMD, and roughly follow the shapes of the stellar isochrones as they flare out in color at higher luminosities.  The outlying points on this diagram largely represent stars for which either the photometry is in gross error or for which the \caii\ triplet line measurements are either computed from a very low S/N spectrum or suffered from an artifact (e.g., poor sky subtraction, residual cosmic ray, wavelength calibration error, etc.).

Although the errors on individual \caii\ triplet based \feh\ measurements are large,  we can bin the spectroscopic measurements to test how they compare in aggregate with our photometric measurements as a function of \feh.   This is illustrated in Figure~\ref{fig:photvspec}.  Despite the large scatter in individual measurements, there is an obvious correlation in the mean \feh\ values, fully consistent with the one-to-one line.  The large points are constructed by creating bins in \fehp\ that include at least 25 stars and a minimum bin size of 0.2~dex, and calculating the median of both \fehp\ and \fehs\ in the bin (bins at high values of \fehp\ have considerably more stars than the minimum).  The error bars represent the error in the median value.  The two measurements are in good agreement with one another over the entire metallicity range.  The effect of implementing the \caii\ triplet measurement quality cuts described in Section~\ref{sec:met_est_spec} is also shown: as indicated in Figure~\ref{fig:mdf_all_spec}, the largest outliers in \fehs\ are removed from the sample.  While there is a slight tendency for the median \fehs\ of the restricted sample to be higher than the median \fehs\ of the full sample, the values are consistent within 1$\sigma$.

Figure~\ref{fig:photvspec} also displays the relation between the photometric and spectroscopic metallicity estimates of the full sample  if isochrones of different age or alpha enrichment are used to compute \fehp\ (Section~\ref{sec:age_afe_variations}).  Changing either the age or alpha enhancement results in a small change in the normalization of the median values, and a negligible change in the slope of the \fehs\,---\,\fehp\ relation.

\section{Metallicity Profile of M31's Stellar Halo}\label{sec:met_halo}

This section explores the metallicity distribution of stars in M31's halo as a function of projected distance from M31's center (\rproj), both with and without tidal debris features.  Section~\ref{sec:innervouter} presents the metallicity distribution functions for 4 radial regions.  Section~\ref{sec:met_grad} analyzes the median metallicity of M31's stellar halo as a continuous function of \rproj.  Section~\ref{sec:biases} quantifies the systematic biases in our analysis.   We defer discussion of the physical implications of the observed trends in M31's stellar halo to Sections~\ref{sec:sims} and \ref{sec:subst_v_smooth}.  A full listing of the fields presented here, including their radial distances from M31's center, the number of slitmasks and M31 RGB stars in the field, and relevant references, can be found in Table~1 of \citet{gilbert2012}.

\subsection{Changes in the MDF as a Function of Radius}\label{sec:innervouter}

\begin{deluxetable}{rrrrrr}
\tablecolumns{6}
\tablewidth{0pc}
\tablecaption{Metallicity of M31's Halo.}
\tablehead{\multicolumn{1}{c}{$R_{\rm min}$} & \multicolumn{1}{c}{$R_{\rm max}$} & \multicolumn{1}{c}{$N_{\rm stars}$} &\multicolumn{1}{c}{Mean} & \multicolumn{1}{c}{Median} & \multicolumn{1}{c}{$\sigma_{\rm [Fe/H]}$}  \\
\multicolumn{1}{c}{(kpc)} & \multicolumn{1}{c}{(kpc)} &  &  \multicolumn{1}{c}{[Fe/H]\tablenotemark{a}} &  \multicolumn{1}{c}{[Fe/H]} & \\
}
\startdata
\sidehead{{\it All M31 Stars}} 
\hline
9 & 180 & 1549	 & $-0.69\pm0.01$ & $-0.56\pm0.02$ & 0.54 \\ 
\hline
\sidehead{{\it Selected Radial Regions}}
\hline
9   &   20 & 837 & $-0.55\pm0.01$ & $-0.47\pm0.02$ &  0.43\\  
20 &   40 & 397 & $-0.70\pm0.03$ & $-0.55\pm0.03$ &  0.53\\ 
40 &   90 & 248 & $-1.01\pm0.04$ & $-1.02\pm0.04$ &  0.56\\ 
90 & 180 &   67 & $-1.32\pm0.08$ & $-1.4\pm0.10$ &  0.63
\enddata
\tablenotetext{a}{[Fe/H] values for individual stars were calculated by comparison of the star's location in the $I, (V-I)$ color-magnitude diagram with 10~Gyr, [$\alpha$/Fe]=0.0, \citet{vandenberg2006} isochrones.  Stars beyond the bounds of the isochrone grid are not included in these measurements.
}
\label{tab:regions}
\end{deluxetable}

\begin{figure*}[tb!]
\centerline{
\includegraphics[width=1.738in]{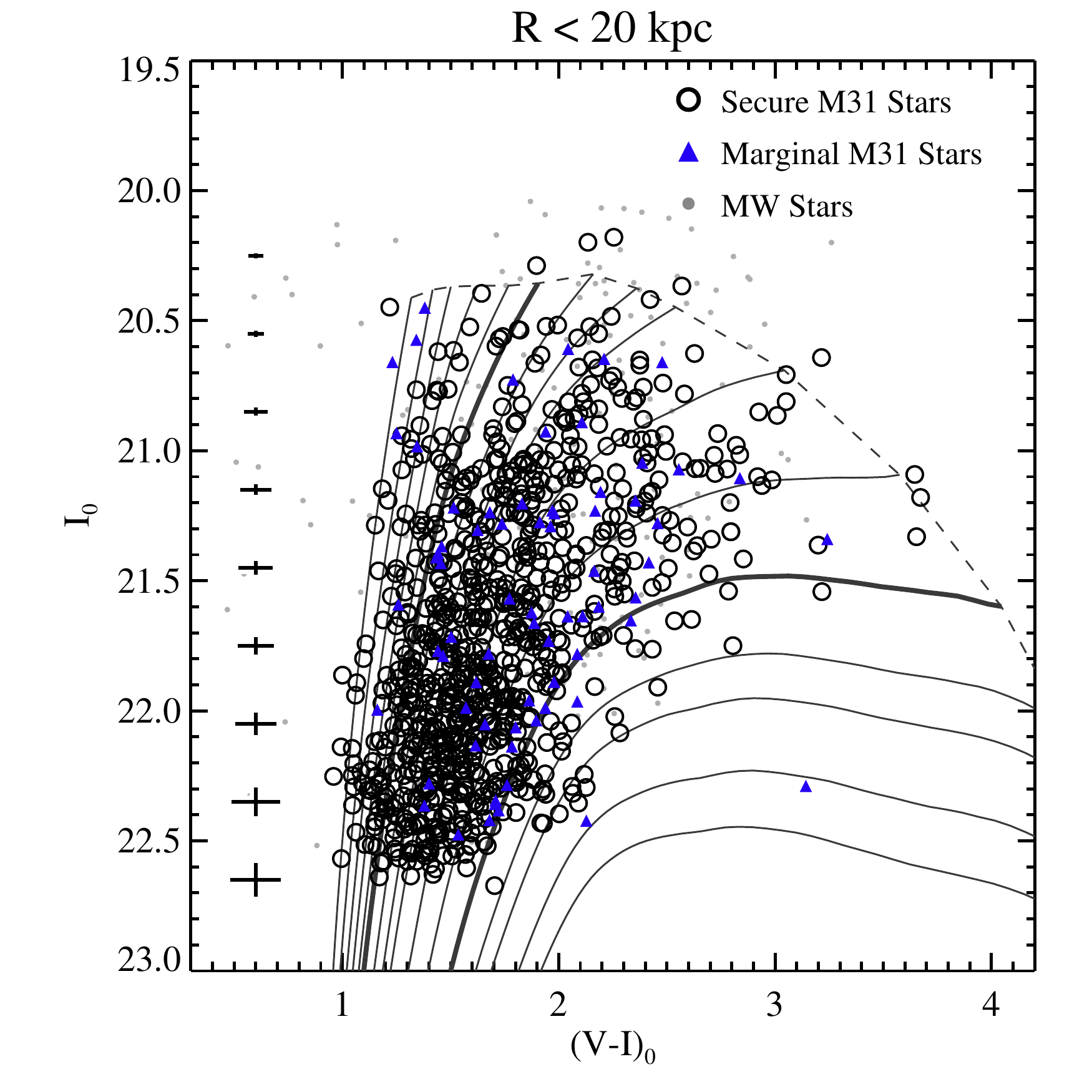}
\includegraphics[width=1.738in]{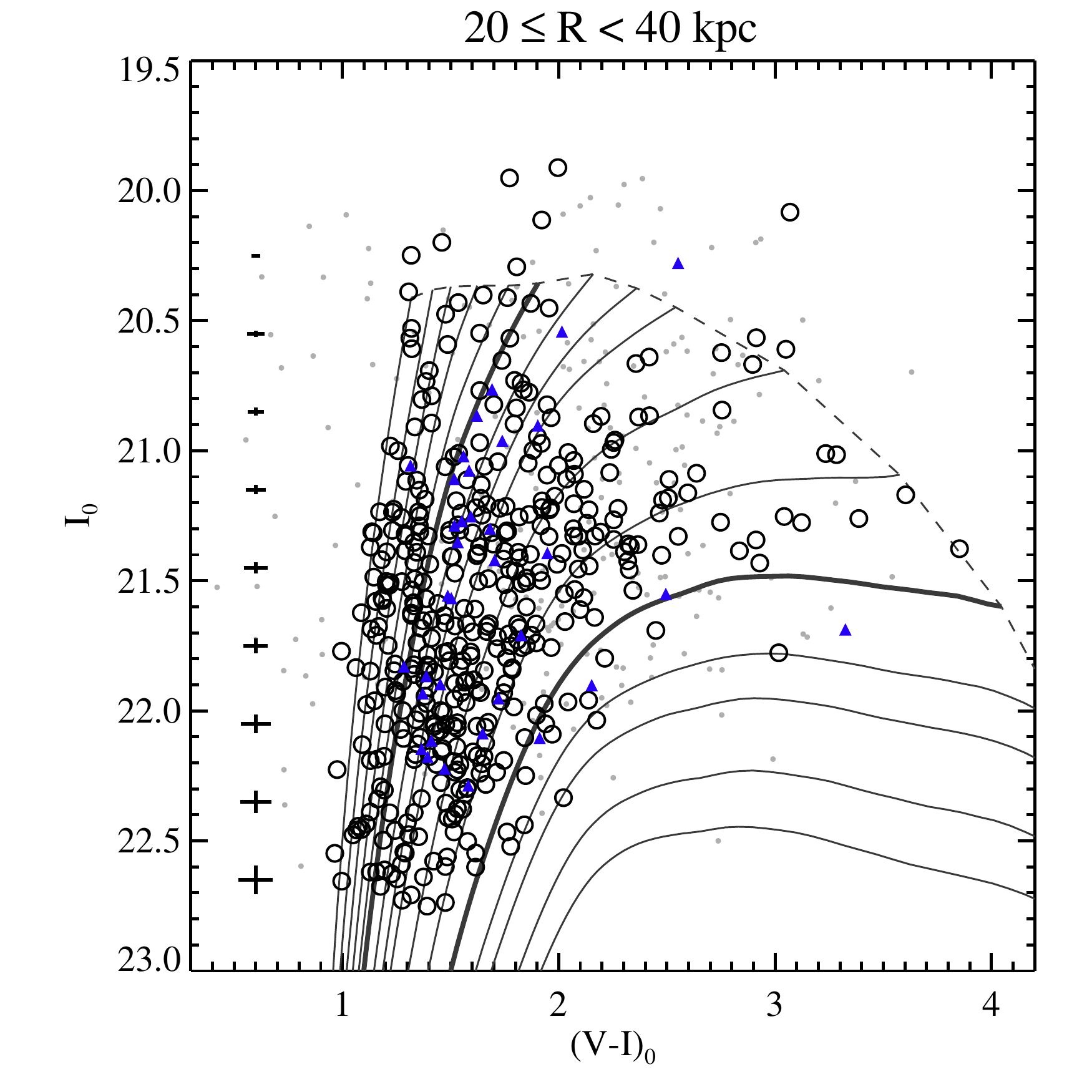}
\includegraphics[width=1.738in]{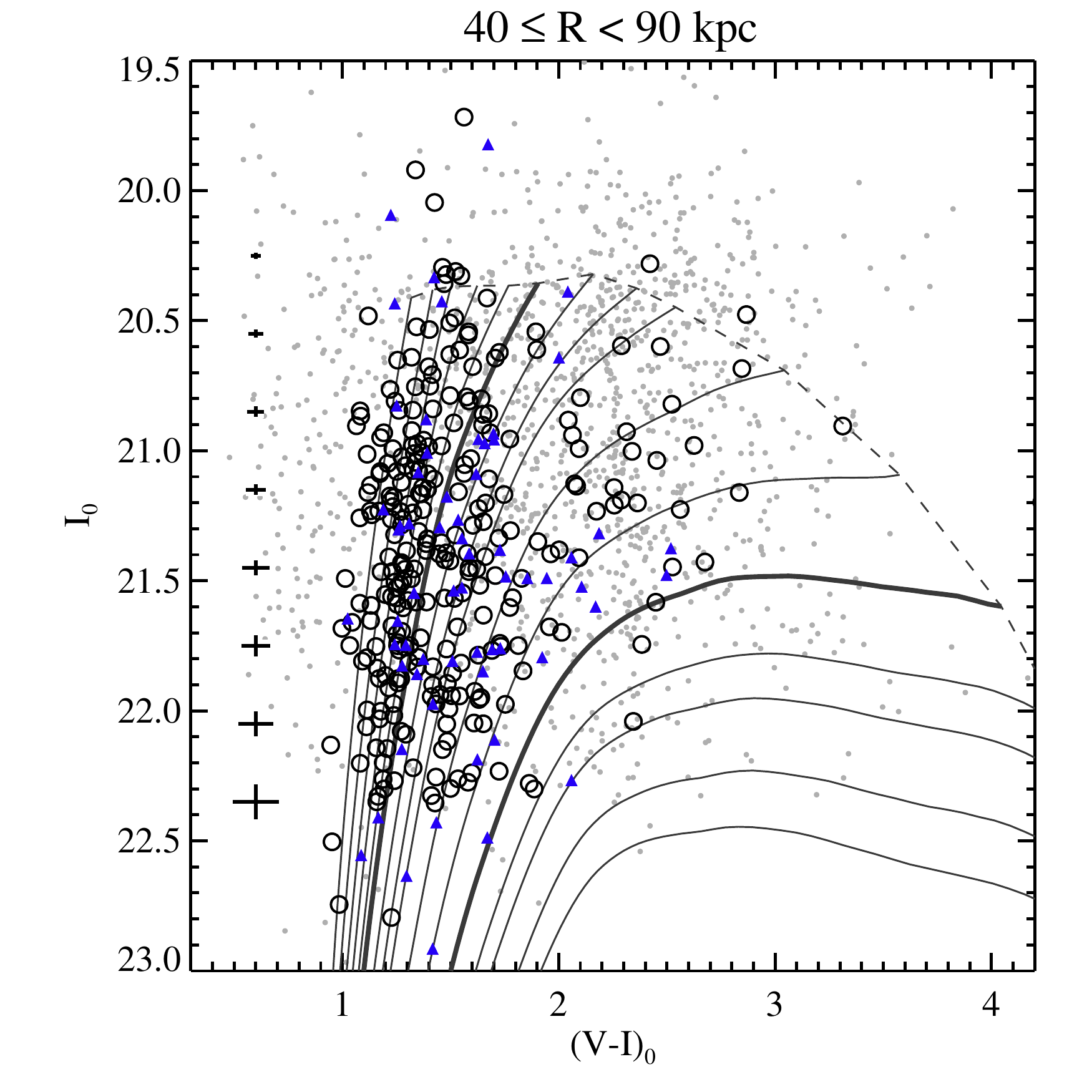}
\includegraphics[width=1.738in]{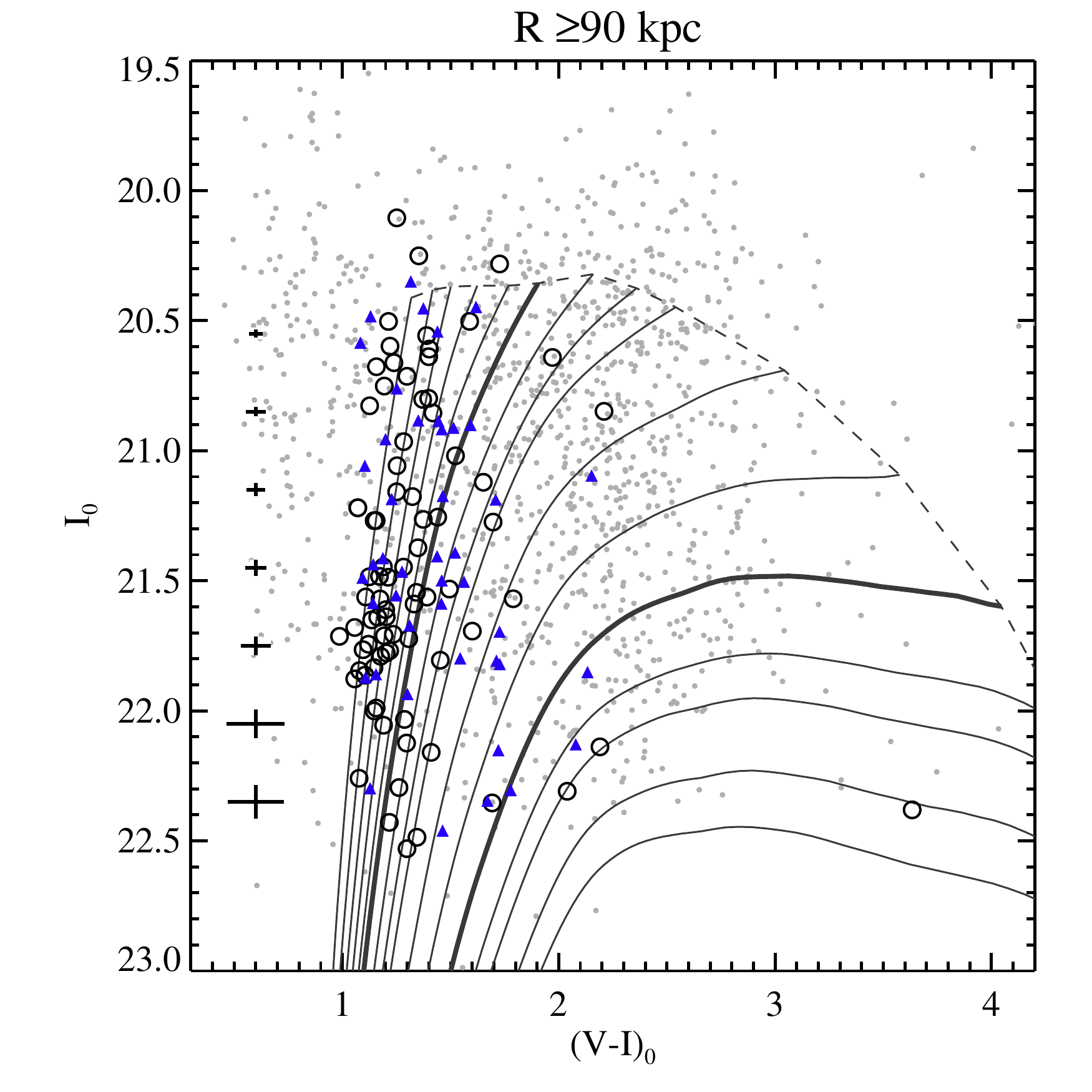}
}
\centerline{
\includegraphics[width=1.738in]{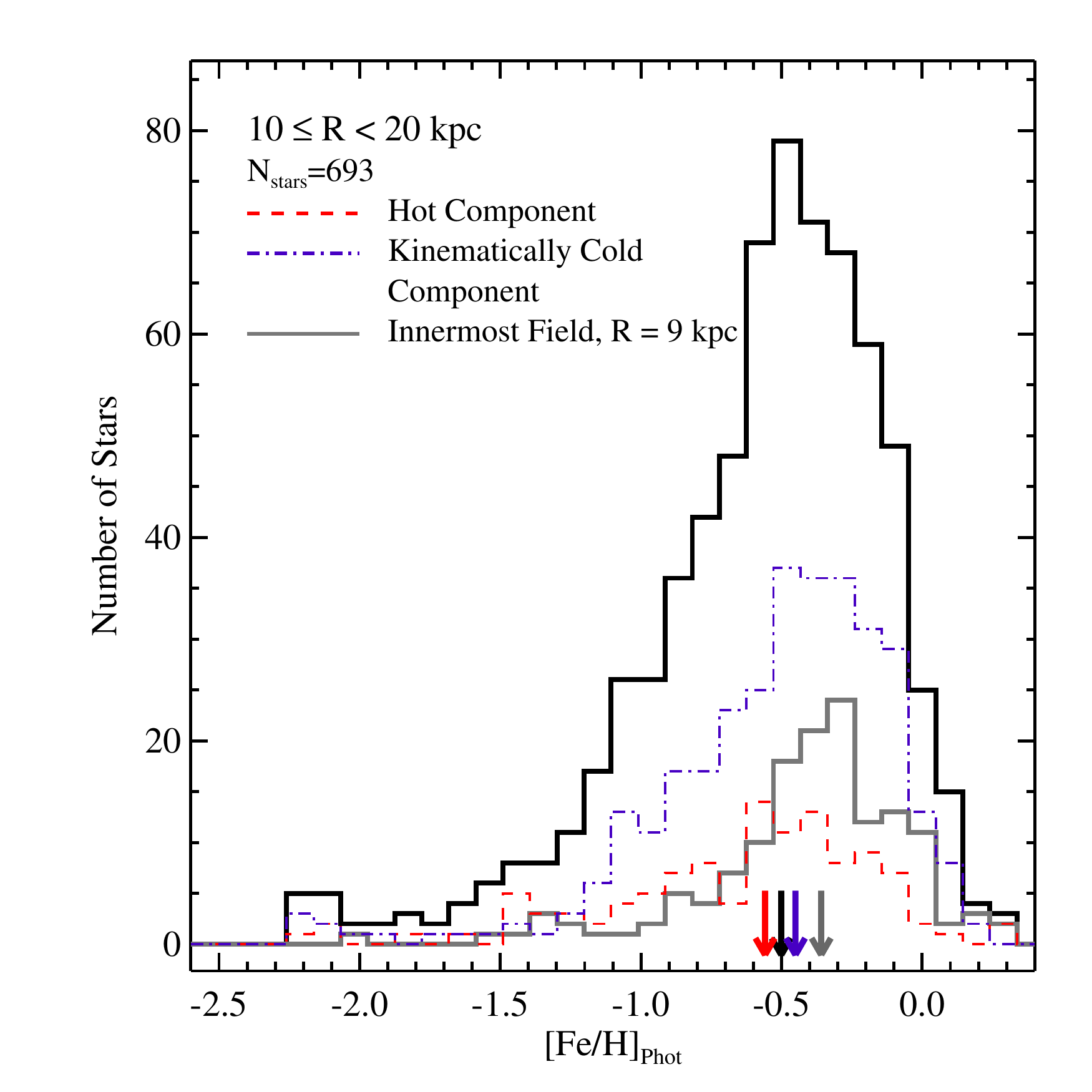}
\includegraphics[width=1.738in]{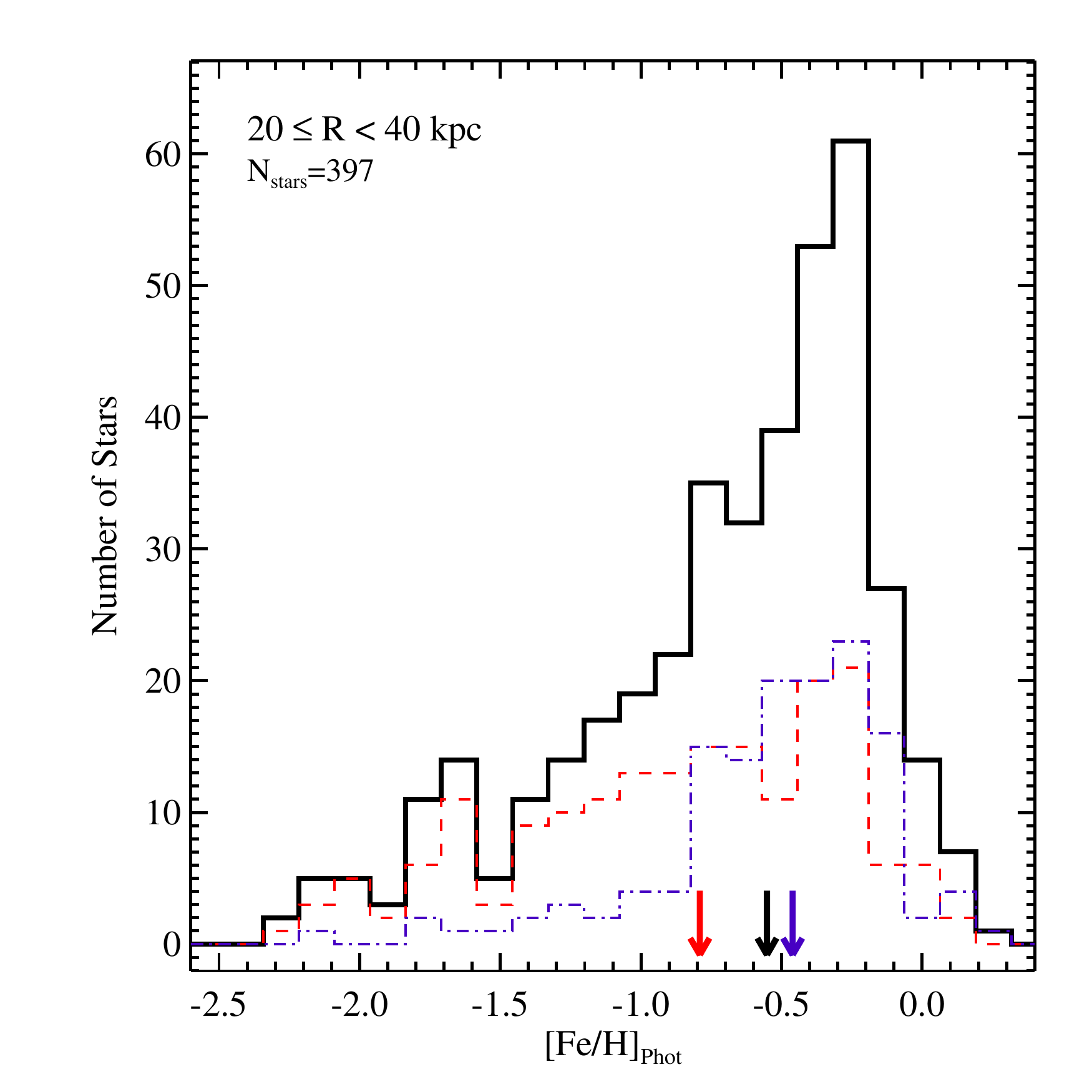}
\includegraphics[width=1.738in]{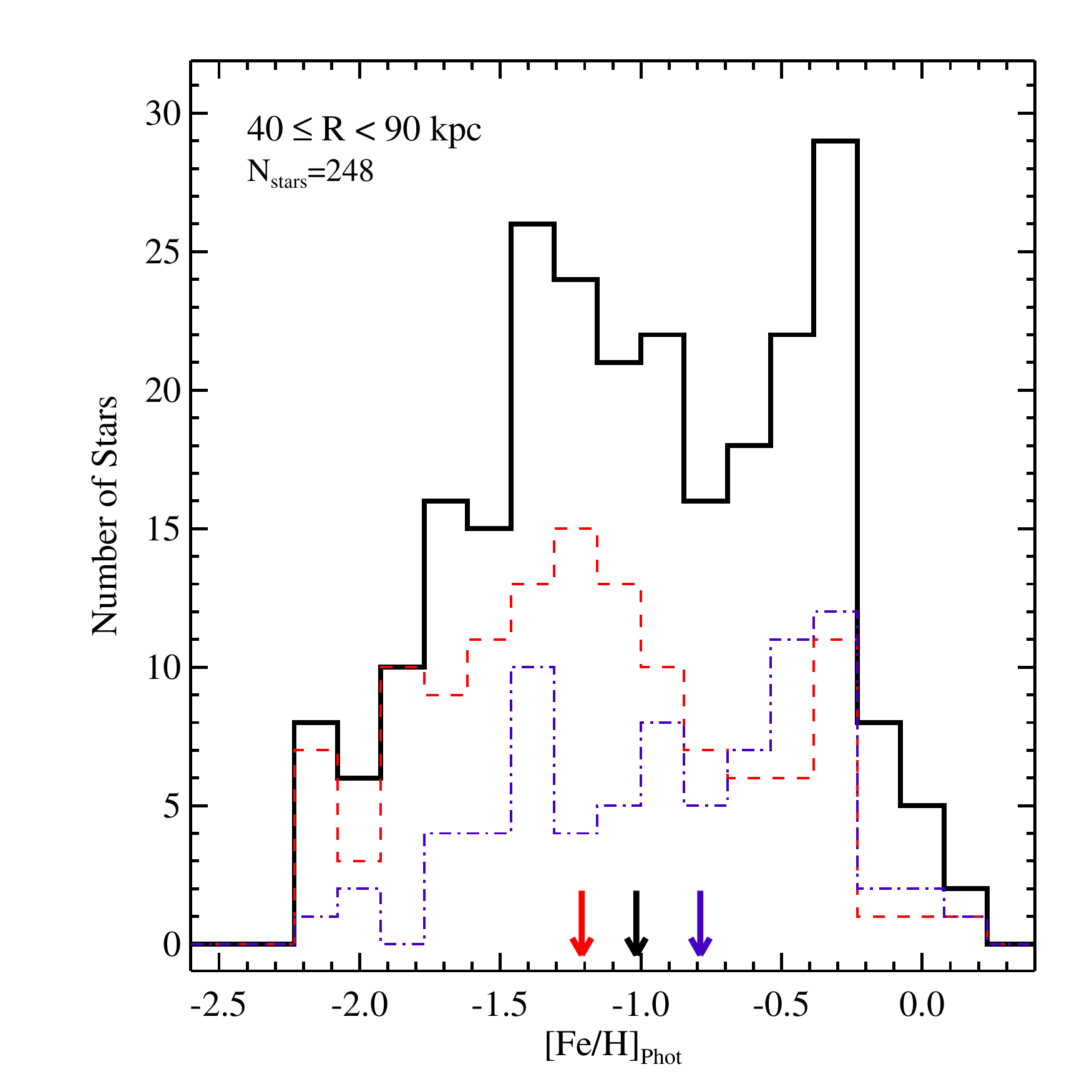}
\includegraphics[width=1.738in]{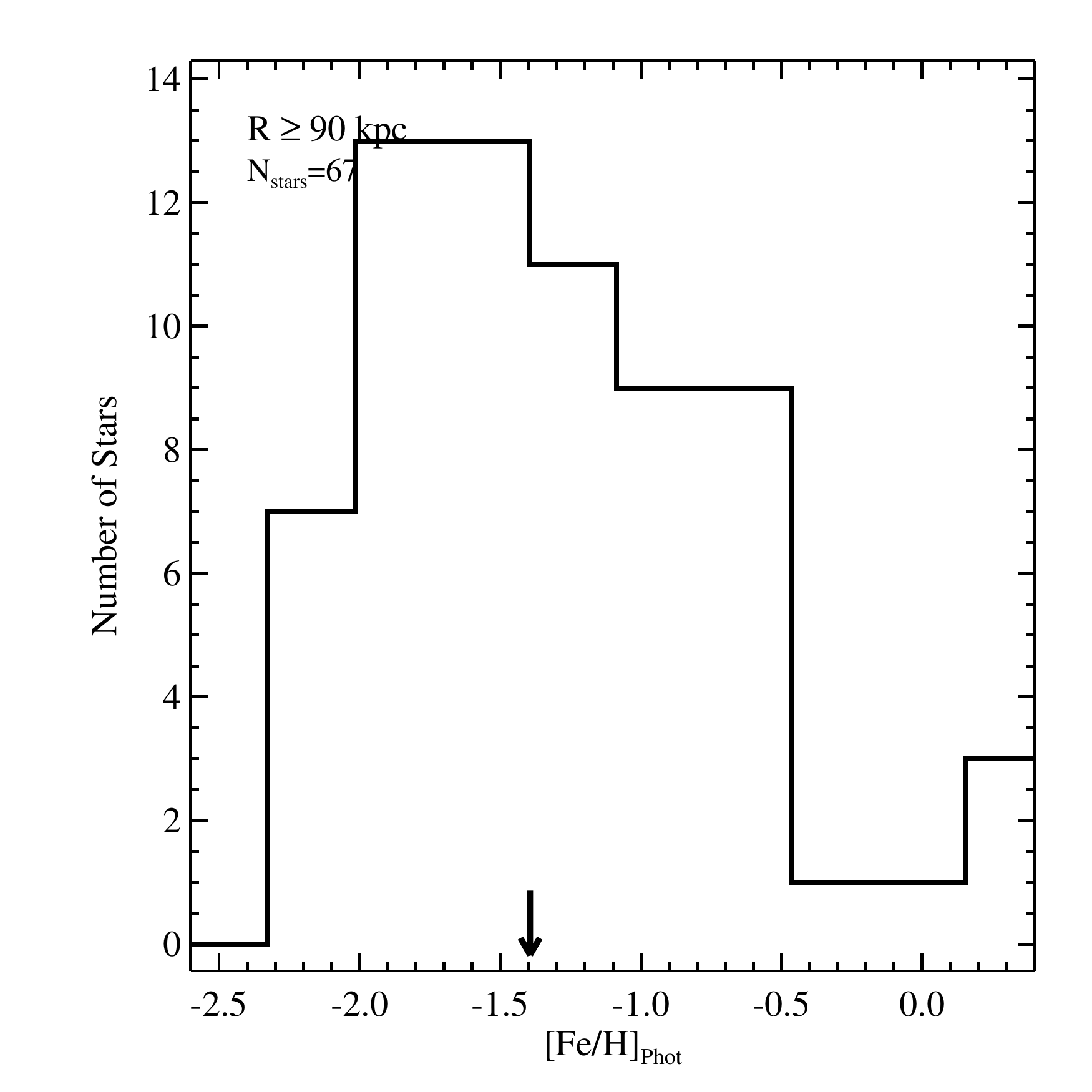}
}
\centerline{
\includegraphics[width=1.738in]{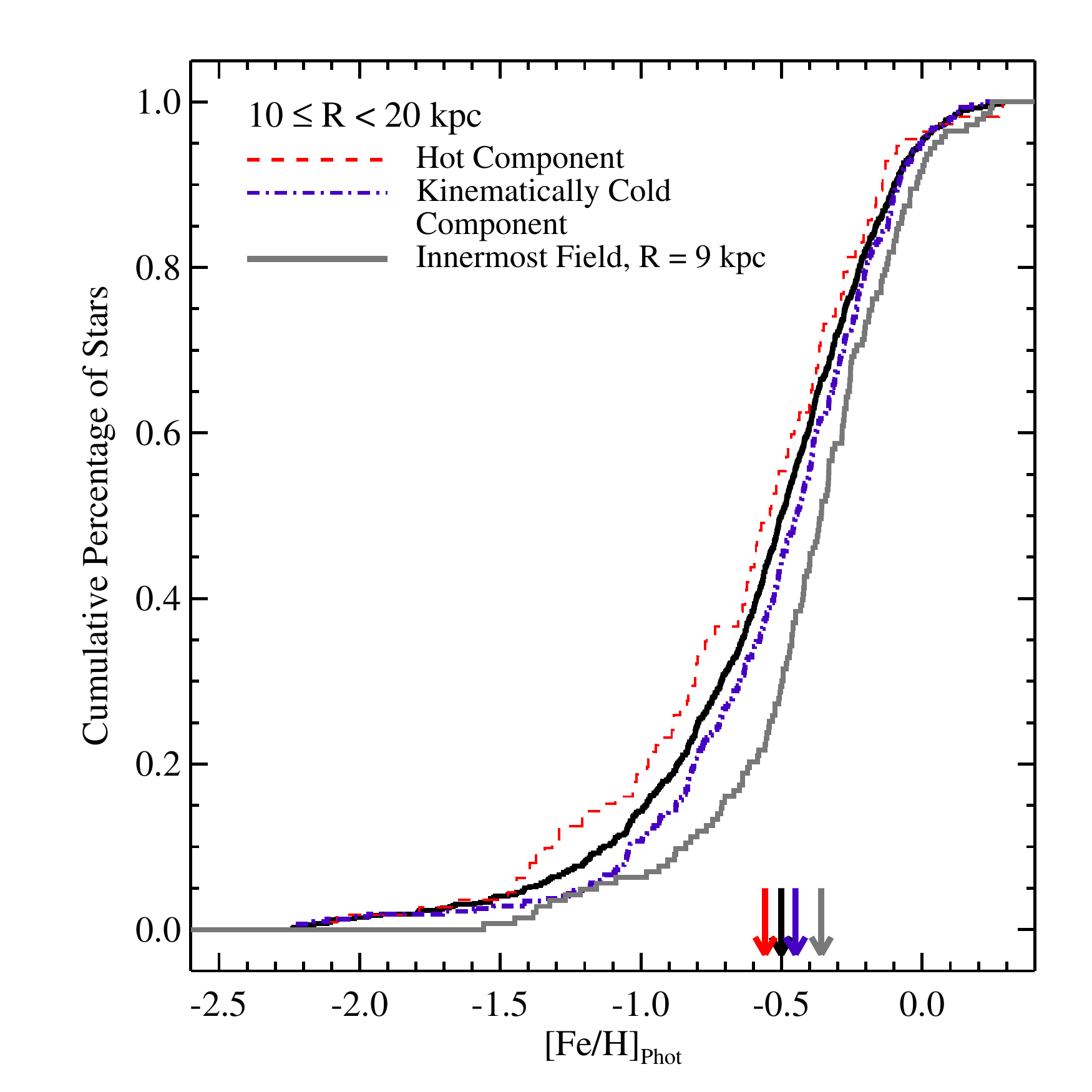}
\includegraphics[width=1.738in]{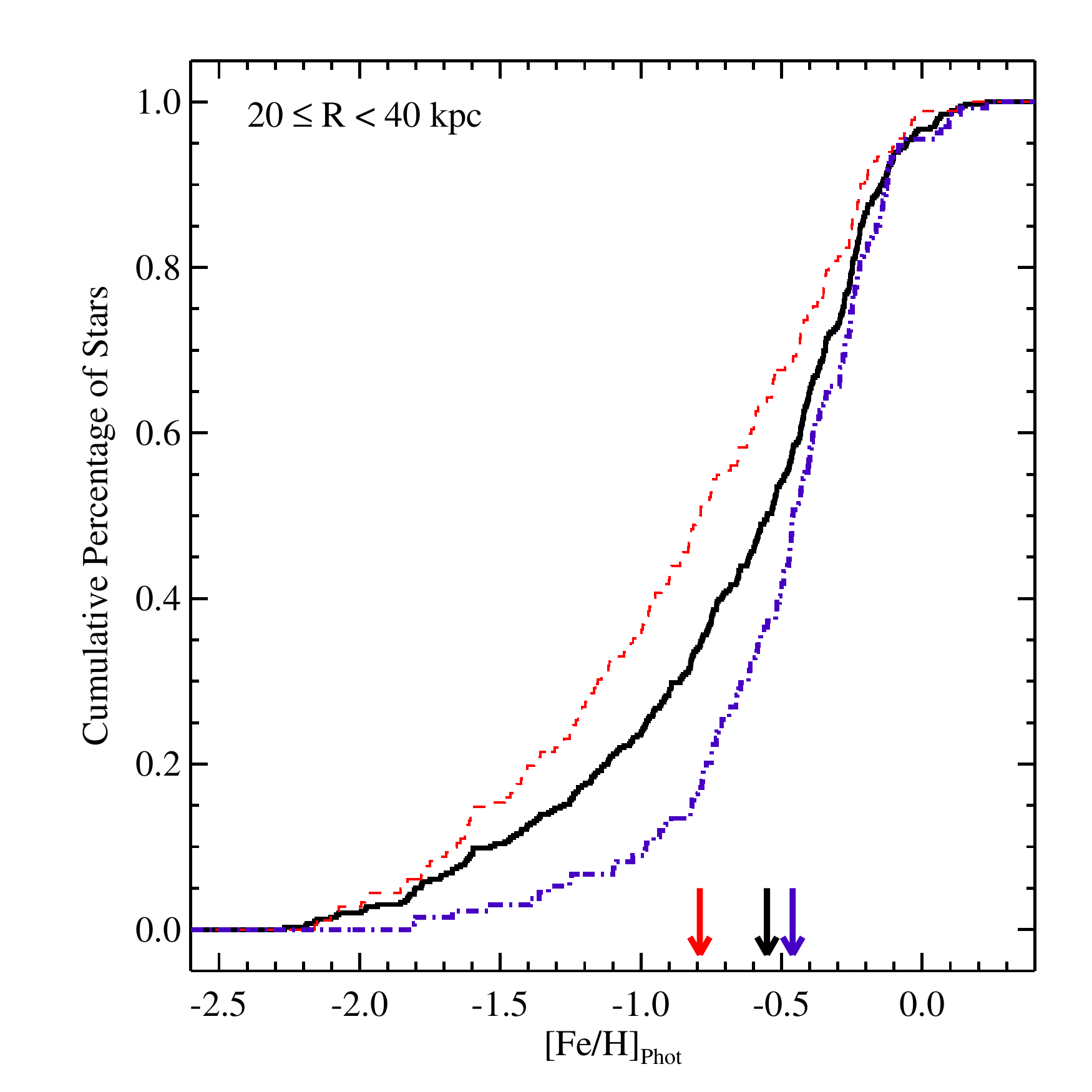}
\includegraphics[width=1.738in]{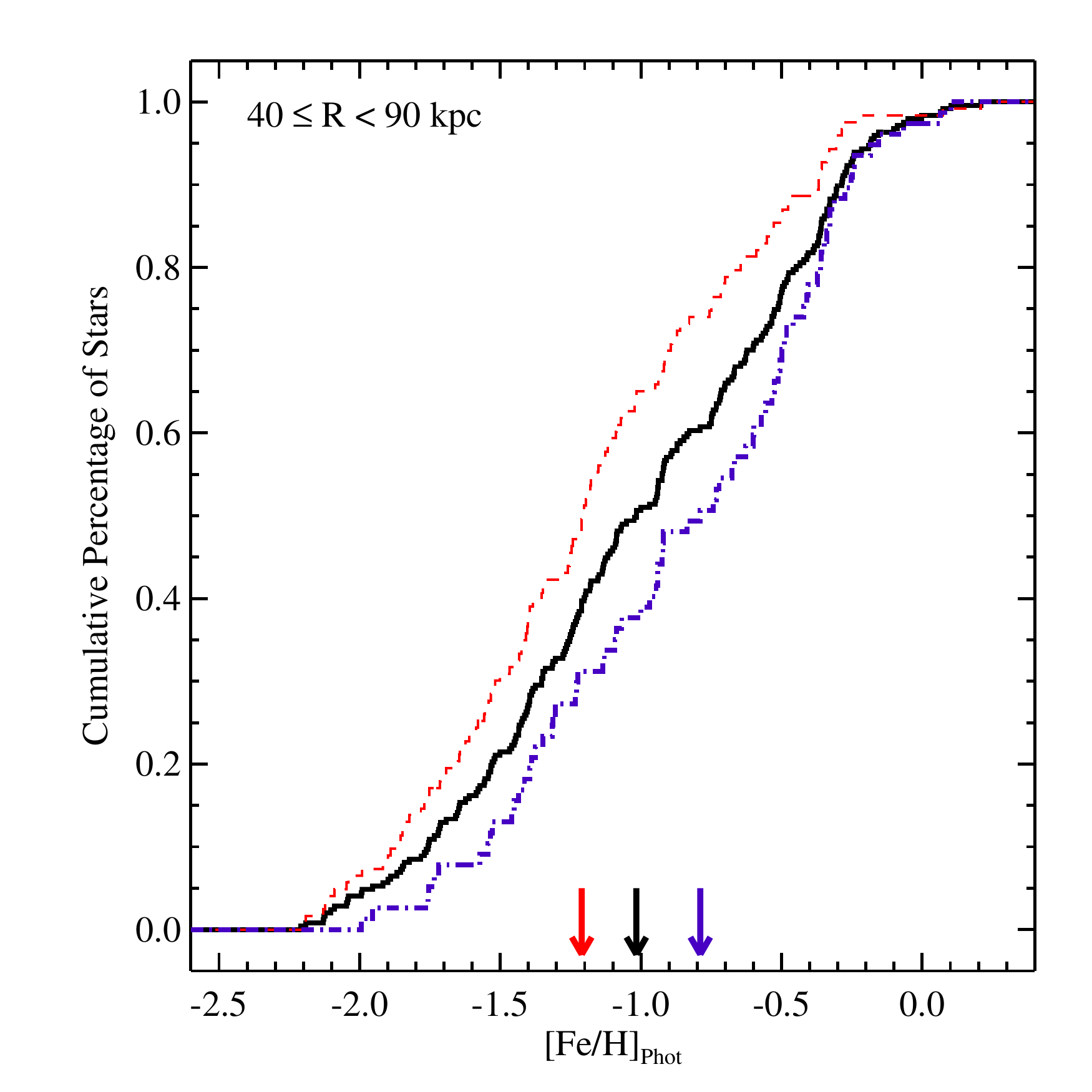}
\includegraphics[width=1.738in]{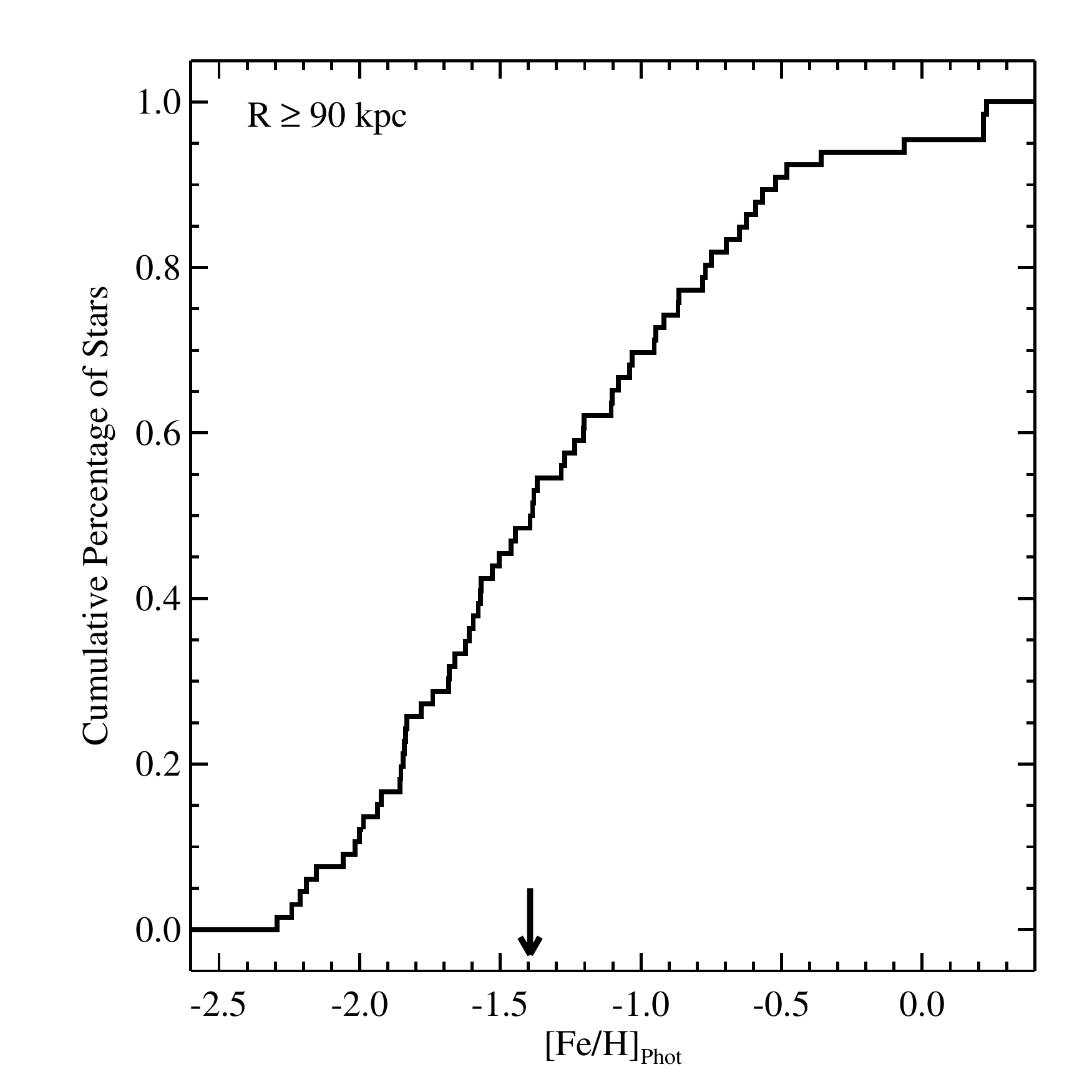}
}
\caption{Properties of the stellar population of M31's halo in four radial ranges (Section~\ref{sec:innervouter}).  The top row shows the location of M31 stars ({\it circles:} secure, {\it triangles:} marginal) and MW contaminants ({\it grey dots}) in the CMD, with theoretical isochrones overlaid as in Figure~\ref{fig:cmd} (10 Gyr, [$\alpha$/Fe]=0, \citet{vandenberg2006}).  The mean photometric errors as a function of magnitude are shown on the left of each panel.  The stellar population gets noticeably bluer as the distance from M31's center increases.  The middle and bottom rows show the MDFs measured by comparison of the star's location on the CMD to the isochrones.  MDFs are shown for all secure M31 stars that lie within the isochrone grid (black curves).   M31's stellar halo becomes increasingly more metal-poor on average as the projected distance from M31's center increases.  In the first three radial bins, some of the fields have kinematically identified tidal debris features (Section~\ref{sec:kccs}).  The red-dashed curves show the MDF of stars that are most likely to belong to the kinematically hot population while the blue-dotted curves show the MDF of stars that are most likely to belong to tidal debris features.  Arrows show the median [Fe/H] for each MDF. On average, the dynamically cold tidal debris is more metal-rich than the dynamically hot halo population, and has a significantly smaller tail to lower metallicities.  
}
\label{fig:cmds_and_mdfs}
\end{figure*}

We first examine the CMDs and MDFs of M31 stars in four broad radial regions in M31's halo
(Figure~\ref{fig:cmds_and_mdfs}), motivated by the results of previous observations (Section~\ref{sec:intro}) and expectations from simulations (Section~\ref{sec:sims}):  
(1) an innermost sample of stars in the region of M31's halo that is known to be relatively metal-rich 
(\rproj\,$<20$~kpc), 
(2) an intermediate region between 
$20\le$\,\rproj\,$<40$~kpc, and two outer halo regions, (3) interior and (4) exterior to 90 kpc ($40\le$\,\rproj\,$<90$~kpc and \rproj\,$\ge90$~kpc).  
This final radius is chosen because at radii larger than 90 kpc, we have too few stars 
per field to identify tidal debris via the kinematics of the stellar population 
\citep[Section ~\ref{sec:kccs};][]{gilbert2012}.  
The CMDs clearly show that the 
stars in the inner regions of M31's halo are on average redder than stars in the outer regions, resulting
in a clear difference in the MDFs as a function of radius.  Metal-rich (\fehp\,$>-1.0$) 
and metal-poor stars (\fehp\,$<-1.0$) are found at all radii in M31's stellar halo.
However, as distance from the center of M31 increases, the peak of the distribution shifts 
to more metal-poor values, 
and the metal-poor tail of the distribution comprises an increasingly large percentage of 
the stellar population, clearly seen in the MDFs (Figure~\ref{fig:cmds_and_mdfs}, middle row).  The median value of the \fehp\ distribution correspondingly shifts to more metal-poor values, which is most easily seen in the cumulative distribution functions (Figure~\ref{fig:cmds_and_mdfs}, bottom row).

\begin{figure*}[tb!]
\centerline{
\includegraphics[width=2.3in]{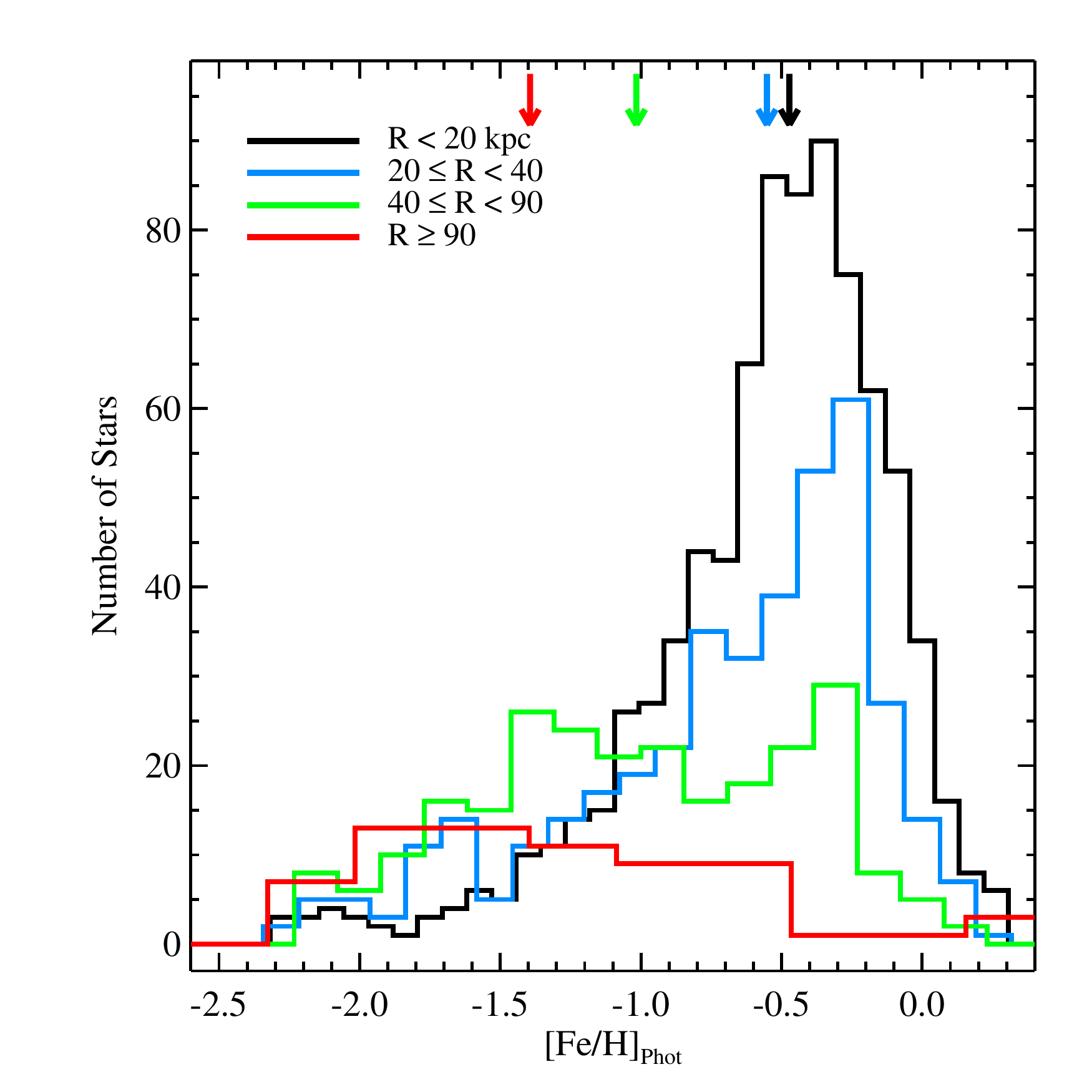}
\includegraphics[width=2.3in]{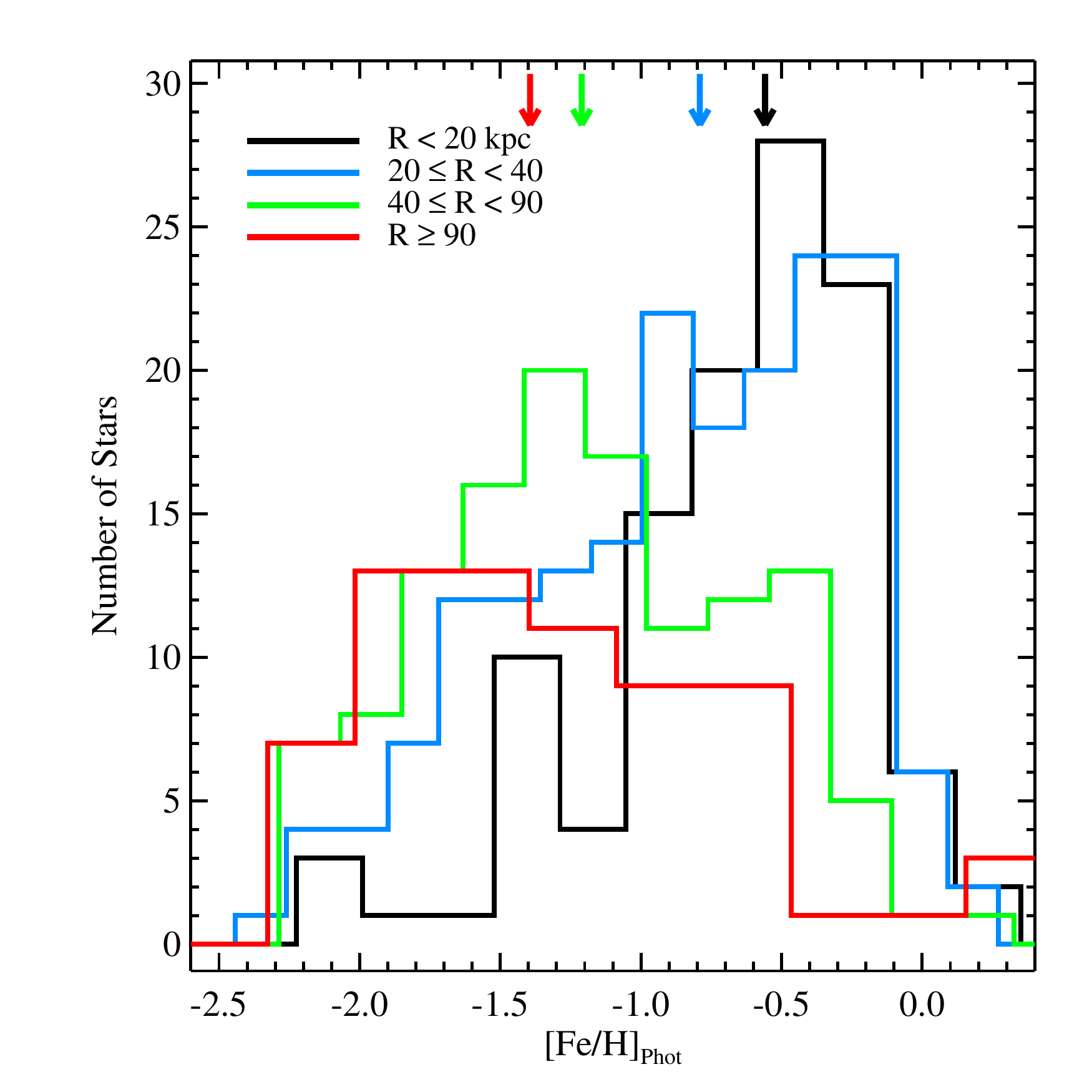}
\includegraphics[width=2.3in]{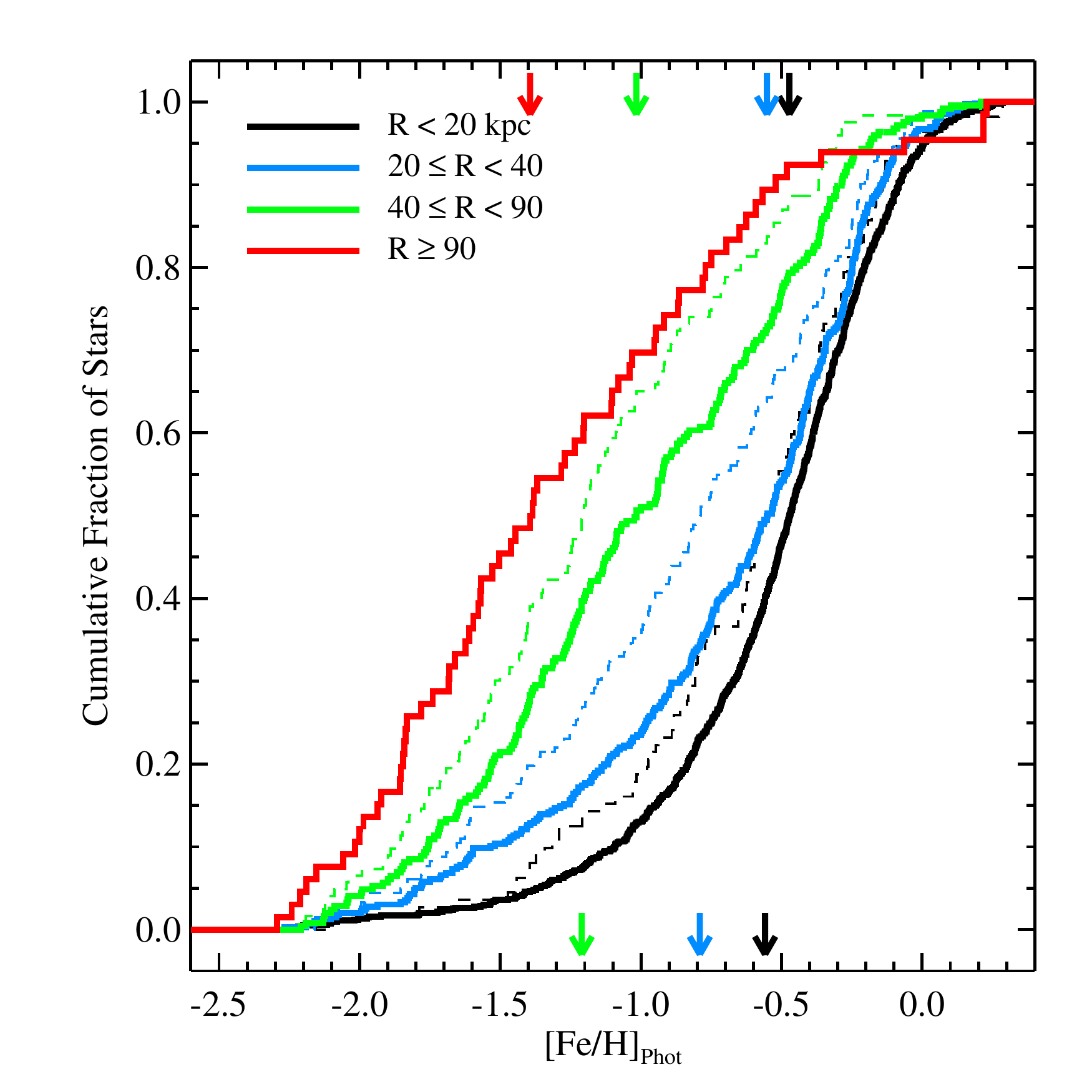}
}
\caption{MDFs of the four radial regions shown separately in Figure~\ref{fig:cmds_and_mdfs}, in differential ({\it left:} all M31 stars; {\it middle:} excluding kinematically cold substructure) and cumulative ({\it right}) form.  Arrows mark the median metallicity of each distribution.  As projected distance from M31 increases, the stellar population becomes increasingly dominated by metal-poor
stars (\fehp$<-1.0$), and the median metallicity of the population decreases significantly, irrespective of whether or not tidal debris features are included.  The cumulative diagram directly compares the MDF of M31's stellar halo including (solid curves; top arrows) and excluding  (dashed curves; bottom arrows) kinematically cold substructure. The difference between the MDFs of adjacent radial bins becomes more pronounced when tidal debris features are removed.  The MDF of the kinematically hot population in the $40\le$\,\rproj\,$<90$~kpc approaches the MDF of the \rproj\,$\ge 90$~kpc bin. 
}
\label{fig:mdfs_cum}
\end{figure*}

The differences in the MDFs of each radial region are more easily
compared in the left panel of
Figure~\ref{fig:mdfs_cum}.   
The inner two radial bins have a strong peak at higher metallicities and a much higher 
fraction of metal-rich stars than the outermost bins.  In the outer two radial bins, there
is no dominant peak at higher metallicities, and metal-poor stars comprise more than half of the 
stellar population.

The cumulative distribution panel shows a clear and significant shift in the median 
metallicity of stars in each bin as a function of radius, with the outer 
radial bins becoming increasingly metal-poor (Table~\ref{tab:regions}).  
Two-sided KS tests on the distributions confirm that there is a statistically 
significant change in the MDF
with radius: the MDF in each radial bin is highly unlikely to be drawn from the 
same distribution as the MDF in the succeeding radial bin.  
The KS-test $p$-values 
of being drawn from the same distribution are 
$p=1\times10^{-4}$ (\rproj\,$<20$~kpc versus $20\le$\,\rproj\,$<40$~kpc), 
$p=1\times10^{-11}$ ($20\le$\,\rproj\,$<40$~kpc versus $40\le$\,\rproj\,$<90$~kpc), 
and $p=0.001$ ($40\le$\,\rproj\,$<90$~kpc versus \rproj\,$\ge 90$~kpc).

\subsubsection{Effect of Tidal Debris on the MDFs}\label{sec:kccs}
Wide-field imaging observations of M31 have demonstrated abundant substructure
extending from \rproj$=10$\,--\,100~kpc \citep[Figure~\ref{fig:roadmap}][]{ibata2001,ferguson2002,ibata2007,mcconnachie2009}.
This substructure reveals that M31 is actively accreting stars from dwarf galaxies. 
The substructure manifests not only as spatial groupings of stars, but can also 
be discerned in velocity and metallicity space.   
One-third of the M31 halo fields we have targeted contain
evidence of spatially coherent, kinematically cold tidal debris streams \citep[Table 1 of][and references therein]{gilbert2012}.  In fact, 
many of the kinematically identified tidal debris features observed in our fields (particularly 
in the inner regions of the halo, Section~\ref{sec:innerfields}) are related to 
a single accretion event that produced 
both the Giant Southern Stream and multiple shell features \citep{ibata2001a,fardal2007,gilbert2007}. 

In general, tidal debris features in M31's stellar halo
have been found to be primarily metal-rich compared to the 
kinematically hot stellar population 
\citep[e.g.,][]{ferguson2002,ferguson2005,guhathakurta2006,gilbert2009a}. 
Therefore, we leverage our spectroscopic data to investigate what effect these tidal debris 
features have on the observed metallicity distribution
of M31's stellar halo, and to determine the metallicity distribution of M31's 
underlying spatially diffuse, kinematically hot spheroid population.

We have quantified all measurable substructure in our survey in 
previous papers  \citep{guhathakurta2006,kalirai2006gss,gilbert2007,gilbert2009gss,gilbert2012}.
In each field with identified kinematically cold components,
we perform maximum-likelihood multi-Gaussian fits to the stellar velocity distribution (see
\citet{gilbert2012} for details). 
The fits include a Gaussian with a large velocity dispersion \citep[$\sigma_{v} = 129$~\kms;][]{gilbert2007}
that corresponds to the spatially diffuse, kinematically hot underlying
spheroid distribution, and additional Gaussians, as supported by the data, that correspond to 
kinematically cold stellar streams ($\lesssim 10$\,--\,$\sim 20$~\kms).
This provides an estimate of the mean velocity ($\langle v_{\rm kcc}\rangle$) and 
velocity dispersion (\sigvkcc) of each kinematically cold component of the velocity distribution, 
as well as an estimate of the percentage of the stellar population in each field in 
kinematically cold tidal debris and in the underlying kinematically hot spheroid.  

The dashed red and blue MDFs in Figure~\ref{fig:cmds_and_mdfs} are computed by 
identifying subsets of stars in each field that are most likely to be associated 
with either the kinematically hot spheroid or with kinematically cold tidal debris features.
For fields with no identified tidal debris features, all the stars are 
included in the kinematically
hot distribution.  In fields with tidal debris features, the samples are defined 
using the maximum-likelihood fits to the velocity distribution.  Stars with velocities removed by more 
than 3\sigvkcc\ from the mean velocities of all tidal debris features in the field are included
in the kinematically hot MDF,  while stars that have velocities within one \sigvkcc\ of 
the mean velocity of any tidal debris features in the field are included in the kinematically cold component.  

These will not be pure samples of the kinematically hot and kinematically cold stellar populations, 
and the true difference in the MDFs can be expected to
be even larger than what is seen in Figure~\ref{fig:cmds_and_mdfs}.
However, in the vast majority of fields the tidal debris features have velocity dispersions of $\lesssim 20$~\kms.   Two fields have kinematically cold features that have \sigvkcc\,$\sim 30$~\kms\ \citep[and1 and a13;][]{gilbert2009gss} and only three fields have kinematically cold components with \sigvkcc$ > 30$~\kms\ (these are three of the innermost fields and are described in more detail in Section~\ref{sec:innerfields}).  
We use the fits to the velocity distributions to estimate the expected level of contamination in both the kinematically hot and cold samples.  The level of contamination varies by field, and depends on the strength of the substructure and its velocity with respect to M31's systemic velocity.  Due to our conservative cut of 3\sigvkcc, the expected contamination in the kinematically hot sample from stars belonging to tidal debris is negligible: $<1$ star per field (the full range is $0.03\le N_{\rm contaminants}\le 0.3$ per field).  The expected contamination of the kinematically cold sample by kinematically hot halo stars is significantly larger, ranging from 8\% to 40\%.  The lowest level of contamination occurs in fields where the kinematically cold tidal debris features have low \sigvkcc\ ($\lesssim 20$~\kms) and are well removed from M31's systemic velocity.  The highest level of contamination is found in the inner minor axis fields, where the tidal debris in the fields has a larger \sigvkcc\ ($>30$~\kms) and is centered on M31's systemic velocity, which is also the mean velocity of the kinematically hot halo.  

The MDFs of the kinematically hot and cold populations are similar in the 
innermost radial region, where the underlying spatially diffuse halo is relatively metal-rich, 
although the kinematically hot component is slightly more metal-poor than the 
kinematically cold component. 
However, in the middle two radial regions the MDF of the kinematically hot population 
is markedly different from the MDF dominated by stars in kinematically cold tidal 
debris features.  The stellar population
associated with the kinematically hot spheroid has a significantly larger
fraction of metal-poor stars and correspondingly lower median metallicity.  
Two-sided Kolmogorov-Smirnov (KS) tests 
confirm 
that the MDFs of the kinematically hot and kinematically cold populations 
are unlikely to be drawn from the same underlying 
distributions in the $20\le$\,\rproj\,$<40$~kpc or $40\le$\,\rproj$<90$~kpc radial bins 
($p=2.2\times10^{-8}$ and $p=0.001$, respectively).
The difference in these two populations will be explored more fully 
in Section~\ref{sec:subst_v_smooth}. 

The middle panel of Figure~\ref{fig:mdfs_cum} displays the MDFs of the kinematically hot sample in each radial region; they are shown in cumulative form by the dashed curves in the right panel.  As in Figure~\ref{fig:cmds_and_mdfs}, the innermost field is excluded from this sample (Section~\ref{sec:innerfields}).
The evolution of the MDF with radius is strongly apparent in the three inner radial bins even after substructure is 
removed.  Figure~\ref{fig:mdfs_cum} also indicates that the MDF of stars with $40\le$\,\rproj\,$<90$~kpc after the removal of substructure
approaches the MDF of all stars with \rproj\,$\ge90$~kpc. 
A two-sided KS test on these two MDFs yields a probability of 13\%, thus it is possible
these two MDFs are drawn from the same distribution.  However, we note that any substructure present cannot be identified in fields at \rproj\,$> 90$~kpc  due to the small number of M31 stars per field at these large radii.

\subsubsection{Contributions to the Stellar Population in the Innermost Fields}\label{sec:innerfields}

Multiple contributions to the stellar population in the innermost regions of M31's halo have been identified in the literature.
These include a disk component extending to a radius of $\sim 40$~kpc in the plane of the disk with an observed velocity dispersion of $\sim 30$~\kms\ \citep{ibata2005}, and an inner halo with a large velocity dispersion \citep[$\sigma_v>100$~\kms;][]{chapman2006,gilbert2007}.  Also present is a large shell system composed of debris from the interaction that produced the Giant Southern Stream, with a velocity dispersion that is highly dependent on position \citep{fardal2007,gilbert2007,fardal2012}.  This shell system presents in two dimensions as a series of `shelf' features: the northeast, west, and southeast shelves.   

Two-thirds of the fields at \rproj\,$\le 25$~kpc are on the southeast minor axis.  This complicates interpretation of the stellar population because the various components described above are all expected to have line of sight velocity distributions centered at M31's systemic velocity on the minor axis.  Given this complexity, we briefly discuss the evidence for contributions from these different stellar populations in our innermost fields, summarizing the results from our previous analysis of these fields in \citet{gilbert2007} and \citet{gilbert2009gss}.

In fact, the shell system produced by the progenitor of the Giant Southern Stream, and predicted by the models of \citet{fardal2007}, was confirmed by our detection of the as-yet unidentified southeast (SE) shelf in our minor axis fields from $10<$\rproj\,$<20$~kpc \citep{gilbert2007}.  The kinematic signature of the southeast shelf observed in our minor axis fields is distinct and clearly identified, centered on the systemic velocity of M31 and with a velocity dispersion that decreases from $55.5$~\kms\ in the 12~kpc field to $\sim 0$~\kms\ at the edge of the shelf at \rproj\,$\sim 19$~kpc.  This feature is removed from the kinematically hot component in Figures~\ref{fig:cmds_and_mdfs}, \ref{fig:mdfs_cum} and \ref{fig:fehvsrad_nosubst} as described in Section~\ref{sec:kccs}.  

The velocity distribution in the innermost field (\rproj$=9$~kpc) was found to be consistent with being drawn from a single Gaussian distribution with a large, halo-like velocity dispersion of $120$~\kms\ \citep{gilbert2007}.  This rules out a substantial disk component in this field, which would be at a distance of 38~kpc in the disk plane (assuming a disk inclination of $77\deg$); \citet{gilbert2007} estimated the disk fraction to be $<10$\% if it is present in this field.  However, the model of the shell system produced by the progenitor of the Giant Southern Stream predicts that debris from two shells will be present in this field, covering a similar range of velocities as expected for the kinematically hot halo.  It is probable that there is significant, kinematically unresolved tidal debris in this field.  Therefore, it is displayed separately in Figure~\ref{fig:cmds_and_mdfs} and not included in the MDFs of the kinematically hot population in Figure~\ref{fig:mdfs_cum}.  The MDF of the stars in this field is more metal-rich than that of the stars in the fields from $10<$\rproj\,$<20$~kpc, in keeping with the trend seen for the rest of the M31 halo. 

Finally, we note that we have found no evidence of disk contamination in any of our fields.  At \rproj$> 20$~kpc on the southern minor axis and in the southeast quadrant, fields in the \citet{ibata2005} and \citet{chapman2006} studies were also found to have no evidence of the extended disk component.  Likewise, we see no evidence of it in the velocity distributions of our fields in this region.  There is also no indication of contamination from Giant Southern Stream debris in our minor axis fields between $20<$\rproj\,$<40$~kpc.  In our fields with \rproj\,$<20$~kpc, there is also no evidence of a contribution to the velocity distribution from an extended disk component \citep{gilbert2007,gilbert2009gss}.  Our 12~kpc minor axis field would be at 51~kpc in the plane of the disk; the expected disk fraction at this radius on the minor axis is only 1\% for a smooth disk \citep{brown2006a}.  The rest of the minor axis fields are more distant, and thus would have an even smaller expected contribution.   We note that any disk contamination would be centered at the systemic velocity of M31, and be expected to have a velocity dispersion smaller than that measured for the SE shelf from 12 to 15~kpc; therefore it would be removed by the velocity cuts used to identify the kinematically hot population.

\subsection{Metallicity Gradient}\label{sec:met_grad}
\begin{figure}[tb!]
\plotone{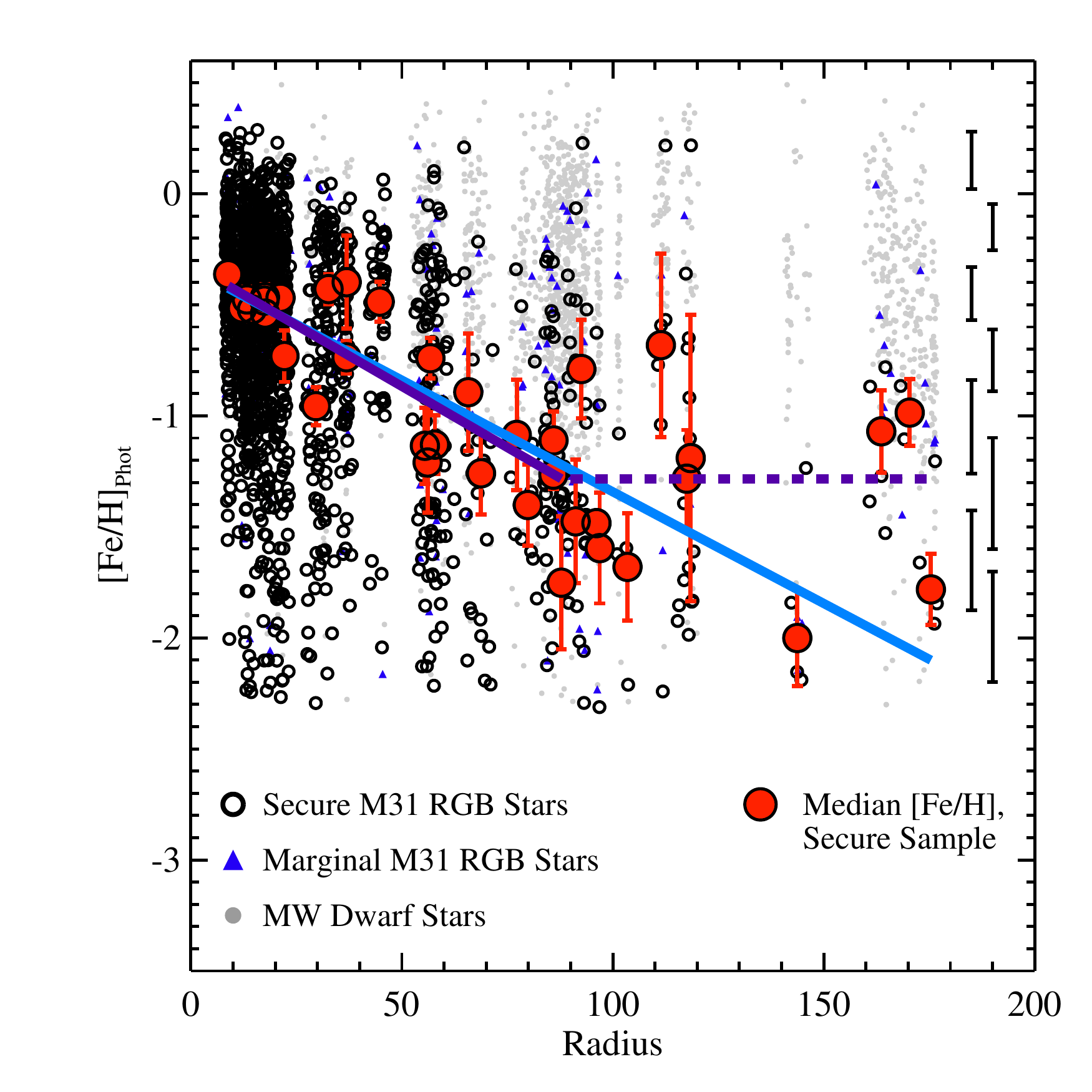}
\caption{Metallicity as a function of radius for our spectroscopic data, using photometric 
metallicity estimates.  
Stars are designated as secure M31 RGB stars ({\it black open circles}), marginal M31
RGB stars ({\it blue triangles}), or MW dwarf stars (secure and marginal, {\it grey dots})
(Section~\ref{sec:cleansample}).  Metallicity is based on comparing the position of the star in the CMD to a grid of isochrones at the distance of M31, and is therefore not physically meaningful for the MW stars. 
The large red points with error bars denote the median [Fe/H] values and the estimated error in the median 
for each spectroscopic field, computed using only the securely identified M31 stars.
Typical errors in  \feh\ for individual stars are shown by the error bars on the right.   
The solid lines show linear least-squares fits to the median field values (Section~\ref{sec:met_grad}); the blue line is fit using the full radial range of the data, while the purple line is fit only to fields within \rproj$<90$~kpc of M31's center.  The purple dashed line shows the median [Fe/H] value for fields at \rproj$>90$~kpc.  It is apparent that the decrease in metallicity with increasing radius seen in the previous figures is due to a continuous gradient in the median metallicity of M31's stellar halo extending to \rproj\,$\gtrsim 100$~kpc. 
}
\label{fig:fehvsrad}
\end{figure}

The previous section demonstrated that the metallicity distribution of 
M31's stellar halo becomes increasingly
more metal-poor with increasing distance from M31's center.    This could be due to 
either relatively sharp transitions in metallicity between different structural components in M31's stelar halo 
or a smoothly varying gradient in metallicity as a function of radius.
To differentiate these two senarios, we analyze the metallicity in 
our M31 halo fields as a continuous function of
\rproj.  Following the structure in Section~\ref{sec:innervouter}, we first measure 
the gradient when all M31 stars are included in the analysis, and then investigate the 
effect that the removal of coherent tidal debris features has on the observed properties in 
Section~\ref{sec:kccs_gradient}.  

Figure~\ref{fig:fehvsrad} displays the metallicity of individual M31 halo stars as a function 
of projected radius, as well as the median \fehp\ in each spectroscopic field.  A continuous 
gradient in the median metallicity of M31's stellar halo is clearly apparent in our fields 
to  \rproj$\sim 100$~kpc.  To quantify the decrease in the metallicity as a function of radius, we perform a 
linear least-squares fit to the median metallicity of each field in M31's halo as a 
function of \rproj.  Because there is significant field-to-field variation in the 
median metallicity values, we employ 3$\sigma$ clipping to remove the largest 
outliers.  In practice, this removes few fields and has a small effect 
on the measured slope.  
The best-fit slope to the full dataset is $-0.0101\pm 0.0005$~dex kpc$^{-1}$.   

However, Figure~\ref{fig:fehvsrad} shows that beyond \rproj$\sim 100$~kpc 
the median metallicity of M31's stellar halo 
may be relatively constant with radius,
although with large intrinsic field-to-field variation. 
Figure~\ref{fig:fehvsrad} also shows the linear least-squares fit restricted to fields 
from \rproj\,$=10$ to 90 kpc. This fit yields a metallicity gradient of 
$-0.0110\pm 0.0007$~dex kpc$^{-1}$, implying a gradient of $\sim 0.9$~dex in [Fe/H] over the 80~kpc included in the fit.  
The median metallicity of fields at \rproj\,$>90$~kpc is shown by the dashed line at \fehp\,$= -1.3$.

We are restricting our analysis
to stars identified securely as M31 RGB stars (Section~\ref{sec:cleansample}).    
Stars marginally identified
as M31 RGB stars have a higher contamination rate of MW dwarf stars \citep{gilbert2007}.   MW dwarf 
stars in our sample are more likely to be located in a region of the CMD that overlaps the metal-rich RGB isochrones, therefore the inclusion of
stars identified marginally as M31 RGB stars by the \citet{gilbert2006} diagnostic method 
often increases the median \feh\ in a given field.  In practice, there is little difference between the median metallicities 
calculated with and without the inclusion of marginal M31 RGB stars: the mean absolute difference is 0.08~dex across all fields.

\subsubsection{Effect of Tidal Debris on the Metallicity Gradient}\label{sec:kccs_gradient}

\begin{figure}[tb!]
\plotone{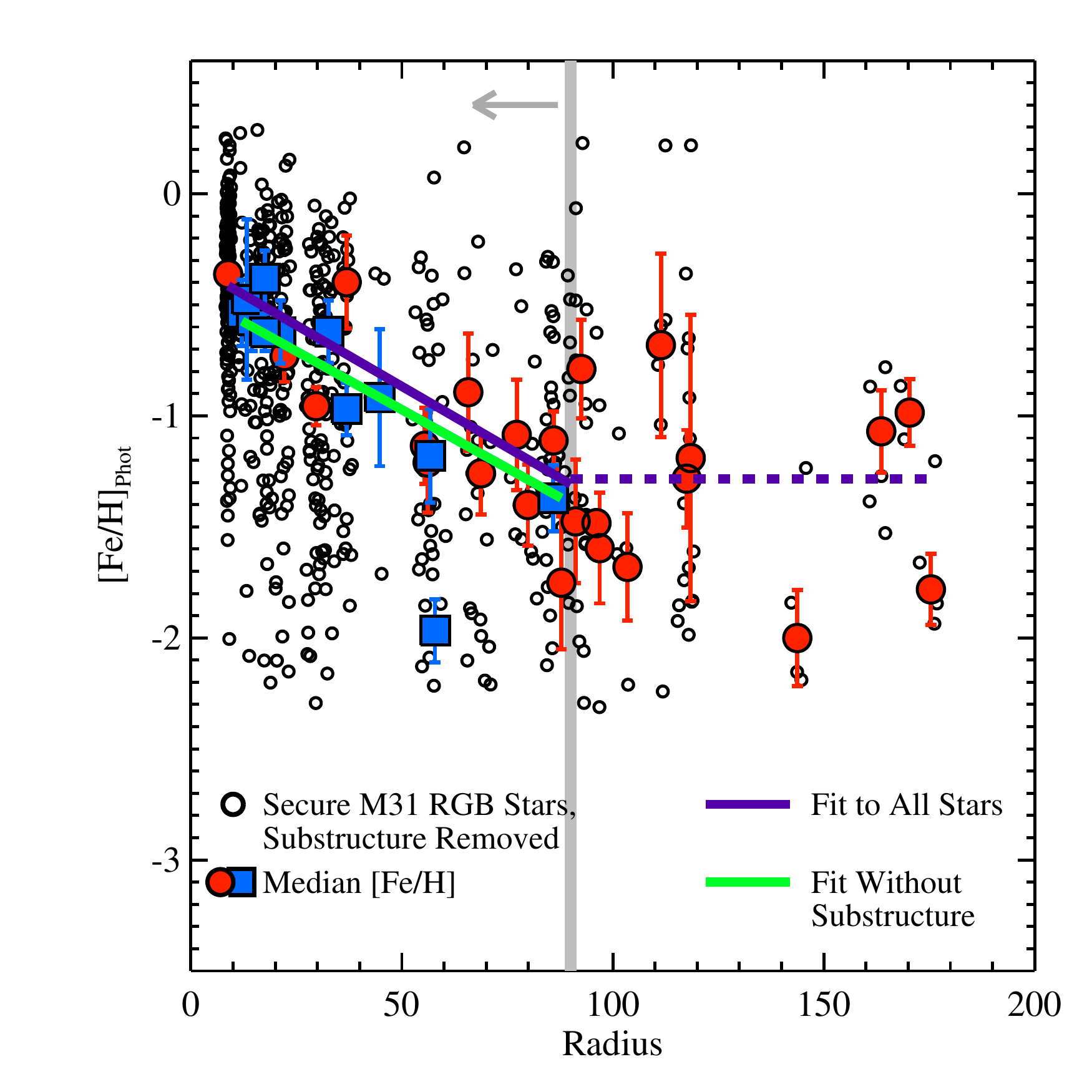}
\caption{Similar to Figure~\ref{fig:fehvsrad}, but showing only M31 halo stars that are not associated with
kinematically distinct tidal debris features. 
In fields with kinematically identified substructure, only stars
more than $3$\sigvkcc\ removed from the mean velocity of kinematic substructures
are included (Section~\ref{sec:kccs});  the median \feh\ values for these fields are shown as large blue squares.  Beyond \rproj\,$=90$~kpc, our spectroscopic fields have too few stars to kinematically identify tidal debris features; this boundary is marked by the light grey line and arrow.  The green line shows the linear least-squares fit to the median \feh\ values in fields with $10<$\rproj$<90$~kpc (Section~\ref{sec:kccs_gradient}).   As in Figure~\ref{fig:fehvsrad}, the purple line shows the fit to the median \feh\ values when all M31 stars are included, while the purple dashed line shows the median \feh\ of all M31 stars with \rproj$>90$~kpc.  When substructure is included the median metallicity of fields within $10\le$\,\rproj\,$\le 30$~kpc is approximately constant (Figure~\ref{fig:fehvsrad}); once substructure is removed, the fields within this radial range are consistent with the observed large scale gradient.
}
\label{fig:fehvsrad_nosubst}
\end{figure}

The gradients we calculate above include all securely identified M31 stars.  However, Section~\ref{sec:kccs} clearly demonstrates that the exclusion of stars most likely associated with tidal debris features has a significant effect on the 
MDFs of M31's stellar halo.

Figure~\ref{fig:fehvsrad_nosubst} again shows the metallicity of M31 stars, and the median \fehp\ in each field, 
this time including only stars that are more than 3\sigvkcc\ removed from kinematically identified tidal debris features (Section~\ref{sec:kccs}).   Beyond \rproj\,$\sim 90$~kpc, our spectroscopic fields have too few M31 stars to identify tidal debris features as multiple kinematic components.  Therefore, no fields beyond 90~kpc are included in the fits discussed below, as it is possible some of them are in fact dominated by substructure \citep{gilbert2012}.  As discussed in Section~\ref{sec:innerfields}, the innermost field at \rproj$\sim 9$~kpc is likely to have significant contamination from tidal debris features that cannot be isolated in the velocity distribution.  Therefore, to measure the gradient of M31's underlying, relatively smooth stellar halo we perform a linear least-squares fit (Section~\ref{sec:met_grad}) to the median \fehp\ in fields at $10<$\rproj\,$<90$~kpc.  The resulting metallicity gradient is 
 $-0.0105\pm 0.0013$~dex kpc$^{-1}$, consistent to within one sigma of the fit to all the stars, but with a lower normalization.   

Close inspection of Figure~\ref{fig:fehvsrad} shows that interior to \rproj\,$\lesssim 20$~kpc, 
the median metallicities of the spectroscopic
fields when all stars are included are almost uniformly high, with an approximately 
constant value of 
\fehp\,$\sim -0.45$ from 9 to 23~kpc (Figure~\ref{fig:fehvsrad}).  
As discussed in Section~\ref{sec:innerfields}, the M31 spectroscopic fields in this radial range include a  
significant population of stars from a single accretion event \citep{gilbert2007,gilbert2009gss}.  
The stellar population associated with this 
accretion event appears to have the same age and metallicity distribution regardless
of whether it is positioned in the Giant Southern Stream or the SE shelf \citep{brown2006a}.
When stars associated with this debris are removed, the median \fehp\ values in 
these fields are fully consistent with a metallicity gradient extending from our innermost 
fields to at least 90~kpc (Figure~\ref{fig:fehvsrad_nosubst}).

Finally, we investigate the significance of the best-fit gradient when the inner fields are removed.  If the range of the fit is restricted to $19.5\le$\rproj$\le 90$~kpc, removing the inner minor-axis fields on the SE shelf as well as the innermost field on the Giant Southern Stream, the best  fit slope is $-0.0089\pm0.0016$~dex kpc$^{-1}$, consistent with the fit to fields from $10<$\rproj\,$<90$~kpc and discrepant by more than 5.5$\sigma$ with no gradient.  If the fit is further restricted to the range $35\le$\rproj$\le 90$~kpc, excluding an additional four fields including one field on the minor axis at \rproj\,$\sim 30$~kpc and two Giant Southern Stream fields, the best fit slope is $-0.0089\pm0.0026$~dex kpc$^{-1}$, still fully consistent with the fit to fields from $10<$\rproj\,$<90$~kpc and discrepant by more than 3$\sigma$ with no gradient. Only by severely restricting the radial range, to $50\le$\rproj$\le 90$~kpc (including only ten fields in the fit) does the best fit slope become consistent within $1\sigma$ with zero ($-0.0048\pm0.0047$~dex kpc$^{-1}$), although we note that due to the large error bar on this fit it is also consistent with the best fit slope to fields from $10<$\rproj\,$<90$~kpc at the $\sim 1\sigma$ level.

\subsection{Discussion of Potential Systematic Effects}\label{sec:biases}
\begin{figure}[tb!]
\plotone{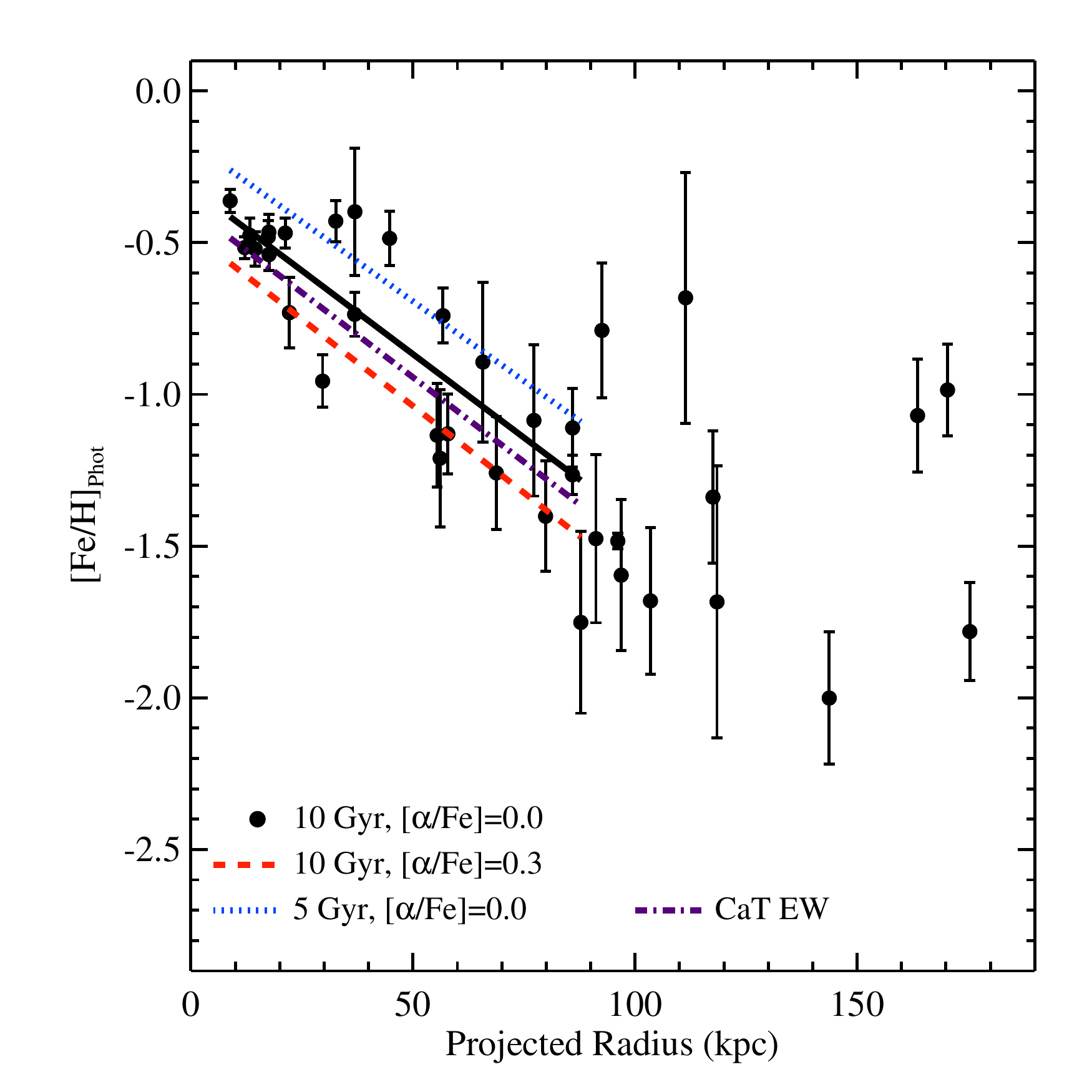}
\caption{  
Median photometric metallicity of all M31 stars in each field, along with the linear fit to the median field values (solid line), calculated using the nominal 10 Gyr, \afe$=0.0$ \citet{vandenberg2006} isochrones.  Error bars show the error in the median value of each field.   
The dashed and dotted lines show the linear least-squares fits to the median metallicity estimates in each field calculated using alternative methods: 10 Gyr, \afe$=0.3$ isochrones (red dashed line), 5 Gyr, \afe$=0.0$ isochrones (blue dotted line), and spectroscopic metallicity estimates based on the \caii\ triplet EW (purple dot-dashed line).  
}
\label{fig:fehvsrad_systematic}
\end{figure}

To measure a metallicity for each M31 RGB star, we must assume a single age and $\alpha$-element abundance for the stellar population.  We also remove from the sample stars that were classified as M31 RGB stars but were bluer than the most metal-poor isochrone.  In the following sections, we investigate the possible systematic effects these assumptions may have on the magnitude of the measured gradient. 
 
\subsubsection{Metallicity Measurement Methodology}\label{sec:biases_measurement}
In using a single set of isochrones to compute [Fe/H] for each star, we are implicitly assuming (in addition to a simple stellar population) a constant age and alpha enrichment throughout M31's stellar halo.  However, deep photometric observations reaching below the ancient main sequence turn-off in fields from 11 to 35~kpc have shown that at any one location there is a range of ages in M31's stellar halo \citep{brown2006,brown2007,brown2008}.   It is certainly possible that M31's stellar halo harbors a gradient in age and it is also likely that stars with different metallicities have different relative abundances of $\alpha$ elements and Fe, as seen in the Milky Way and its satellites \citep[e.g.,][]{venn2004,ishigaki2012,bensby2014}.  

In Figure~\ref{fig:fehvsrad_systematic} we explore the systematic effect a gradient in mean age and alpha abundance might have on our measured metallicity gradient.  Figure~\ref{fig:fehvsrad_systematic} shows the median metallicity in each field with radius, calculated using the nominal 10~Gyr, [$\alpha$/Fe]$=0$ isochrones 
along with several fits to the data using alternate metallicity estimates.  Assuming a younger age or higher [$\alpha$/Fe] in estimating the metallicity of stars results in gradients with slopes that are consistent with each other and the nominal fit by well within the $1\sigma$ errors of the fits.  However, the normalizations are significantly offset.  Thus a systematic change from a dominantly old and high [$\alpha$/Fe] population in the outer regions to a dominantly younger and low [$\alpha$/Fe] population in the inner regions would result in the true gradient being steeper than we have measured.  However, we know that while M31's halo fields do have a population of stars with ages $\sim 7$~Gyr, these stars comprise a small part of the overall population, which is primarily older than $\sim 10$~Gyr \citep{brown2008}.  Furthermore, \citet{brown2008} did not find evidence of a monotonic increase in age with radius in their three HST fields at 11, 21, and 35~kpc (although they did find evidence for a monotonic decrease in [Fe/H] with increasing radius).  Therefore, any effect 
on the metallicity gradient from a systematic change in the star formation history of the population as a function of radius will be significantly smaller than indicated in Figure~\ref{fig:fehvsrad_systematic}.  

We can also compare the gradient measured from our photometric metallicity estimates with the gradient measured 
from the spectroscopic \caii\ EW metallicity estimates (Section~\ref{sec:met_est_spec}).  
From the earlier comparison of spectroscopic and photometric metallicities (Section~\ref{sec:met_est_comp}), 
we expect a similar gradient if we fit to the median spectroscopic metallicity per field.   
The linear least-squares fit to the median spectroscopic metallicity in each field 
yields 
a slope of  $-0.0112\pm 0.0016$~dex kpc$^{-1}$, nearly identical  to the slope measured 
from the nominal, 10~Gyr, [$\alpha$/Fe]$=0$ photometric metallicity estimates.  
We have imposed a spectral S/N cut of 3 pixel$^{-1}$, as that is the S/N value below which \citet{ho2014} found the \caii\ EW measurements to be unreliable.  

We have explored the effect on the observed gradient of assuming different ages or [$\alpha$/Fe] abundances for the theoretical stellar models (which largely affects the normalization), and using \caii\ triplet-based spectroscopic metallicities. It is notable that while the details of the gradient change, a gradient is seen regardless of the methodology used to calculate [Fe/H].   
Furthermore, since stars that form within host stellar halos are expected to be preferentially found in the inner regions, the most probable change in the age and chemical abundance properties of the population with radius would be from a younger, less $\alpha$-enhanced stellar population in the inner regions to an older, more $\alpha$-enhanced population in the outer regions (Section~\ref{sec:sims}).  This would result in a steeper gradient.  

\subsubsection{Inclusion of Stars Bluer than the Isochrone Grid}\label{sec:baises_bluestars}
The top panels of Figure~\ref{fig:cmds_and_mdfs} show that in all radial regions, there are stars identified as M31 RGB stars whose blue colors place them outside the isochrone grid in the CMD.  We have removed these stars from the above analysis, since it would require an extrapolation of the isochrone grid to assign a photometric metallicity to these stars.  However, as noted above, while a small fraction of these stars could be asymptotic giant branch stars or MW main sequence turn-off stars, most are likely M31 RGB stars shifted beyond the isochrone grid due to photometric errors or modeling errors in the isochrones, or stars that are indeed more metal-poor than the most metal-poor isochrone in the grid (\feh\,$=-2.3$).  

Under this assumption, removal of these stars from the MDFs and measurement of the metallicity gradient would introduce a systematic bias that could cause our measured gradient to be flatter than the true gradient.   
While stars bluer than the most metal-poor isochrone comprise only 0.5\% of the stars within \rproj\,$<20$~kpc, and 1.5\% of the stars within $20\le$\,\rproj\,$<40$~kpc range, they comprise 7\% and 14\% of the stars within the $40\le$\,\rproj\,$<90$~kpc and \rproj\,$\ge 90$~kpc ranges, respectively.  Given the observed metallicity gradient in M31, the increasing percentage of metal-poor stars with radius is indicative that these stars are in fact M31 stars that are intrinsically bluer than the most metal-poor isochrone.  Furthermore, the photometric errors as a function of magnitude of these blue stars are fully consistent with the full sample of M31 stars. 

We can easily estimate the effect removal of these stars has on the measured gradient by including these stars in our estimates of the median \feh\ in each field (assigning them a value of \feh$=-2.3$) and measuring the gradient using these median field values.  The resulting gradients are $-0.011\pm0.0007$~dex kpc$^{-1}$ for all M31 stars and $-0.0105\pm0.0013$~dex kpc$^{-1}$ when tidal debris features are removed, each of which are fully consistent with the slopes found when stars bluer than the isochrones are excluded. 
We therefore conclude that removal of the M31 stars bluer than the isochrone grid is not significantly biasing our results.

\section{Comparison with Previous Metallicity Measurements of M31's Halo}\label{sec:litcompare}
The gradient observed in our spectroscopic fields is largely consistent with results in the literature that also measured a gradient in M31's stellar halo \citep[Section~\ref{sec:intro};][]{kalirai2006halo,koch2008,tanaka2010,ibata2014}.   We have presented a large increase in fields and number of M31 halo stars over the data included in \citet{kalirai2006halo} and \citet{koch2008}, greatly increasing the number of stars beyond 40~kpc, gaining more contiguous radial coverage, and including fields located beyond the southeast quadrant and southern minor axis.  The difference in the median metallicity at large radii between the data presented here and by \citet{koch2008} is likely due primarily to differences in sample selection, as well as differences in the metallicity measurement methodology.

This work, as well as the work cited above, stands in contrast to the \citet{chapman2006} result,  which measured an approximately constant mean metallicity of [Fe/H]$\sim -1.4$ out to $\sim 60$~kpc in 23 slitmasks, primarily located near the major axis of M31.  The \citeauthor{chapman2006} study therefore probed a different region of M31's halo than that probed by the above studies or the current study.  They employed a conservative cut in velocity to remove any stars associated with the disk of M31 \citep{ibata2005}: only stars with velocities removed by more than 160~\kms\ from the disk velocity were included in the halo sample.  However, they specifically note that their result ``should not be interpreted as claiming that all components of the $R < 70$~kpc stellar halo are metal-poor, but simply that a non-rotating metal-poor component to M31 exists'' \citep{chapman2006}.  If a component of the M31 stellar halo rotates, it would have been preferentially removed (compared to a non-rotating component) from the \citeauthor{chapman2006} sample by their velocity cuts.  This effect would be maximized along the major axis, where the majority of the \citeauthor{chapman2006} fields lie.  This would introduce a systematic bias against the rotating population.  The larger the rotation velocity, the larger the systematic bias would be (assuming the rotation is in the same direction as the disk rotation).    

Indeed, \citet{dorman2012} found significant rotation ($|v-v_{\rm M31}|\sim 50$~\kms) in the innermost regions of M31's halo.  A large fraction of these stars would have been preferentially removed from \citeauthor{chapman2006}'s study.   This rotating component is likely dominated by metal-rich stars.   \citet{dorman2013} present evidence that a fraction of the stars with halo kinematics have a disk-like luminosity function, indicating they have been dynamically heated from the disk.  Furthermore, they found a large core radius (10.6~kpc) for the density profile of the stellar halo, while a study of the density of resolved, metal-poor blue horizontal branch stars found a significantly smaller core radius ($<3$~kpc) \citep{williams2012}.  These two results can be reconciled only if the metal-poor population in M31's stellar halo follows a a different density profile in the inner regions than the dominant, metal-rich population (Williams et al., in prep).  This interpretation is consistent with the different power-law indices measured for metal-rich and metal-poor stars in the outer regions by \citet{ibata2014}.  It is also similar to the differences between inner and outer halo populations in the Milky Way inferred from the work of \citet{carollo2010}, \citet{an2013} and \citet{allende-prieto2014}.

However, the \citet{dorman2012} sample is interior to the majority of the \citet{chapman2006} fields, as well as the fields presented here.  We thus do not know if the rotation observed in the innermost regions of M31's halo extends into the \citeauthor{chapman2006} fields.  If it does, \citeauthor{chapman2006} would have preferentially removed a significant fraction of stars in the rotating halo component.  This could explain the discrepancy between the \citeauthor{chapman2006} study and other studies, including this one, that find significant gradients in M31's stellar halo.

\section{Comparison of M31's Metallicity Gradient with Expectations from Simulations}\label{sec:sims} 
Section~\ref{sec:met_halo} demonstrates that there is a significant and continuous gradient in M31's stellar halo from our innermost fields to $\sim 90$~kpc, and that this gradient becomes more pronounced when tidal debris features are removed.  In this section we discuss the physical interpretation of M31's metallicity profile using the results of cosmological simulations of stellar halo formation.

There are two physical mechanisms that have been shown to produce metallicity gradients in stellar halos.  
In stellar halos built entirely of accreted satellites, a metallicity gradient as strong as 
$\sim -1.5$~dex over the radial 
range 10\,--\,100~kpc can be produced when a small
number (1 or 2) of more massive satellites dominate the bulk of the stellar halo population \citep{cooper2010}.
The magnitude of the gradient, and even whether or not a gradient is produced, 
depends sensitively on the merger history of the host galaxy.  Many
stellar halos produced entirely via accretion are not expected to have a 
large scale metallicity gradient \citep{font2006b,cooper2010}.

Additionally, metallicity gradients are produced when a fraction of the stellar halo is composed of stars that 
form in the host halo (i.e., in the dominant dark matter halo progenitor). 
These stars have been shown to be a 
generic feature of simulations performed using smoothed particle hydrodynamics, which follow the evolution of the baryonic components as well as the dark matter particles.  These stars are primarily formed in the primordial disk of the galaxy \citep{zolotov2009,mccarthy2012,tissera2012} and subsequently dynamically heated into the stellar halo population.
These kicked-out stars generally have higher metallicities
than the majority of stars accreted in satellite systems and contribute preferentially to the inner regions of the stellar halo,
naturally forming a gradient in metallicity.  The resulting magnitude of the large-scale metallicity gradient is however still dependent on the merger history of the galaxy \citep{tissera2014}, because that is a significant factor in determining the relative fraction of stars formed in the host halo and accreted.

\begin{figure*}[tb!]
\plottwo{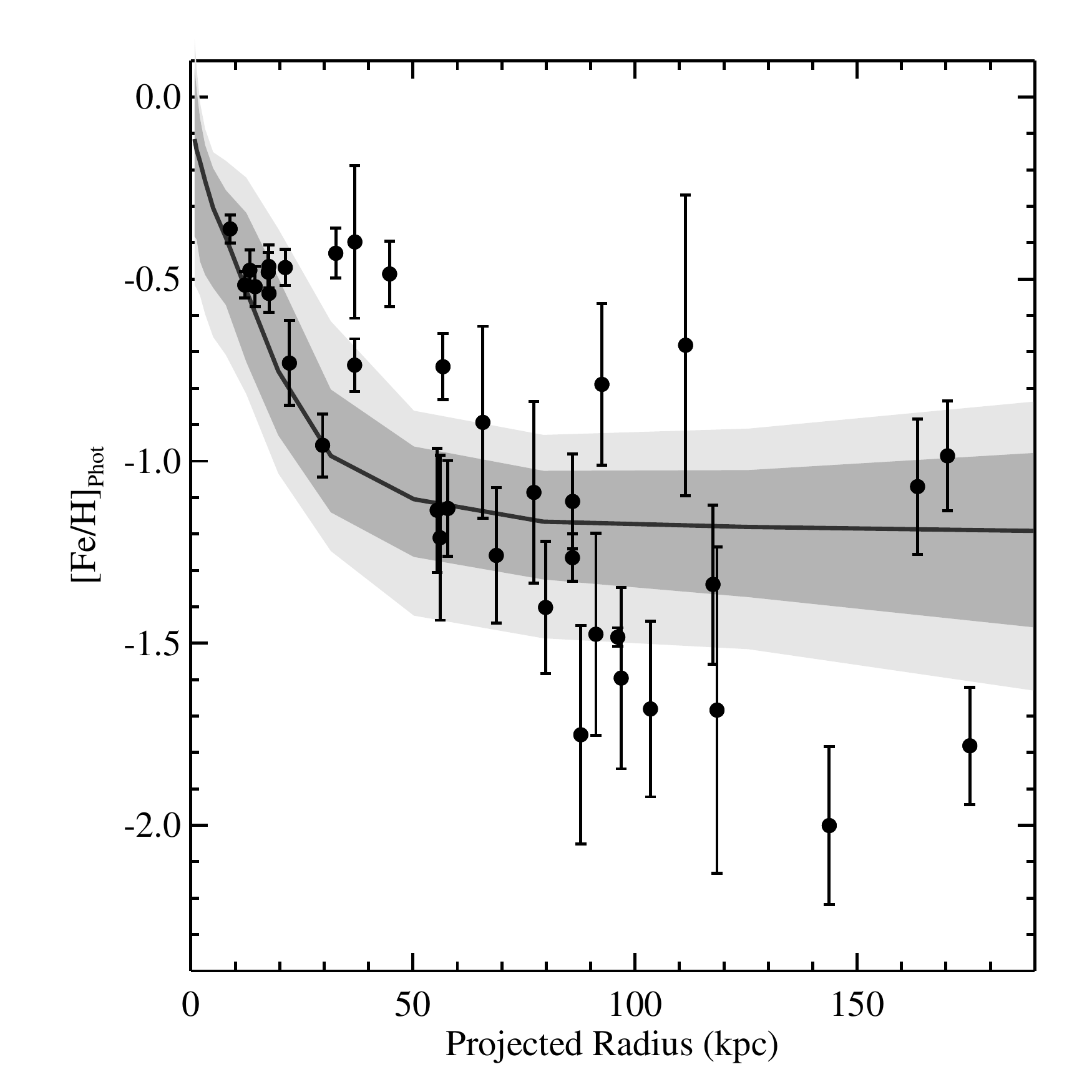}{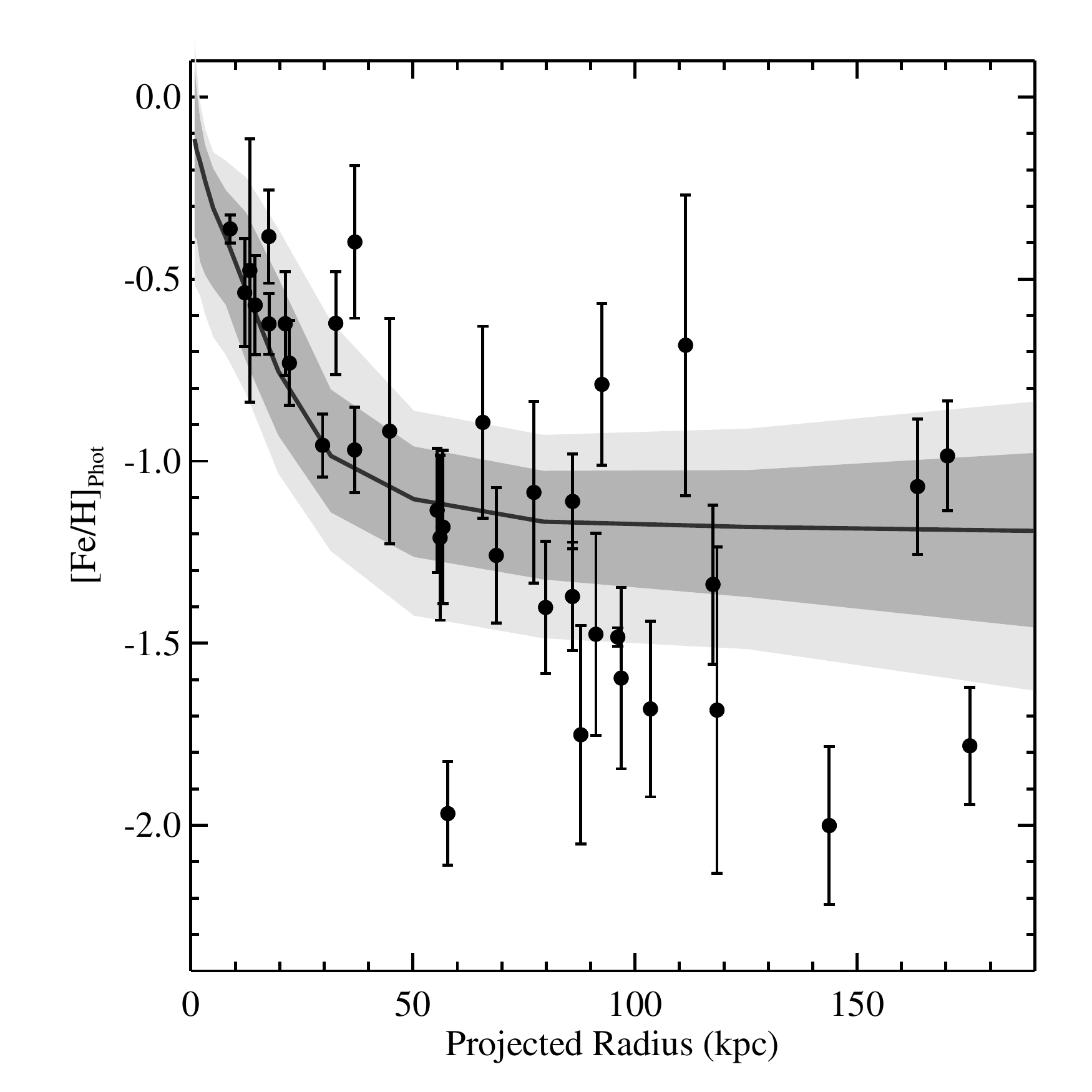}
\caption{Comparison of the median metallicity of the M31 halo fields, calculated with ({\it left}) and without ({\it right}) tidal debris features, with the metallicity profile of the stellar halos in the simulations of \citet{font2011} .    
The black curve shows the median spherically averaged metallicity as a function of radius for all 400 galaxies in the simulation, while the dark (light) grey shaded region encloses the 14th and 86th (5th and 95th) percentiles.  
}
\label{fig:fehvsrad_simulation}
\end{figure*}

Because stellar halo properties depend on the individual merging history, the best practice is to compare observations to a large suite of simulations.  Only one large suite currently exists, that of \citet{font2011}, which used the GIMIC suite of cosmological hydrodynamical simulations to study the formation of stellar spheroids in a sample of $\approx 400$ $L_*$ disk galaxies.  In these simulations,  stars formed in the host halo dominate stellar spheroids by mass for $R\lesssim 30$~kpc, while accretion dominates at larger radii.    
The stacked minor-axis surface brightness profile of the simulated halos agrees well with the minor axis surface brightness profile observed in M31 \citep[Figure~4 of][]{font2011}.  In Figure~\ref{fig:fehvsrad_simulation}, the median metallicity in each of our M31 halo fields is compared to the median spherically averaged metallicity profile (and the range) of the \citet{font2011} sample.  The simulations are compared to the median field values computed using the full halo dataset, as well as the median field values when tidal debris features are excluded.  The removal of tidal debris features in effect allows us to compare with the simulations a psuedo-version of M31's stellar halo drawn from an alternate evolutionary path: i.e., a quieter recent merger history.

The data are in rough agreement with the average results of the simulations, however M31's metallicity profile appears, for the most part, to continue to decrease with radius past the point at which the \citet{font2011} profile flattens out.   
The best agreement is in the inner $\sim 80$~kpc after removal of kinematically identified tidal debris features in the data.  A $\chi^2$ analysis confirms this; for fields at \rproj\,$< 90$~kpc, the calculated $\chi^2$ values of the data with respect to the simulation profile is more than a factor of three lower when tidal debris features are removed.  We remind the reader that the majority of the tidal debris removed in the right panel of Figure~\ref{fig:fehvsrad_simulation} comes from a single recent accretion event: the event that produced the Giant Southern Stream and extended shell system.  Thus, the comparison between the left and right figures implies that M31 has experienced a more massive, recent accretion event than expected for the average stellar halo in the \citet{font2011} simulations.  

\citet{tissera2012} simulated a sample of six stellar halos at a higher resolution than the \citet{font2011} work, and found similar behavior in the metallicity profiles and the stellar population characteristics of stars formed in the host halos.  The metallicity profiles are in general quite steep in the first 20~kpc, and then flatten to shallower gradients, which the authors fit in the range 25\,--\,150~kpc \citep{tissera2014}.  They found that the slopes in this range are largely flatter than what is found in M31's stellar halo, but a couple of their halos do have slopes comparable to the measured gradient in M31.  The halos with steeper metallicity gradients have a larger fraction of halo stars originating from massive satellite progenitors (dynamical masses $>10^9$\Msun) than the halos with shallower metallicity gradients.   
The \citet{tissera2014} metallicity profiles show an increase in the dispersion of the metallicity profiles within a single halo as a function of radius, a feature also seen in the M31 metallicity profile.  

The M31 metallicity profile is consistent with the expectations for simulated halos with significant fractions of stars formed in the innermost regions of the host halo and subsequently kicked out to more eccentric orbits.  However, these populations are only expected to be present in any significant fraction within the first 20\,--\,30~kpc of the stellar halo.  
Thus, some other mechanism must generate large scale gradients at larger distances.  A
possible mechanism is the mass-dependence in the accretion history, as
noted above.  Dynamical friction drives massive satellites to the
center of their host, and these massive satellites also tend to have
higher metallicities, which can create a large-scale metallicity gradient like
that observed in M31's stellar halo.  Exactly this effect has been
observed in semi-analytic $N$-body models of stellar halo formation
\citep{cooper2010} and SPH simulations \citep{tissera2012}, both based
on the Aquarius project \citep{springel2008}.  While it is not
certain that the star formation and feedback in these simulations
represents the real universe, they provide support for the basic
argument.  Furthermore, if the simulations are interpreted at face
value, M31's halo was likely built with more massive (e.g.,
$>10^9$\Msun) satellite progenitors than the average halo in the
simulations.   The observed metallicity gradient in M31 implies
that in addition to a population of stars formed within the inner regions of the host halo, 
something else must generate and maintain the gradient found in the outer halo,
and a plausible mechanism is an accretion history featuring more massive satellites 
than the typical halo.

Finally, we note that beyond 100~kpc, the field-to-field variation in the median metallicity increases dramatically.   \citet{gilbert2012} noted that the scatter in the surface brightness of individual lines of sight was larger than expected based on the observational errors (Figure~\ref{fig:sb}), and was roughly consistent with the field-to-field variation in star counts expected for stellar halos built via the accretion of many satellite progenitors \citep{bullock2001}.  The field-to-field variation in the median metallicities in M31's halo, especially at large radii, reinforce the implication from the surface brightness profile that in addition to the larger progenitors, the outer halo is composed of many small accreted objects that are not resolvable into kinematically cold or photometrically distinct streams with the available data.  

\section{Comparison of the kinematically hot and cold stellar populations in M31's halo}\label{sec:subst_v_smooth}

\begin{figure}[tb]
\plotone{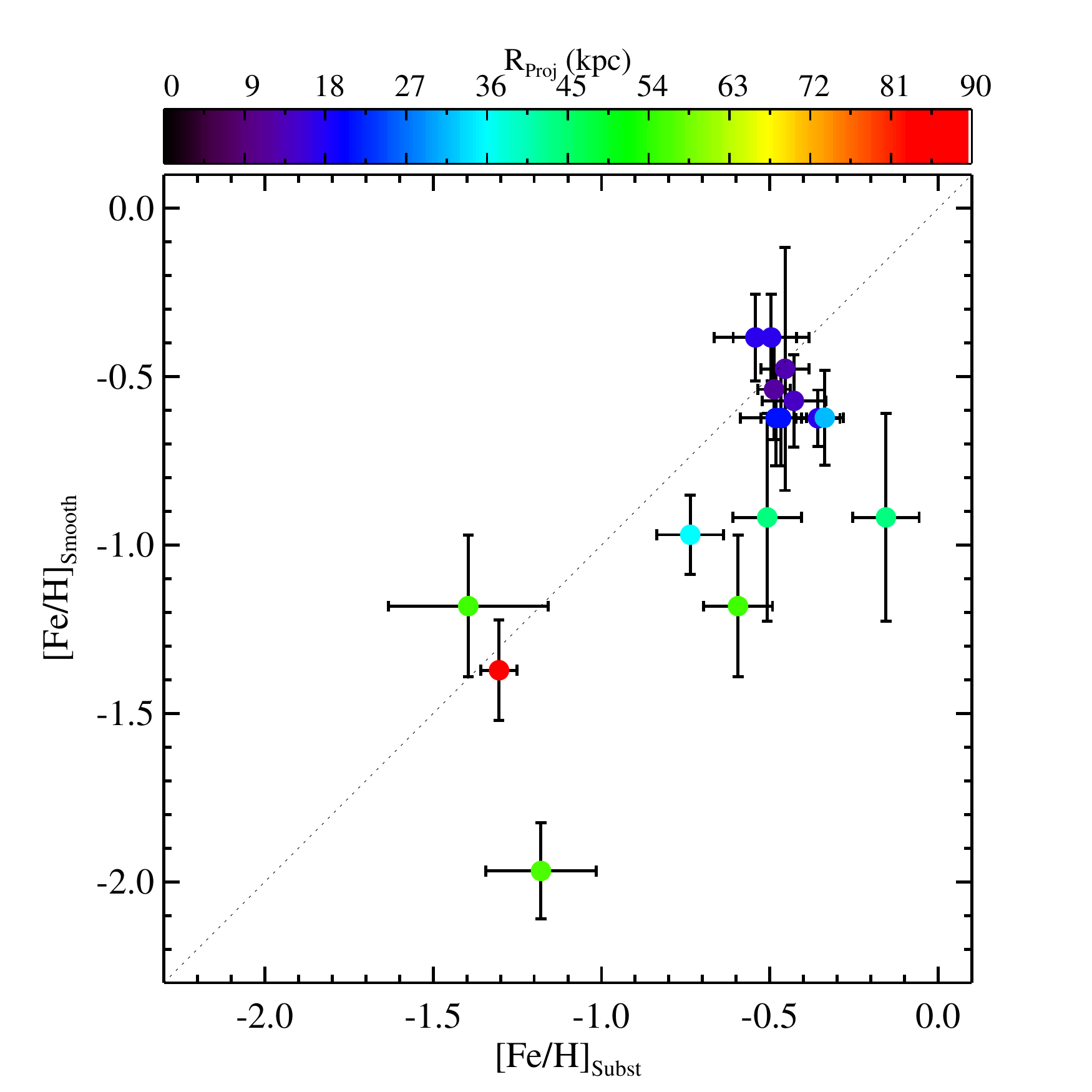}
\caption{  
Comparison of the median metallicity of stars associated with each kinematically identified cold tidal debris feature and stars associated with the comparatively smooth, kinematically hot stellar component in individual lines of sight in M31's stellar halo.  Error bars show the error in the median value of each field.     
Points are color-coded by the field's mean projected distance from M31's center.  The dotted line shows the one-to-one line. The majority of tidal debris features have a higher median metallicity than the spatially diffuse halo in the same location, even though most of the kinematically identified substructure is found in fields within $\sim 35$~kpc of M31's center, where the metallicity of the dynamically hot population is relatively high.  The largest differences between the median metallicity of stars associated with the kinematically hot stellar population and the kinematically cold substructure is seen in fields at larger radius.  
}
\label{fig:subst_v_smooth_rproj}
\end{figure}

The previous section demonstrates that in aggregate there is a significant difference in the mean metallicity of stars most likely associated with the kinematically cold tidal debris features in M31's stellar halo and stars most likely associated with the underlying kinematically hot halo (e.g., Figure~\ref{fig:cmds_and_mdfs}; Section~\ref{sec:kccs}).  In this section, we investigate the difference in metallicity of kinematically hot and cold components on a field-by-field basis, as a function of projected distance from M31's center, strength of the tidal debris feature, and surface brightness.   These observational comparisons are then used to motivate a discussion of the likely physical origins of the kinematically hot and cold subpopulations in M31's stellar halo.

The magnitude of the aggregate difference between the metallicity of tidal debris features and the underlying stellar halo shown in Figure~\ref{fig:cmds_and_mdfs} clearly has some dependence on projected distance from M31's center.  Figure~\ref{fig:subst_v_smooth_rproj} compares, for each M31 halo line of sight, the median metallicity of stars most likely associated with tidal debris features and stars most likely associated with the kinematically hot population, with points color-coded by the projected distance of the field from M31's center.    
Fields that contain more than one kinematically cold component in the stellar velocity distribution are represented in this figure with one point for each separate cold component.  As expected from the metallicity distribution functions (Figure~\ref{fig:cmds_and_mdfs}) and the analysis of the metallicity gradient (Section~\ref{sec:kccs_gradient}; Figure~\ref{fig:fehvsrad_nosubst}), the majority of the tidal debris features are more metal-rich than the underlying kinematically hot, spatially diffuse stellar population, even in the relatively metal-rich inner regions of M31's stellar halo.   
However, the largest metallicity differences between kinematically identified substructure and M31's underlying stellar halo are seen in fields farther than $\sim 35$~kpc from M31's center, where M31's stellar halo becomes increasingly metal-poor.

\begin{figure}[tb]
\plotone{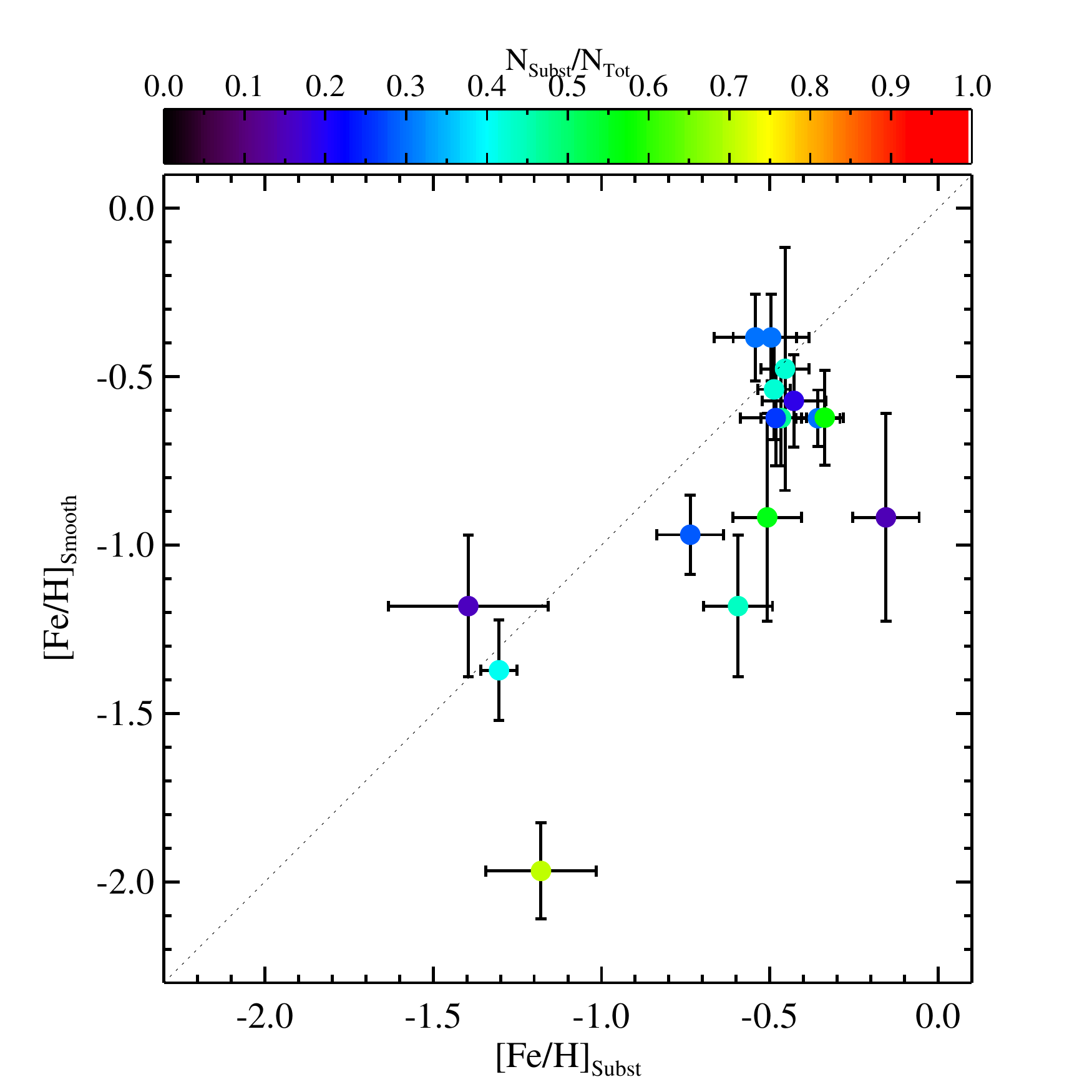}
\caption{  
Same as Figure~\ref{fig:subst_v_smooth_rproj}, with points color-coded by the fraction of stars in the field that are estimated to belong to the kinematical substructure.  The difference in metallicity between the two populations is generally largest when the tidal debris dominates the stellar population in the field.}
\label{fig:subst_v_smooth_frac}
\end{figure}

In addition to distance of the field from M31's center, the strength of a tidal debris feature may impact the magnitude of the observed difference in metallicity between stars associated with the tidal debris feature and the stellar halo; this effect will be discussed further below.    Figure \ref{fig:subst_v_smooth_frac} once again shows the median metallicity of kinematically cold tidal debris features and the kinematically hot population in each field, with each point color-coded by the fraction of stars in the field that are part of the kinematically cold component, as determined from fits to the stellar velocity distribution in each field (Section~\ref{sec:kccs}).    
The largest metallicity differences are typically seen in fields where the tidal debris feature comprises $\gtrsim$ 50\% of the stellar population.    

\begin{figure}[tb]
\plotone{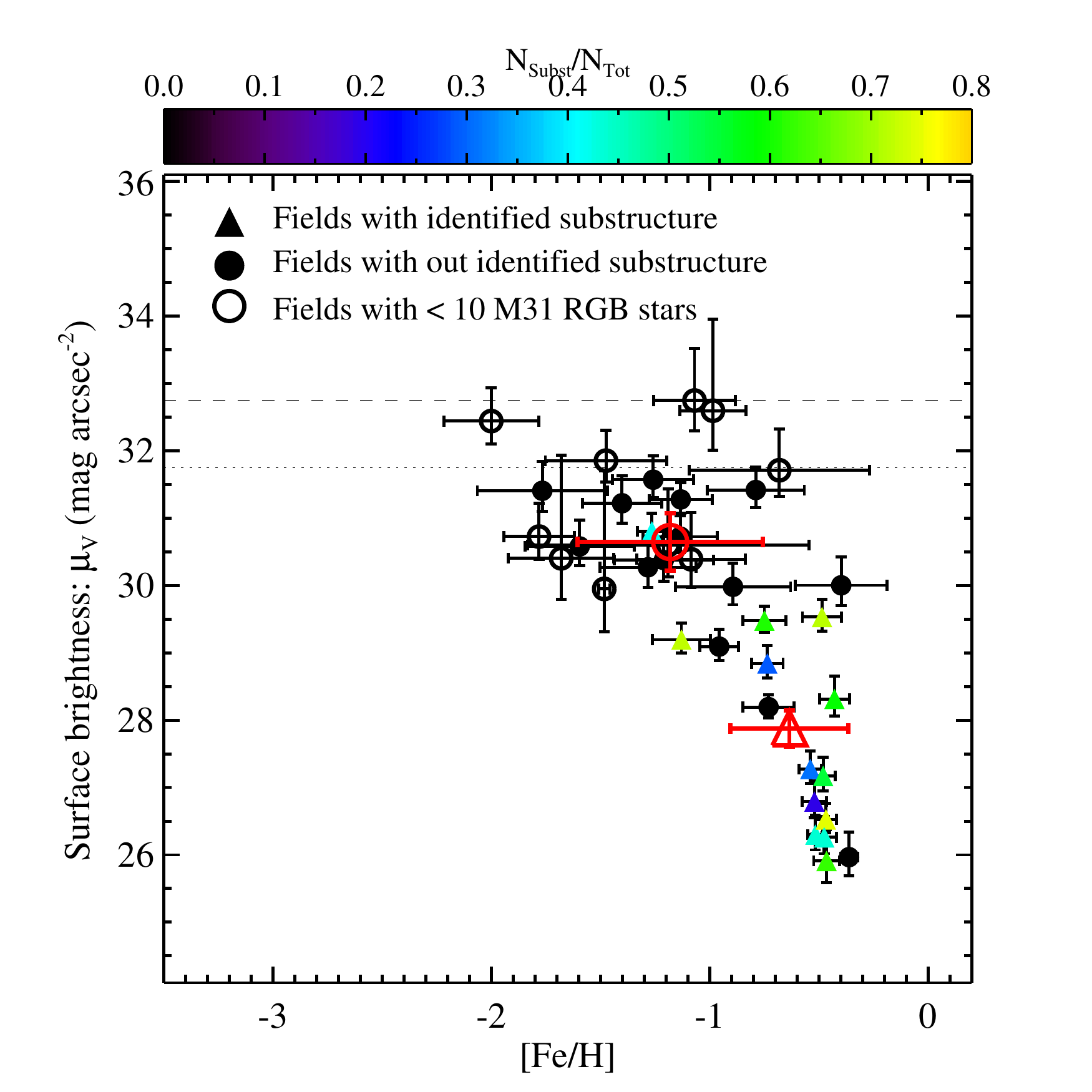}
\caption{ 
The surface brightness of lines of sight through M31's stellar halo as a function of the median metallicity.  Values are computed using all M31 halo stars in the field, and points are color-coded by the fraction of M31 halo stars associated with tidal debris features.
The large red points show the mean values for fields with and without identified substructure.   The dotted line denotes the faintest tidal debris feature identified in our survey, while the dashed line denotes the faintest field observed.  Fields with identified substructure are preferentially more metal-rich and higher surface brightness than fields without.
}
\label{fig:subst_v_smooth_g09a}
\end{figure}

In stellar halos built largely through the accretion of smaller stellar systems, there is a fairly complex interplay expected between the surface brightness of tidal debris features, their location in the stellar halo, and their metallicity, compared to the surface brightness and metallicity of the underlying stellar halo population.  \citet{gilbert2009a} explored these themes through a comparison of a subset of the M31 observations presented here and simulations of stellar halo formation via accretion presented by \citet{robertson2005} and \citet{font2006}.  To provide a physical context for the trends observed in our dataset and discussed above, we will briefly summarize the discussion of \citet{gilbert2009a}, referencing updated versions of the main figures that include the expanded dataset presented here.

Figure~\ref{fig:subst_v_smooth_g09a} displays the surface brightness of each of our spectroscopic fields in M31's stellar halo as a function of the median metallicity of stars in the field.   
Each point is color-coded by the total fraction of stars in tidal debris features in the field.  Black points have no detected tidal debris features.  As would be expected from the above discussion, fields with tidal debris features are preferentially more metal-rich than fields without substructure.  They are also preferentially at higher surface brightness, as would be expected given both the enhancement of stellar density provided by the tidal debris feature, and the large fraction of spectroscopic fields with kinematically identified tidal debris features in the inner, higher surface brightness regions of M31's stellar halo.

\begin{figure}[tb]
\plotone{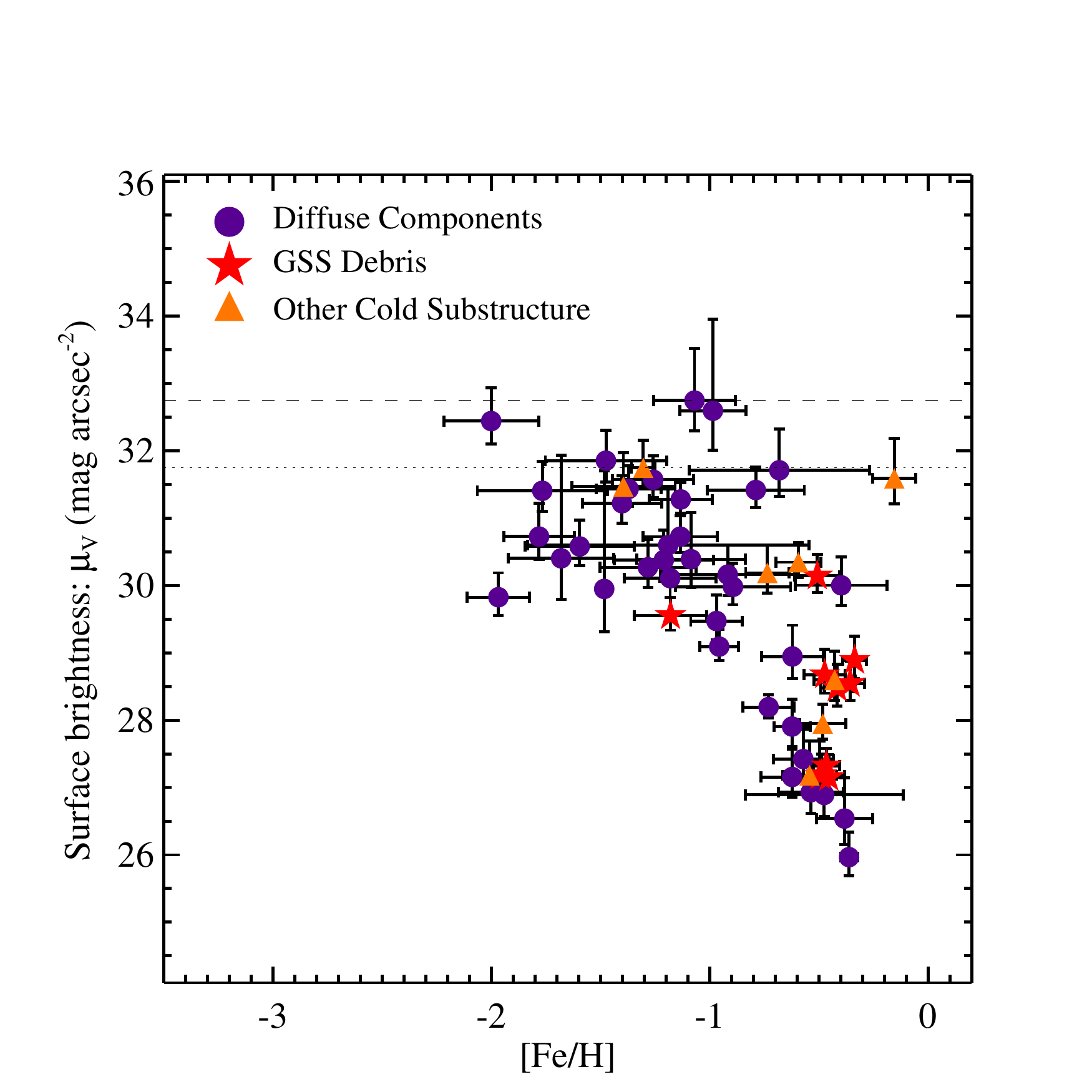}
\plotone{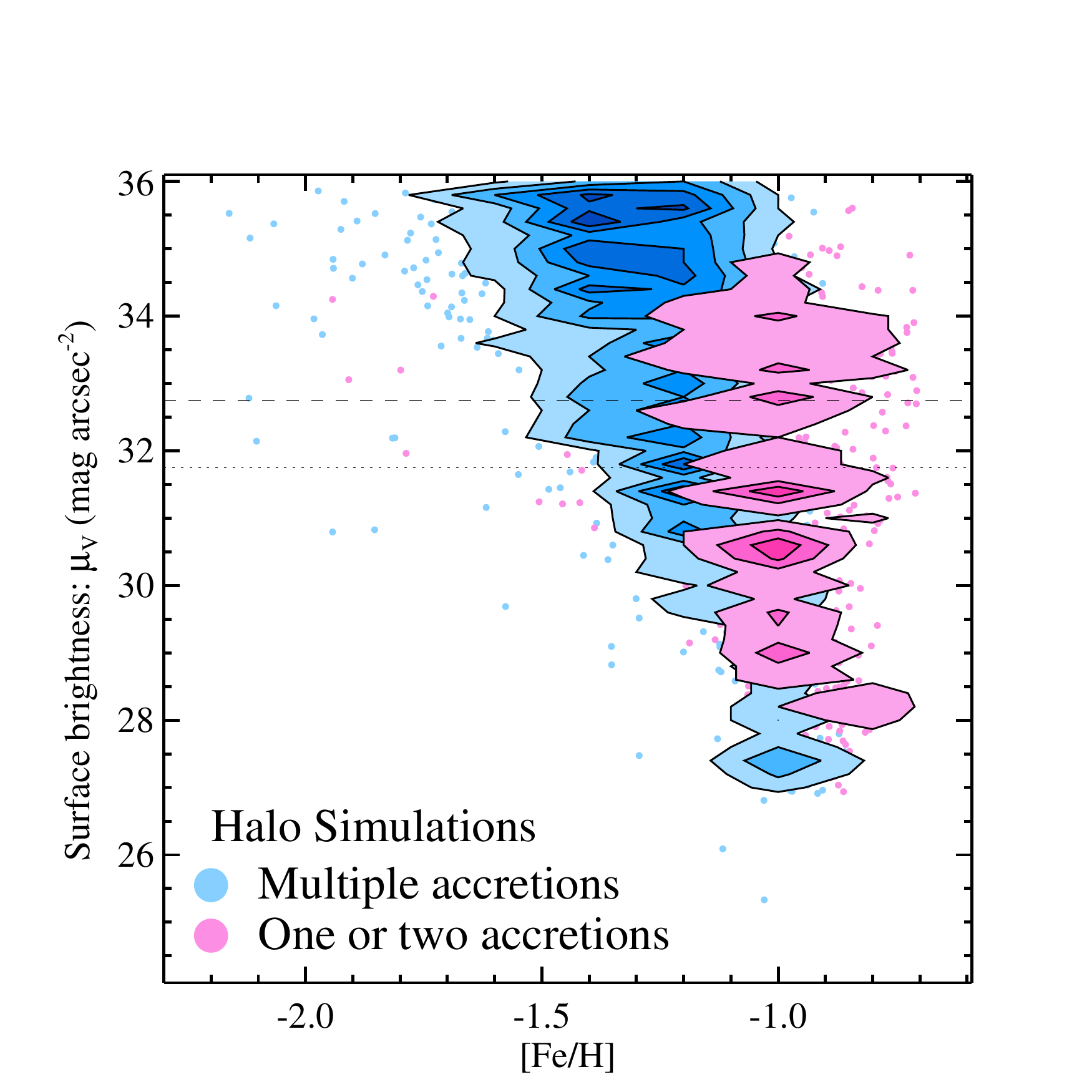}
\caption{ {\it Top:} Surface brightness vs.\ median metallicity of individual kinematic components in each of the M31 halo fields.  Purple points denote measurements of stars belonging to the kinematically hot component in each field. The red stars and orange triangles are for individual kinematically cold components: red stars denote the values for lines of sight along the length of M31's Giant Southern Stream and its associated debris, while orange triangles are used for all other tidal debris features.   
{\it Bottom:} The surface brightness and metallicity of lines of sight in the \citet{robertson2005,font2006} simulations, color coded by the prominence of individual tidal streams along the line of sight.  
Lines of sight where the stellar population is dominated by one or two accretion events (stars and triangles in the top panel, pink contours  in the bottom panel) are systematically more metal-rich at a given surface brightness than
lines of sight with multiple progenitors contributing roughly equally to the stellar population (purple points in the top panel, blue contours in the bottom  panel).
}
\label{fig:subst_v_smooth_g09b}
\end{figure}

The top panel of Figure~\ref{fig:subst_v_smooth_g09b} also displays surface brightness vs.\ \feh, however this time the M31 data has been parsed to show the surface brightness and metallicity of the separate individual kinematic components in each field.  
On average, at a given surface brightness kinematically cold tidal debris features are more metal rich than M31's diffuse, kinematically hot stellar halo.  This panel can be qualitatively compared to the bottom panel, which shows the surface brightness and metallicity of individual lines of sight through simulations.  The simulations are based on the \citet{bullock2005} $N$-body simulations of the formation of stellar halos around MW-mass galaxies built entirely via accretion.  The metallicities are computed by semi-analytic modeling of the $N$-body simulations \citep{robertson2005,font2006}.  Dividing the observed M31 fields into kinematically cold and hot components more closely approximates lines of sight dominated by one or two accretions or by multiple accretions, respectively, as defined in the simulations \citep[see][]{gilbert2009a}.  
Both the M31 halo observations and the simulations show the same tendency for lines of sight dominated by one or two strong tidal debris features to be more metal-rich at a given surface brightness than other lines of sight.

The observed and simulated distributions seen in Figure~\ref{fig:subst_v_smooth_g09b} have a fairly straightforward physical interpretation.  More massive satellites naturally produce brighter stellar streams upon tidal disruption in the host halo.  Given the observed mass-metallicity relation of dwarf galaxies \citep[e.g.,][]{kalirai2010,kirby2013}, more massive satellites will also produce more metal-rich stellar streams, thus the surface brightness and metallicity of tidal debris features are correlated.  This also implies that lines of sight dominated by debris from one or two accretion events (e.g., relatively bright stellar streams) will tend to be more metal-rich than lines of sight without a single dominant progenitor to the stellar population (e.g., lines of sight which are a superposition of many, relatively faint stellar streams).   

However, the present surface brightness of tidal debris features is determined by multiple physical mechanisms, including the orbit of the progenitor and the time since accretion.  Therefore, a significant spread in metallicities at a given surface brightness among different lines of sight is expected.   While the highest surface brightness features are only produced by the most recently accreted massive progenitors, faint stellar streams can be formed by either the early accretion of a satellite whose tidal debris has spatially dissipated, or by the recent accretion of a low-mass progenitor, or both.  Thus, an increasing spread in chemical properties with decreasing surface brightness among individual lines of sight, as seen in Figure~\ref{fig:subst_v_smooth_g09b}, is also expected.  

These effects are discussed in length by \citet{gilbert2009a}.  Figures~\ref{fig:subst_v_smooth_g09a} and \ref{fig:subst_v_smooth_g09b} nicely reproduce the earlier results, which are based on a small subset of the fields included here.  The M31 observations and $N$-body simulations are in good agreement with respect to the relative differences in the metallicity and surface brightness properties of distinct tidal debris features and the underlying stellar population.  This provides further evidence that the observed properties of M31's stellar halo are consistent with being built largely through the accretion of dwarf galaxies.

\section{Conclusions}\label{sec:conclusion}
We have shown that the median metallicity of M31's stellar halo decreases by an order of magnitude over $\sim 100$~kpc, and that the metallicity profile shows a continuous decrease in metallicity over the range $10\le$\,\rproj\,$\le 100$~kpc (Section~\ref{sec:met_halo}).  
The metallicity distribution function of M31 stars has a clear peak at \feh\,$\sim -0.4$ for stars within 20~kpc in projected distance from M31's center, with a median \feh\,$= -0.5$, while the median metallicity of M31 halo stars beyond 90~kpc is \feh\,$= -1.4$. 
While metal-poor stars are present at all radii, their relative contribution to the population increases significantly with projected radius.  The removal of stars likely associated with kinematically identified tidal debris features does not reduce the gradient in metallicity, and in fact makes the observed gradient in the inner $\sim 30$~kpc more pronounced.    

The 37 lines of sight analyzed here are spread throughout three quadrants of M31's halo and over a radial range of 9\,--\,177~kpc, and include over 1500 spectroscopically confirmed M31 halo stars.  This is the most radially extended M31 halo metallicity study to date, presenting over a factor of two increase in data compared to previous studies based on spectroscopically confirmed M31 stars. 
Careful spectroscopic selection significantly decreases the contamination rate of MW stars in the sample, something that is a struggle for purely photometric studies that must statistically subtract a MW foreground population \citep[e.g.,][]{richardson2009} or significantly restrict the color range of stars to limit contamination \citep[e.g.,][]{tanaka2010,ibata2014}.  In the low density regions of M31's outer halo, MW foreground stars greatly outnumber M31 RGB stars, thus the ability to identify M31 stars is paramount for studies of the extended regions of M31's halo.  

The metallicity profile of M31's stellar halo is largely consistent with the results of the most recent simulations of stellar halo formation, which include baryonic processes and produce a stellar halo composed both of accreted stars and stars formed in the host galaxy.  In the simulations, a large-scale gradient of the magnitude and radial range seen in M31's halo is most likely to form from the accretion of relatively massive progenitors. 
However, the field to field variation seen in both the surface brightness and metallicity profiles indicate that multiple smaller progenitors are also likely to have contributed substantially to the outermost regions of M31's halo.  

\acknowledgments

Support for this work was provided by NASA through Hubble Fellowship 
grants 51273.01, and 51316.01 awarded to K.M.G. and E.J.T. by the Space Telescope 
Science Institute, which is operated by the Association of Universities for 
Research in Astronomy, Inc., for NASA, under contract NAS 5-26555.
 P.G., J.S.B., S.R.M., and R.L.B. acknowledge support from collaborative NSF grants AST-1010039, AST-1009973, AST-1009882, and AST-0607726. This project was also supported by NSF grants AST03-07842, AST03-07851, AST06-07726, AST08-07945, and AST10-09882, NASA grant HST-GO-12105.03 through STScI, NASA/JPL contract 1228235, the David and Lucile Packard Foundation, and the F. H. Levinson Fund of the Peninsula Community Foundation (S.R.M., R.J.P., and R.L.B.). E.N.K acknowledges support from the Southern California Center for Galaxy Evolution, a multicampus research program funded by the University of California Office of Research, and partial support from NSF grant AST-1009973.  E.J.T. acknowledges support from a Graduate Assistance in Areas of National Need (GAANN) fellowship. 
 R.L.B. acknowledges receipt of the Mark C. Pirrung Family Graduate Fellowship from the Jefferson Scholars Foundation and a Fellowship Enhancement for Outstanding Doctoral Candidates from the Office of the Vice President of Research at the University of Virginia. M.T. acknowledges support from Grant-in-Aid for Scientific Research (25800098) of the Ministry of
Education, Culture, Sports, Science, and Technology of Japan.  The analysis pipeline used to reduce the DEIMOS data was developed at UC Berkeley with support from NSF grant AST-0071048.

The authors recognize and acknowledge the very significant cultural role and reverence that the summit of Mauna Kea has always had within the indigenous Hawaiian community. We are most fortunate to have the opportunity to conduct observations from this mountain.


\bibliography{m31}

\end{document}